\documentclass{aa}  
\usepackage{amsmath}
\usepackage{amssymb}
\usepackage{graphicx}
\usepackage{graphics}
\usepackage{epsfig}
\usepackage{blindtext}
\usepackage{float}
\usepackage{siunitx}
\usepackage{bm}
\usepackage{placeins}
\usepackage{longtable}
\usepackage{threeparttable}
\usepackage{adjustbox}
\usepackage{lscape}
\usepackage{enumitem}
\usepackage[usenames, dvipsnames]{color}
\usepackage{flafter}

\usepackage{txfonts}
\usepackage[
pdfstartview=XYZ,
bookmarks=true,
colorlinks=true,
linkcolor=blue,
urlcolor=blue,
citecolor=blue,
pdftex,
bookmarks=true,
linktocpage=true, % makes the page number as hyperlink in table of content
]{hyperref}

\newcommand\T{\rule{0pt}{2.6ex}}       % Top strut
\newcommand\B{\rule[-1.2ex]{0pt}{0pt}} % Bottom strut

\newcommand{\Lya}{Ly$\alpha$\xspace}
\newcommand{\Ha}{H$\alpha$\xspace}
\newcommand{\Hb}{H$\beta$\xspace}
\newcommand{\farc}{\hbox{$.\!\!^{\prime\prime}$}}

\newcommand{\kms}{\,km\,s$^{-1}$\xspace}
\newcommand{\Msunyr}{\hbox{M$_{\rm{\odot}}\,\rm{yr^{-1}}$}\xspace}

\newcommand{\ergscmA}{erg\,s$^{-1}$\,cm$^{-2}$\,\AA$^{-1}$\xspace}
\newcommand{\ergscm}{erg\,s$^{-1}$\,cm$^{-2}$\xspace}
\newcommand{\lognhicm}{$\log(N_{\rm H{\tiny I}}$/cm$^{-2})$\xspace}
\newcommand{\HI}{H{\tiny I}\xspace}
\newcommand{\HII}{H{\tiny II}\xspace}
\newcommand{\NHI}{$\rm N_{H{\tiny I}}$\xspace} 
\newcommand{\NHIOA}{$\rm N_{H{\tiny I}}^{OA}$\xspace} 
\newcommand{\fescLyC}{$f_{\rm esc}(\rm LyC)$\xspace} 
\newcommand{\fescLya}{$f_{\rm esc}$(\Lya)\xspace} 

\newcommand{\logNHI}{\log (N_{\rm HI}/\mathrm{cm}^{-2})}

\usepackage{natbib,twoopt}
\bibpunct{(}{)}{;}{a}{}{,}             %% natbib format for A&A and ApJ
\makeatletter
  \newcommandtwoopt{\citeads}[3][][]{\href{http://adsabs.harvard.edu/abs/#3}%
    {\def\hyper@linkstart##1##2{}%
     \let\hyper@linkend\@empty\citealp[#1][#2]{#3}}}
  \newcommandtwoopt{\citepads}[3][][]{\href{http://adsabs.harvard.edu/abs/#3}%
    {\def\hyper@linkstart##1##2{}%
     \let\hyper@linkend\@empty\citep[#1][#2]{#3}}}
  \newcommandtwoopt{\citetads}[3][][]{\href{http://adsabs.harvard.edu/abs/#3}%
    {\def\hyper@linkstart##1##2{}%
     \let\hyper@linkend\@empty\citet[#1][#2]{#3}}}
  \newcommandtwoopt{\citeyearads}[3][][]%
    {\href{http://adsabs.harvard.edu/abs/#3}
    {\def\hyper@linkstart##1##2{}%
     \let\hyper@linkend\@empty\citeyear[#1][#2]{#3}}}
\makeatother
\begin{document}

   \title{Gamma-ray bursts as probes of \\
   high-redshift Lyman-$\alpha$ emitters and radiative transfer models}

\author{
J.-B.~Vielfaure\inst{1,2},
S. D.~Vergani\inst{2},
M.~Gronke\inst{3}\thanks{Hubble fellow},
J.~Japelj\inst{4},
J.~T.~Palmerio\inst{2}, 
J.~P.~U.~Fynbo\inst{5,6},
D.~B.~Malesani\inst{7},
B.~Milvang-Jensen\inst{5,6},
R.~Salvaterra\inst{8},
N.~R.~Tanvir\inst{9}
}

\institute{
\inst{1} Université de Paris, CNRS, AstroParticule et Cosmologie, F-75013, Paris \\
\inst{2} GEPI, Observatoire de Paris, PSL University, CNRS, 5 Place Jules Janssen, 92190 Meudon, France\\
\inst{3} Department of Physics \& Astronomy, Johns Hopkins University, Baltimore, MD 21218, USA\\
\inst{4} Anton Pannekoek Institute for Astronomy, University of Amsterdam, Science Park 904, 1098 XH Amsterdam, The Netherlands \\
\inst{5} Cosmic Dawn Center (DAWN)\\
\inst{6} Niels Bohr Institute, University of Copenhagen, Jagtvej 128, 2100 Copenhagen \O, Denmark\\
\inst{7} DTU Space, National Space Institute, Technical University of Denmark, DK-2800 Kongens Lyngby, Denmark \\
\inst{8} INAF – IASF/Milano, via Corti 12, I-20133 Milano, Italy \\
\inst{9} Department of Physics \& Astronomy and Leicester Institute of Space \& Earth Observation, University of Leicester, University
Road, Leicester LE1 7RH, UK \\
}

   \date{Received ... / Accepted ... }
 
  \abstract
  {}
{We present the updated census and statistics of Lyman-$\alpha$ emitting long gamma-ray bursts host galaxies (LAE-LGRBs). We investigate the properties of a sub-sample of LAE-LGRBs and test the shell model commonly used to fit Lyman-$\alpha$ (\Lya) emission line spectra.}
{To perform the census we use all the LGRB host galaxies with relevant information presented in the literature or publicly available. Among the LAE-LGRBs detected to date, 
we select a {\it golden sample} of four LAE-LGRBs (GRBs: 011211, 021004, 060926, 070110) allowing us to retrieve information on the host galaxy properties and of its interstellar medium gas, through the combination of the analysis of their afterglow and host galaxy observations. We fit their \Lya spectra using the shell model, and constrain its parameters with the observed values.}
{The census results in 29 LAE-LGRBs detected to date. Among them, we present five new \Lya emission detections in host-galaxy spectra and the corresponding unpublished VLT/X-shooter data (GRBs: 060926, 070110, 081121, 081222 and 100424A). From the comparison of the statistics and properties of LAE-LGRBs to
those of LAE samples in the literature,
we find evidences of \Lya suppression in dusty systems, and 
a fraction of LAE-LGRBs among the overall LGRB hosts lower than that found for Lyman-break galaxy (LBG) samples at similar redshift range. This result can be explained by the different selection criteria of the parent samples and by the shallower spectral observations of LGRB samples compared to LBG ones.
However, we find that LAE-LGRBs are representative of \Lya emission from the bulk of UV-selected galaxies at $z\sim2$. 
We find that the {\it golden sample} of LAE-LGRBs studied here are complex systems characterized by multiple emission blobs and by signs of possible galaxy interactions.
The fitting procedure recovers the \HI column densities ($\rm N_{HI}$) measured from the afterglow spectra, and the other properties described by the shell-model parameters in the two low-$\rm N_{HI}$ cases, but it fails in doing so in the other two with high $\rm N_{HI}$.
The afterglows of most LGRBs and LAE-LGRBs show $\logNHI>20.3$, implying that statistically the bulk of \Lya photons expected to be produced by massive stars in the star-forming region hosting the GRB will be surrounded by such opaque lines of sight.
We therefore interpret our results in the context of more sophisticated models and of different dominant \Lya emitting regions. 
We also compare LAE-LGRBs to LAE Lyman continuum (LyC) leakers in the literature in terms of properties identified as possible indirect indicators of LyC leakage. We find that only one LGRB (GRB\,021004) would be a likely strong LyC leaker and discuss the validity of such indicators at high redshift.}
{}

\keywords{gamma-ray burst: general – galaxies: star formation – galaxies: ISM – galaxies: evolution – galaxies: high-redshift - line: profiles}

\titlerunning{LAE-LGRBs and models}
\authorrunning{J-B. Vielfaure et al.}
   \maketitle
%
%-------------------------------------------------------------------
\section{Introduction} \label{Intro}

Due to its brightness and rest-frame wavelength, the Lyman-$\alpha$ (\Lya) emission line is one of the most used features to detect high-redshift galaxies \citep[e.g.,][]{Ouchi2009,Sobral2015,Zitrin2015,Bagley2017}.
The natural connection of this line with the UV emission from star-forming regions
makes it an interesting proxy to study the escape of Lyman continuum (LyC; <912 \AA). Indeed, recent studies, such as those of \citet{Verhamme2015,Verhamme2017}, show that this line is one of the most reliable indirect indicators of ionizing photon leakage. 

To escape a galaxy, the \Lya photons produced in star-forming regions have to pass through the gas where they are embedded. As this radiation resonantly scatters in the presence of neutral hydrogen and is easily absorbed by dust, the journey of photons in the interstellar and circumgalactic medium (ISM and CGM, respectively) can be complex. Nevertheless, different properties can favour their escape such as low neutral-hydrogen (\HI) column densities, low dust content or suitable ISM geometries and kinematics \citep[e.g.,][]{Kunth1998,Shapley2003,Verhamme2008,Wofford2013,Henry2015,RiveraThorsen2015}. 
As a consequence, the \Lya line flux and profile reflect the signatures of the physical and dynamical properties of the gas and dust content of the \Lya emitters and their surrounding environment. 
Interpreting the line is complex, and radiative transfer models which take into account the different sources of distortion of the intrinsic profile are necessary to recover the physical meaning of the observed line.
A simple and successful model, commonly used to reproduce the \Lya shape, is the shell model \citep[e.g.,][]{Ahn2004,Verhamme2006,Schaerer2011,Gronke2015}. 
It consists of an homogeneous expanding shell of neutral hydrogen and dust surrounding a central emitting source. 

While successful in reproducing the line profile \citep[see e.g.,][]{Verhamme2008, Lidman2012, Yang2017, Gronke2017}, it is important to test whether the best-fit parameters values of the shell model correspond to the real characteristics of the \Lya-emitting galaxies.
\citet{Orlitova2018} highlighted discrepancies between modelling results and observed double-peaked \Lya lines, by constraining independently 5 out of the 7 shell-model parameters with ancillary data,
for twelve Green Pea (GP) galaxies at $z \sim$\,0.2.
In particular, the constrained model neither reproduces the observed blue peak of the line correctly nor, in half of the cases, the red peak. For the prediction of the parameters in the unconstrained case, the main discrepant values are the redshift, the $\rm FWHM_i(Ly\alpha)$ and the velocity expansion of the shell. 
Similar discrepancies for the $\rm FWHM_i(Ly\alpha)$ were also found by \citet{Hashimoto2015} 
for double peak \Lya profiles of galaxies at $z \sim$\,2.2.
This work emphasises that the use of the shell model to interpret the \Lya line and retrieve physical properties, such as $\rm N_{HI}$, must 
proceed cautiously to avoid misinterpretation. 
It also suggests that considering an homogeneous shell to describe star-forming regions and their surrounding gas could be too simplistic.

The simultaneous availability of the information needed to constrain the model parameters is rare, especially at high redshift. Two individual studies of lensed galaxies at redshift $z=2.7$ allowed to interpret the \Lya line using partially constrained shell model \citep{Schaerer2008,DessaugesZavadsky2010}. The fitting of the \Lya line is in good agreement with the observation in the study of \citet{DessaugesZavadsky2010}, while it requires different expansion velocities for the front and the back of the modelled shell in \citet{Schaerer2008}.
 
Gamma-ray bursts (GRBs) can be a useful additional tool to investigate \Lya emission and test the shell model at high redshift.
GRBs are the most extreme cosmic electromagnetic phenomena (see \citealt{Gehrels2013a} for a review). Their brightness makes them powerful probes through the cosmic history since they can be detected up to the highest redshifts (the spectroscopic record holder is GRB\,080423 at $z = 8.2$; \citealt{Salvaterra2009,Tanvir2009}).
In the case of long GRBs (LGRBs), the energy powering the bursts is released during the core-collapse of massive stars \citep[e.g.,][]{Hjorth2003a}. 
In addition, several studies indicate that LGRBs have the tendency to occur in dwarf galaxies with high specific star-formation rates and prefer low-metallicity environments, typically sub-solar \citep[e.g.,][]{Perley2016b,Japelj2016a,Graham2017,Vergani2017,Palmerio2019}. 
This makes LGRB hosts likely representative of the common galaxies at high redshift,
including during the epoch of reionization \citep{Salvaterra2011, Salvaterra2013, Tanvir2019}.

The bright afterglows associated with LGRBs provide ideal background lights to probe systematically and at any redshift the ISM, CGM and inter-galactic medium (IGM) along the line of sight of this population of faint galaxies.
The absorption present in the afterglow spectra directly traces the environment of the star-forming regions and also outflows or inflows even for the faintest objects. Once the afterglow has faded, the host galaxy can be directly observed through photometry and spectroscopy.
This offers the interesting possibility of combining information on the cold and warm gas with the emission properties of the GRB host galaxy
\citep[e.g.,][]{Vergani2011a,Chen2012,Friis2015,Wiseman2017b,Arabsalmani2018b}.

In this work, (i) we update the statistics of \Lya-emitting (LAE\footnote{In the literature \Lya-emitting galaxies are usually defined as \Lya emitters (LAEs) when their rest-frame \Lya emission equivalent width ($\rm EW_{0}$(\Lya)) is above a certain threshold (typically 20~\AA) because historically they were selected from narrow band observations. In this study, we qualify \Lya-emitting galaxies as LAEs independently of their ($\rm EW_{0}$(\Lya)).}) 
LGRB host galaxies and compare their properties with those of LGRB hosts in general, and of LAEs and LyC leakers in the literature; (ii) we use a {\it golden sample} of four LAE-LGRBs at  $2 < z < 3.2$ allowing the combination of the emission properties of the host galaxies
with the information on the ISM probed by the afterglow, to investigate the properties of such systems and test the \Lya radiative transfer modelling.   

The paper is organized as follows. In Section \ref{LAEdetections}, we present a statistical study on LAEs in LGRB hosts. We describe the physical properties of the host galaxies of our {\it golden sample} in Section \ref{GoldSample} and the \Lya radiative transfer model results in Section \ref{LyaLine}. In section \ref{Discussion}, we discuss the differences between model predictions and observations, and we compare LAE-LGRBs to LGRB hosts in general and to LAEs and LyC-leaker galaxies in the literature. We draw our conclusions in Section \ref{Conclusions}.

All errors are reported at 1$\sigma$ confidence unless stated otherwise.
We consider a $\rm \Lambda CDM$ cosmology with the cosmological parameters provided in \citet{Planck2016}: $\rm H_0 = 67.8 km\ s^{-1} Mpc^{-1}$, $\rm \Omega_m = 0.308$ and $\rm \Omega_{\Lambda} = 0.692$.

\section{LAE detections in LGRB systems} \label{LAEdetections}

\subsection{Previous studies and approach}

The early studies of high-redshift GRB host galaxies, based on the first sample of five objects, \citep{Kulkarni1998,Fynbo2002,Moller2002,Fynbo2003,Vreeswijk2004}, seemed to indicate that all GRB hosts were LAEs.
A subsequent systematic study, based on the larger TOUGH sample (\citealt{Hjorth2012}; sixty-nine LGRB host galaxies), has been carried out by \citet{MilvangJensen2012}. 
They targeted a sub-sample of twenty LGRB hosts in the redshift range $z=1.8-4.5$ 
for VLT/FORS1 \citep{Appenzeller1998} spectral observations. They found seven LGRB hosts with significant \Lya emission (3$\sigma$ detection), corresponding to 35\% LAEs.

The first step of our work is to update the census of LAE-LGRBs. To this end we considered two approaches: (i) a determination of the LAE statistics considering spectroscopic samples of LGRB host galaxies or afterglows; (ii) the search for LAE-LGRBs in the literature and from the host-galaxy spectra available in the ESO archive\footnote{\url{http://archive.eso.org/cms.html}}. In both cases, we consider a minimum value of $z = 1.6$ for the LGRBs, which corresponds to the atmospheric UV cutoff at 310\,nm and is the lower-redshift limit allowing the detection of the \Lya line in the VLT/X-shooter spectra\footnote{In our study the VLT/X-shooter spectrograph \citep{Vernet2011} is particularly interesting for its wide spectral coverage (from $\sim300$\,nm to 2500\,nm) allowing the simultaneous detection of absorption lines in the ISM of the host galaxy and associated nebular emission lines for a large range of redshifts, with a medium spectral resolution of R $\sim 5000$.}.
The samples considered for point (i) are the TOUGH, the {\it X-shooter host-galaxy sample} (\citealt{Kruhler2015}, XHG in the following) and the {\it X-shooter afterglow sample} (\citealt{Selsing2019}; XAFT in the following).

We stress that the census presented in the following sections (especially in Sect.\,\ref{overall}) may not reflect the general statistics of LAEs among LGRB host galaxies.
In addition to issues concerning the completeness of the samples and of the observations, the spectra are generally not homogeneous in terms of exposure times,  instruments and observing conditions. Nonetheless, sometimes it is possible to define a common flux limit, as in the case of \cite{MilvangJensen2012} sample.
The case of afterglow spectra is even more complex as, in addition to inhomogeneous follow-up and target brightness, the \Lya absorption along the GRB line of sight could affect the \Lya emission detection and profile.

\subsection{Data reduction} \label{Data_reduction}

To perform the census, we reduced archival data of several X-shooter spectra of GRB host galaxies. We describe here the method applied for the data reduction.
All observations of the host galaxies, the telluric stars, and the spectrophotometric standards were reduced in the same way using the version 2.8.5 of the X-shooter data reduction pipeline \citep{Modigliani2010}. Before processing the spectra through the pipeline, the cosmic-ray hits and bad-pixels were removed following the method of \cite{VanDokkum2001}.
Then, we subtracted the bias from all raw frames and divided them by the master flat field. We traced the echelle orders and calibrated the data in spatial and wavelength units using arc-line lamps. The flux calibration was done using spectrophotometric standards \citep{Vernet2009} and a correction for flexure was applied. Lastly, the sky-subtraction and the rectification and merging of the orders was done to obtain the final two-dimensional spectra.
Additionally, the spectra were corrected for the Milky Way (MW) extinction using the extinction curve from \citet{Pei1992}. The $\rm A_V$ values are obtained from the NASA Extragalactic Database (NED) and correspond to the extinction map of \citet{Schlafly2011}.
The wavelengths of the extracted 1D-spectra were converted in the vacuum reference and corrected for the Earth's rotation and revolution around the Sun (heliocentric correction).
To optimally select the extraction regions we chose the spatial extension of the brightest emission line and applied this 1D extraction throughout the whole spectrum. 
When the \Lya line is detected, we selected the extraction region according to the spatial extension of this line which can be larger than Balmer lines due to resonant scattering.
Emission line fluxes were determined by fitting a Gaussian function to the data, setting the continuum flux density in a region close to the emission line. We also numerically integrated the flux over the line width as a comparison to control the consistency of the values and uncertainties. For the asymmetric line-profile of the \Lya line, we use a skewed Gaussian parametrized as described in \citet{Vielfaure2020}.
When lines of interest were not detected, we estimated a 3$\sigma$ upper limit. 
For upper limits of nebular emissions, we use a FWHM in agreement with other nebular lines detected in the spectrum. 
For \Lya line, similarly to \cite{MilvangJensen2012}, we select the same width for all upper limits which is 900 \kms centered at 300 \kms.
The fluxes have been corrected for slit loss by 
calculating the flux difference between the observation of a telluric star (close in time and space to the observation of the GRB host, and with the same instrumental setup) to the tabulated values\footnote{The tabulated values of the magnitudes for the telluric stars have been taken from \url{https://www.eso.org/sci/facilities/paranal/decommissioned/isaac/tools/spectra/Bstars.txt} and \url{http://simbad.u-strasbg.fr/simbad/}} expected to be measured.

\subsection{The TOUGH sample} \label{TOUGH_sample}

We first focus on the TOUGH sample of 69 GRBs. The TOUGH \Lya study \citep{MilvangJensen2012} targeted the 20 GRBs with a redshift in the range $z=1.8-4.5$, as known at the time of the \Lya observing campaign. 
Subsequently as part of other TOUGH campaigns \citep{Jakobsson2012, Kruhler2012} and later work \citep{Kruhler2015} the redshift completeness of TOUGH has increased to 60/69 \citep{Kruhler2015}. With respect to \citet{MilvangJensen2012}, this includes adding redshifts for eleven TOUGH GRBs in the range $z = 1.8-4.5$ (GRB: 050714B, 050819, 050915A, 051001, 060805A, 060814, 070103, 070129, 070224, 070328 and 070419B); all these hosts have VLT/X-shooter spectra.

We reduced the X-shooter data, as described in Sect. \ref{Data_reduction}, to look for the detection of \Lya emission.
We find no LAEs among these eleven additional host galaxies. The new statistic of LAEs among LGRB host galaxies of the TOUGH sample is therefore $23\% \pm 7\%$.

Taking into account the fact that the spectra have different flux limits (see Fig. \ref{FluxLimit} and \ref{FluxLimitcumul}), we determine the statistics of \Lya detection above a flux cut of $\rm 1.12 \times 10^{-17}\ erg\ s^{-1}\ cm^{-2}$ (corresponding to the lowest \Lya detection of the TOUGH sample), a luminosity cut of $\rm 5.6 \times 10^{41}\ erg\ s^{-1}$ (corresponding to the flux cut at the median redshift of the TOUGH sample, $z=2.45$), and the fraction of LAEs among the sample with a rest-frame \Lya EW $>20$ \AA. 
To estimate the uncertainty on these statistics, based on the sample sizes, we perform a bootstrap method employing $10^6$ random resamples with replacement of the number of LAEs among the LGRBs considered for each cut.
The results are reported in Table \ref{Tab_fesc}.

In principle, we can look for \Lya emission also at $z>4.5$. This would add three further objects from the TOUGH sample. They lack host galaxy spectral observations but these have afterglow spectra available in the literature (GRB\,050904: \citealt{Totani2006}; GRB\,060522: \citealt{Tanvir2019}; GRB\,060927: \citealt{Fynbo2009}). They show no \Lya emission but formal limits have not been determined.

\subsection{The XHG and XAFT samples} \label{XHG_XAFT_samples}
\label{XHG}

\begin{figure}[!ht]
\centering
    \includegraphics[width=0.9\hsize]{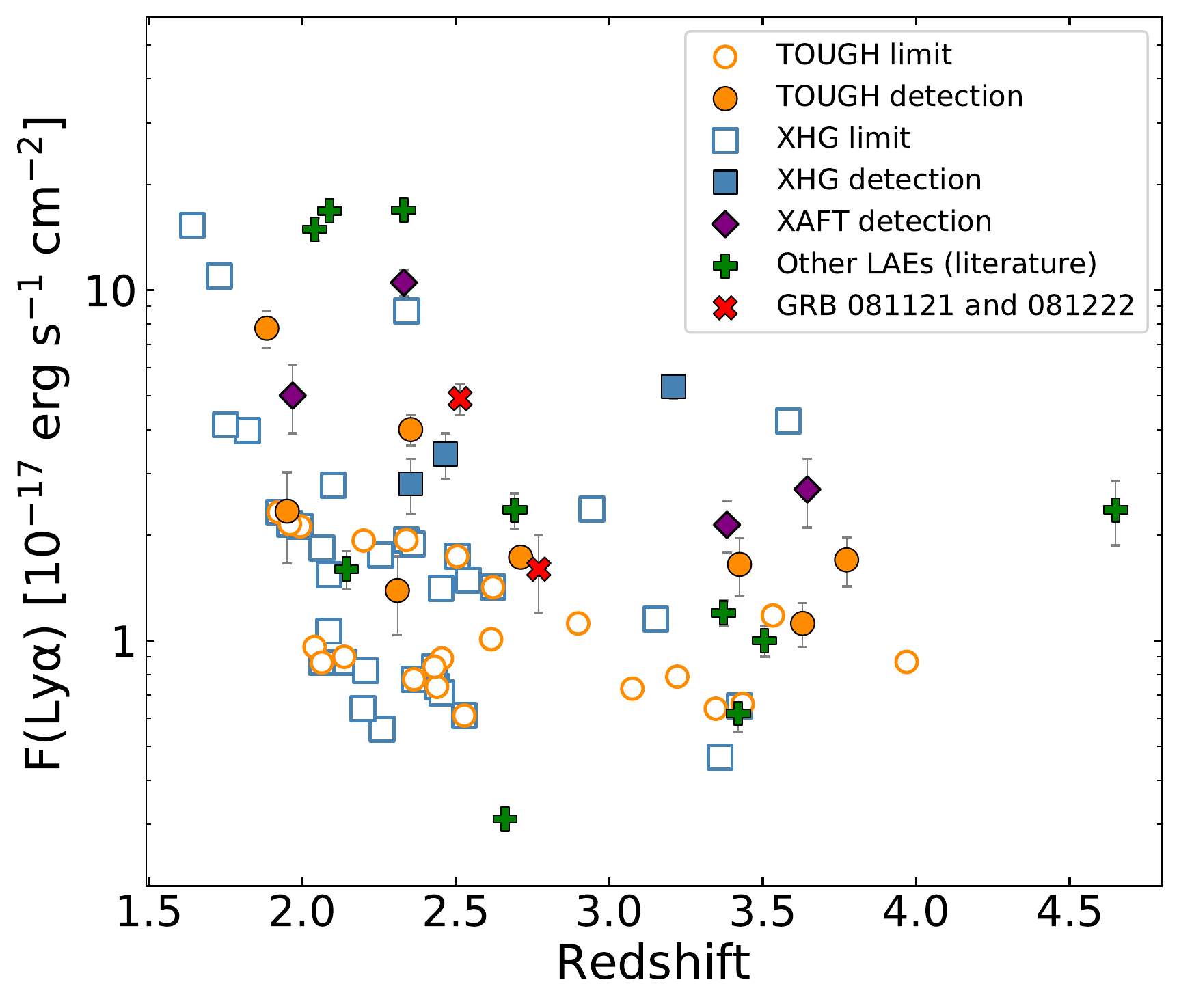}
    \includegraphics[width=0.9\hsize]{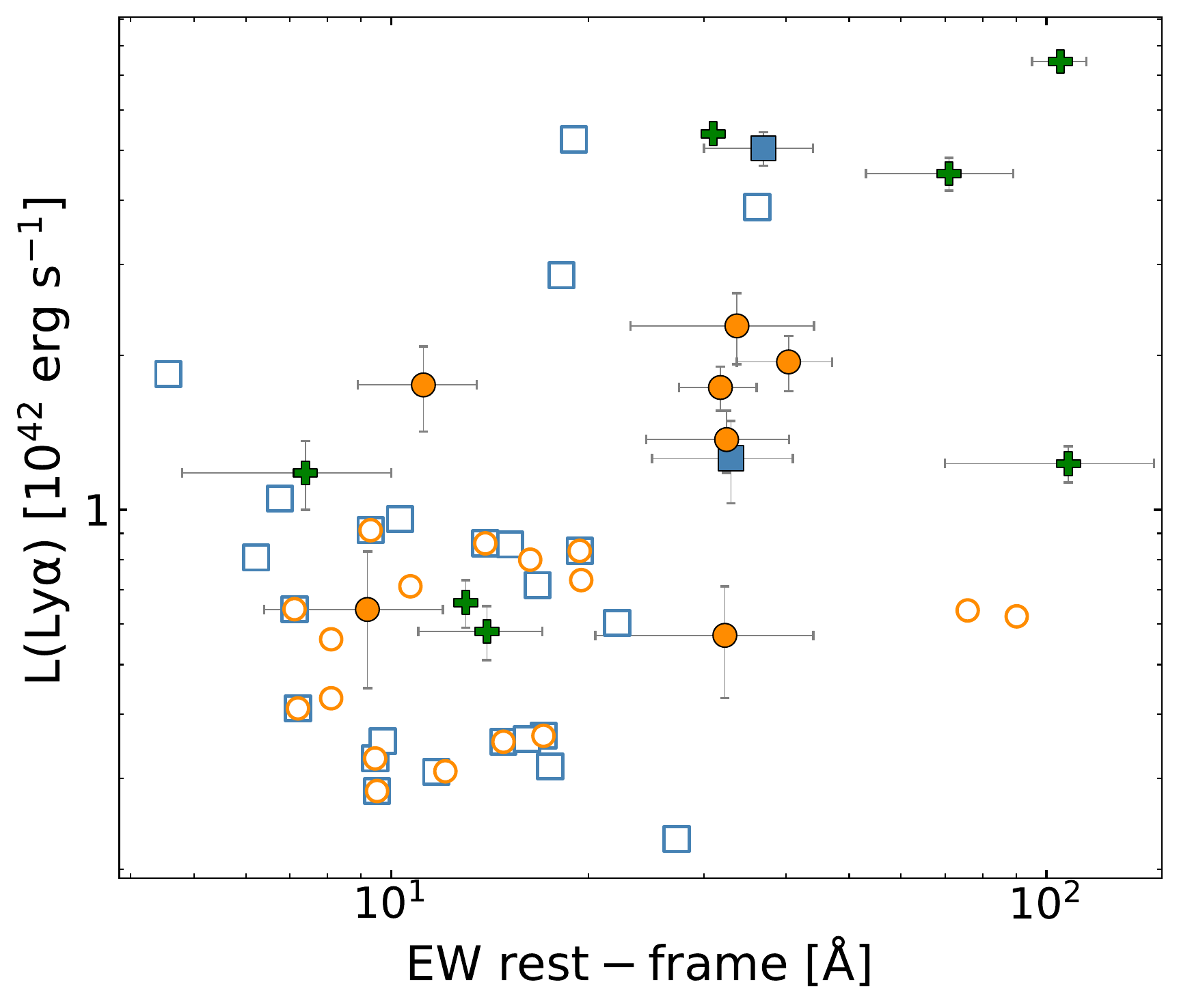}
 \caption{{\it Top Panel}: \Lya fluxes of LAE-LGRBs as a function of redshift. We report the fluxes retrieved from the literature or determined in this paper (see the tables in Appendix \ref{tablerecap}). For the TOUGH and XHG samples, we plot also the upper limits of the host galaxies with no \Lya emission detection (empty symbols). {\it Bottom Panel}: \Lya luminosity of the LAE-LGRBs as a function of rest-frame \Lya equivalent width ($\rm EW_{0}$(\Lya)). The sample and symbols used are the same as in the top panel. }
 \label{FluxLimit}
\end{figure}

\begin{figure}[!ht]
\centering
\includegraphics[width=\hsize]{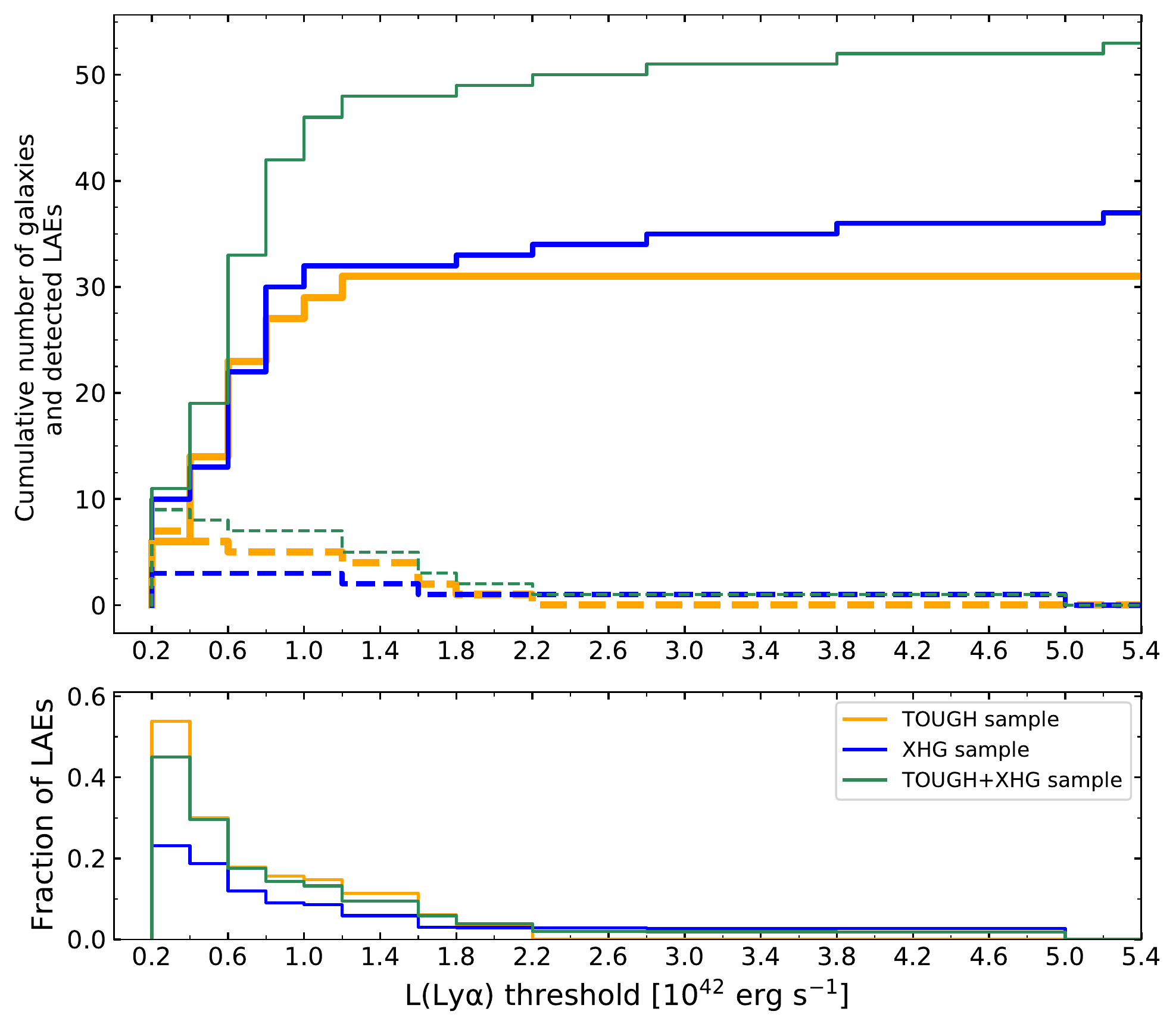}
 \caption{{\it Top Panel}: cumulative number of galaxies in the TOUGH (orange line), XHG (blue line), and merged samples (TOUGH+XHG, green) without overlapping GRBs, as a function of \Lya luminosity detection threshold (3$\sigma$) of the spectra. 
 The dashed lines represent the number of detected LAEs, above the \Lya luminosity threshold, in each sample. {\it Bottom Panel}: fraction of detected LAEs for each sample as function of \Lya luminosity threshold. }
 \label{FluxLimitcumul}
\end{figure}

\citet{Kruhler2015} presented the UVB-arm spectra for only three objects at $z > 1.6$ of their XHG sample.
Therefore, we reduce all the UVB spectra of the sample (following the procedure described in Sect. \ref{Data_reduction}) and inspect them to look for \Lya emission. We find three LAEs among the 37 host galaxy spectra at $z > 1.6$, corresponding to $8\% \pm 6\%$ (see also Fig. \ref{FluxLimit}).
They are GRB\,060926, GRB\,070110 and GRB\,100424A (also reported in \citealt{Malesani2013}). We will focus in more details on GRB\,060926 and GRB\,070110 in Sections \ref{GRB060926} and \ref{GRB070110}. 
The unpublished 1D and 2D spectra of the host of GRB\,100424A, showing its \Lya detection, are reported in Fig. \ref{2DLya}. The flux of the line corrected for Galactic extinction and slit loss is $F_{\rm Ly\alpha}= (3.4\pm0.5) \times 10^{-17}$\ erg\ s$^{-1}$\ cm$^{-2}$. 
The flux limits are much less homogeneous than for the TOUGH sample. 
We apply the same cuts as for the TOUGH sample and summarize the statistics in Table \ref{Tab_fesc}.

Finally, focusing on the XAFT sample, \citet{Selsing2019} report the detection of four LAEs (GRBs\,121201A, 150915A, 151021A, 170202A) among their X-shooter afterglow sample of 41 LGRBs, corresponding to $10\% \pm 5\%$.

\subsection{Overall detections} \label{overall}

\begin{table*}[h!]
\begin{center}
\caption{Fraction of LAEs among GRB host galaxies from the TOUGH and XHG samples.}
\label{Tab_fesc}
\centering
\small
\begin{tabular}{ c c c c c } \hline \hline
Sample &  f$_{tot}$    &  f$_{f}$ &  f$_{L}$ & f$_{EW_0}$  \T\B \\
\hline
\rule[0.2cm]{0cm}{0.2cm}TOUGH   & $23\% \pm 7\%$ (7/31)       &  $28\% \pm 8\%$ (7/25)    &  $39\% \pm 11\%$ (7/18)  & $17\% \pm 7\%$ (5/30)     \B \\
XHG             & $8\% \pm 6\%$ (3/37)     &  $19\% \pm 13\%$ (3/16)    &  $19\% \pm 13\%$ (3/16)    &   $7\% \pm 4\%$ (2/28)  \B \\
TOUGH + XHG     & $17\% \pm 6\%$ (9/53)     & $27\% \pm 9\%$ (9/33)     &   $36\% \pm 8\%$ (9/25)   &   $14\% \pm 5\%$ (6/42)   \B \\
\hline
\end{tabular}%
%}
\tablefoot{The first column corresponds to the name of the sample: TOUGH, XHG or merged samples without overlapping GRBs (TOUGH + XHG). 
 f$_{tot}$: fraction of LAEs among the whole sample; 
 f$_{f}$: fraction of LAEs among the sample with a flux cut of $\rm 1.12 \times 10^{-17}\ erg\ s^{-1}\ cm^{-2}$
 f$_{L}$: fraction of LAEs among the sample with a luminosity cut of $\rm 5.6 \times 10^{41}\ erg\ s^{-1}$ 
 f$_{EW_0}$: fraction of LAEs among the sample with a rest-frame \Lya EW $>20$ \AA.
Due to undetected continuum, seven \Lya $\rm EW_{0}$ measurements are missing for the TOUGH sample and eleven for the XHG sample. The fractions between brackets correspond to the number of LAEs over the size of the sample after applying the corresponding cuts. The uncertainty on the percentage of LAEs is calculated by bootstrap method (see Sect. \ref{TOUGH_sample}). }
\end{center}
\end{table*}

%% 2D Lya GRB 081121 and 081222
\begin{figure}[]
\centering
\includegraphics[width=\hsize]{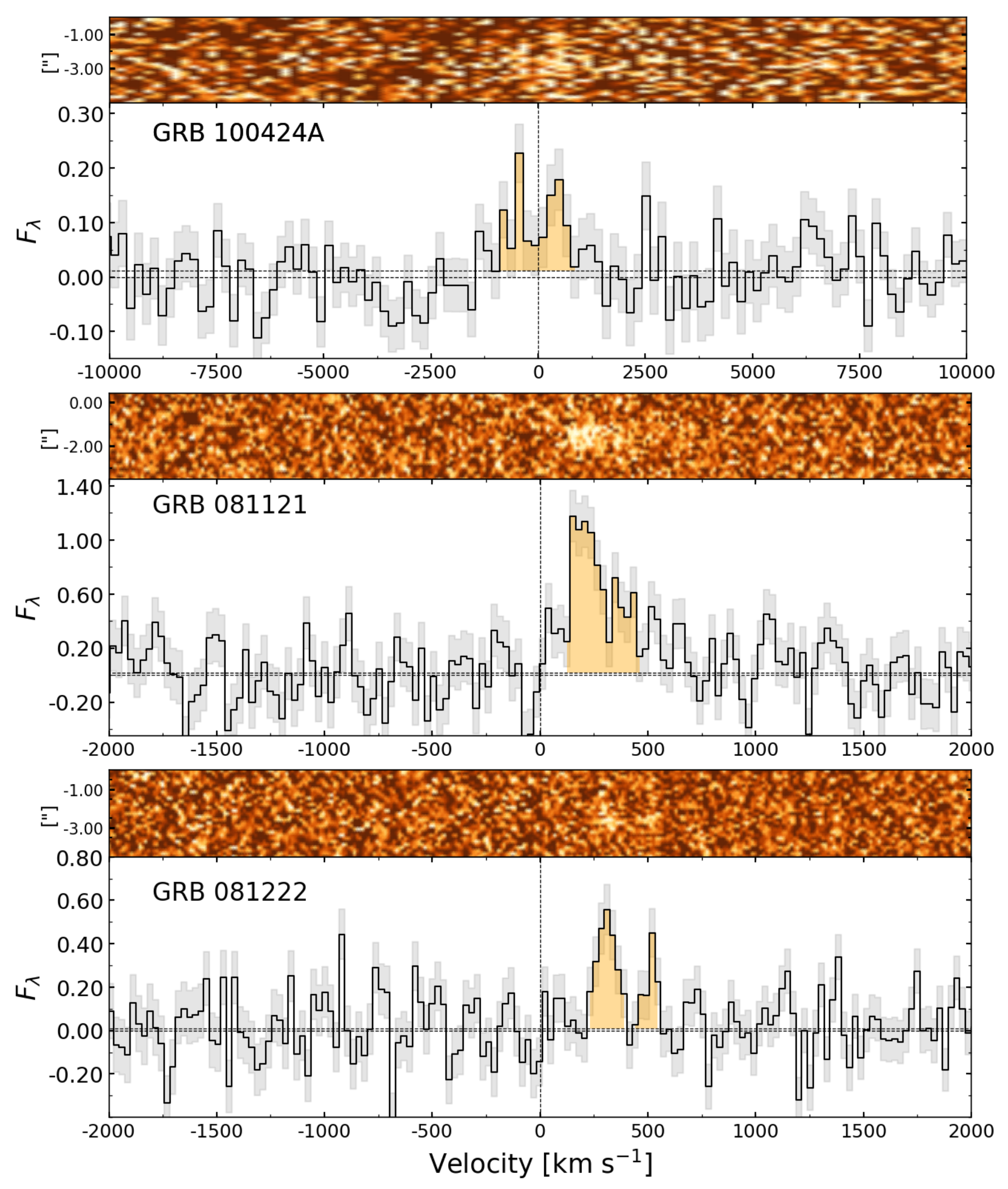}
 \caption{Section of the 2D and 1D X-shooter-UVB spectra, not previously published, showing the \Lya emission line from the host galaxy of GRB\,100424A (top), GRB\,081121 (middle) and GRB\,081222 (bottom). The lines are plotted in velocity frame centered at the systemic redshift of the galaxy (except for GRB\,081222, centered on the redshift determined from the GRB afterglow absorption lines, as no other emission lines are detected). The flux density $F_{\lambda}$ is in units of $\rm 10^{-17} erg\ s^{-1}\ cm^{-2}$ \AA$^{-1}$.  }
 \label{2DLya}
\end{figure}

%% HISTOGRAM
\begin{figure}[!ht]
\centering
\includegraphics[width=\hsize]{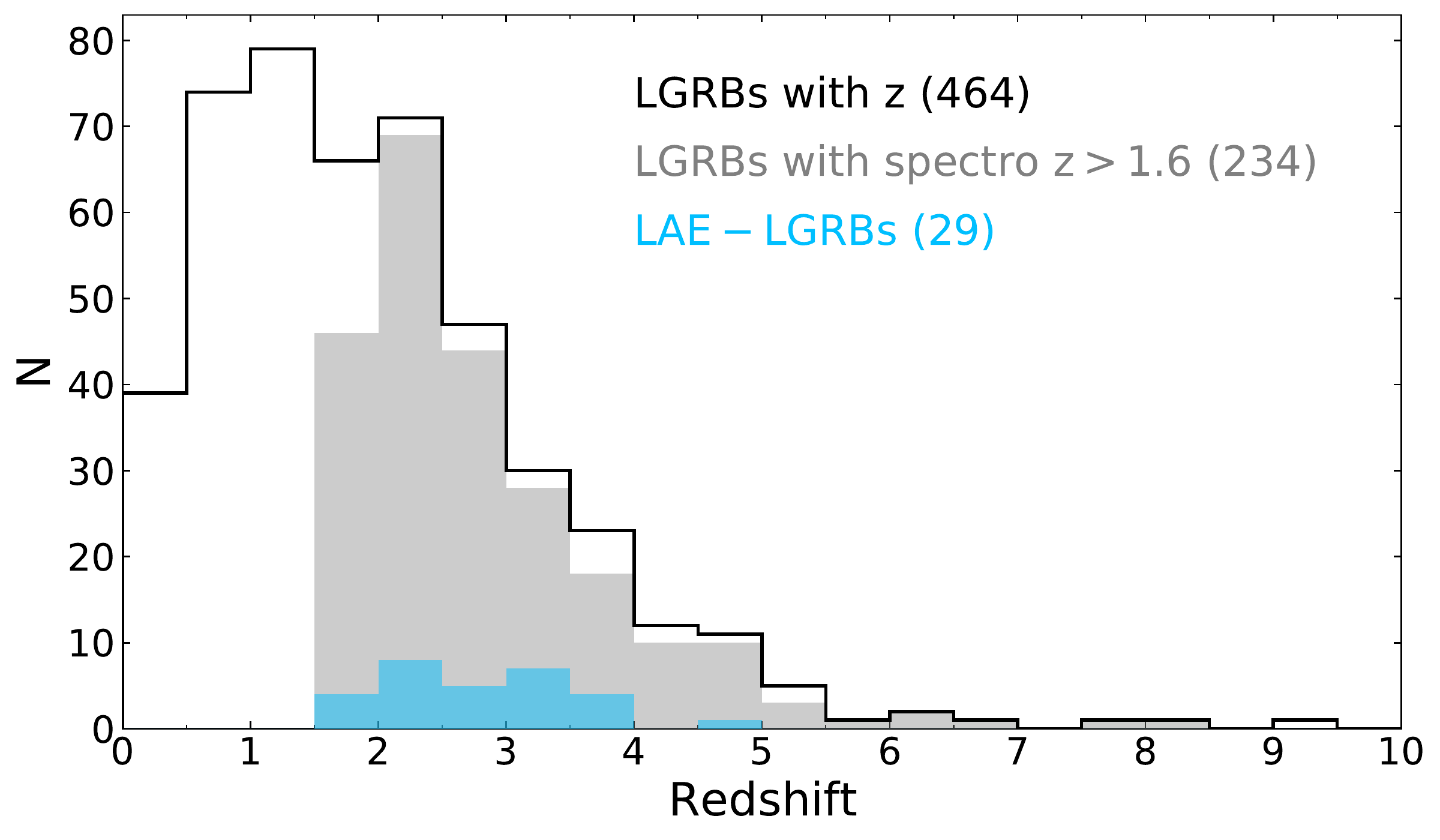}
 \caption{Redshift distribution of the 29 LAE-LGRBs (blue) 
among the 464 LGRBs with a spectroscopic or photometric redshift measurement (black line; as up to December 1 2020). 
The histograms are superimposed and not cumulative.
}
 \label{histo}
\end{figure}

\begin{figure*}[]
    \centering
    \includegraphics[width=6cm]{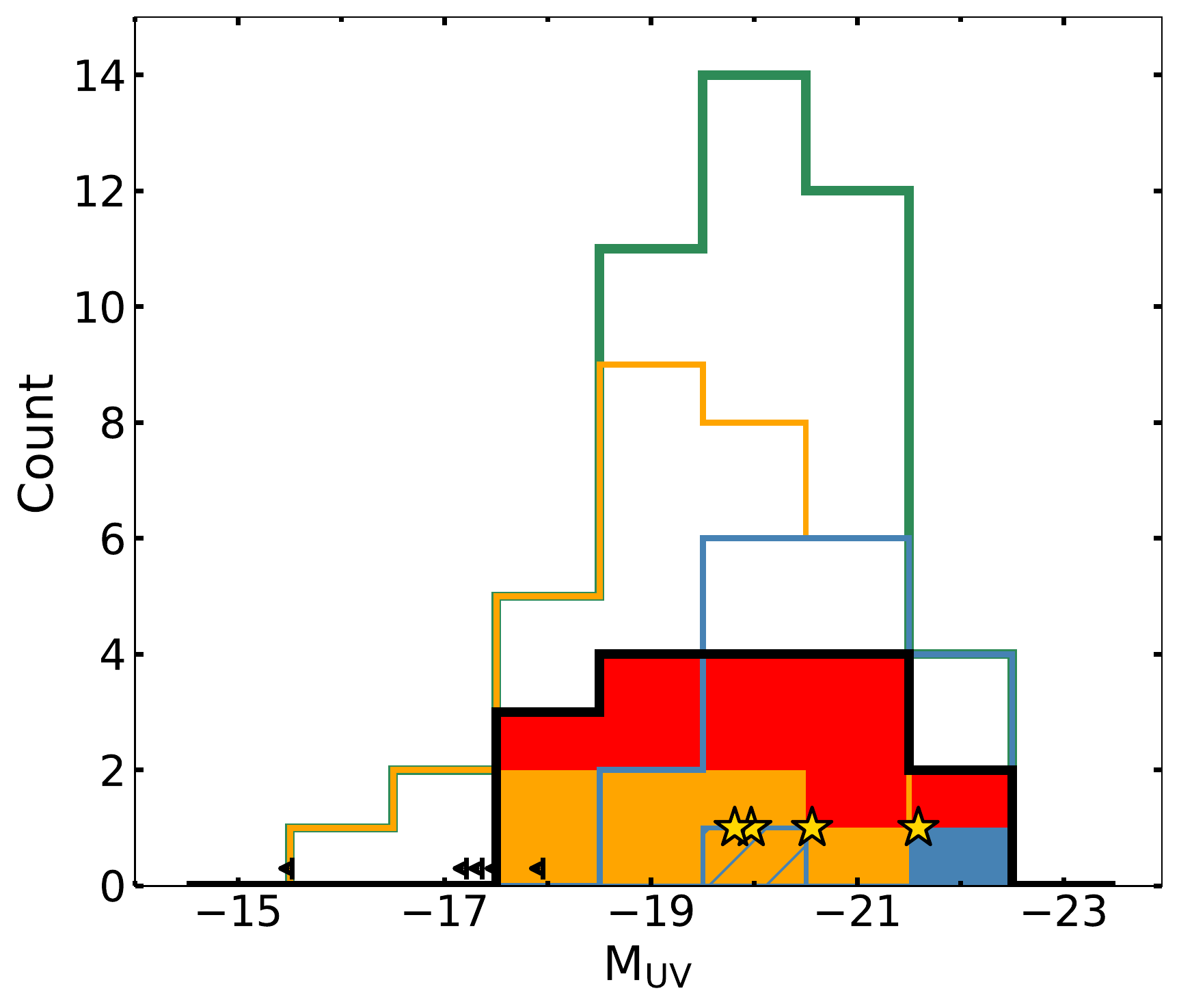}
    \includegraphics[width=6cm]{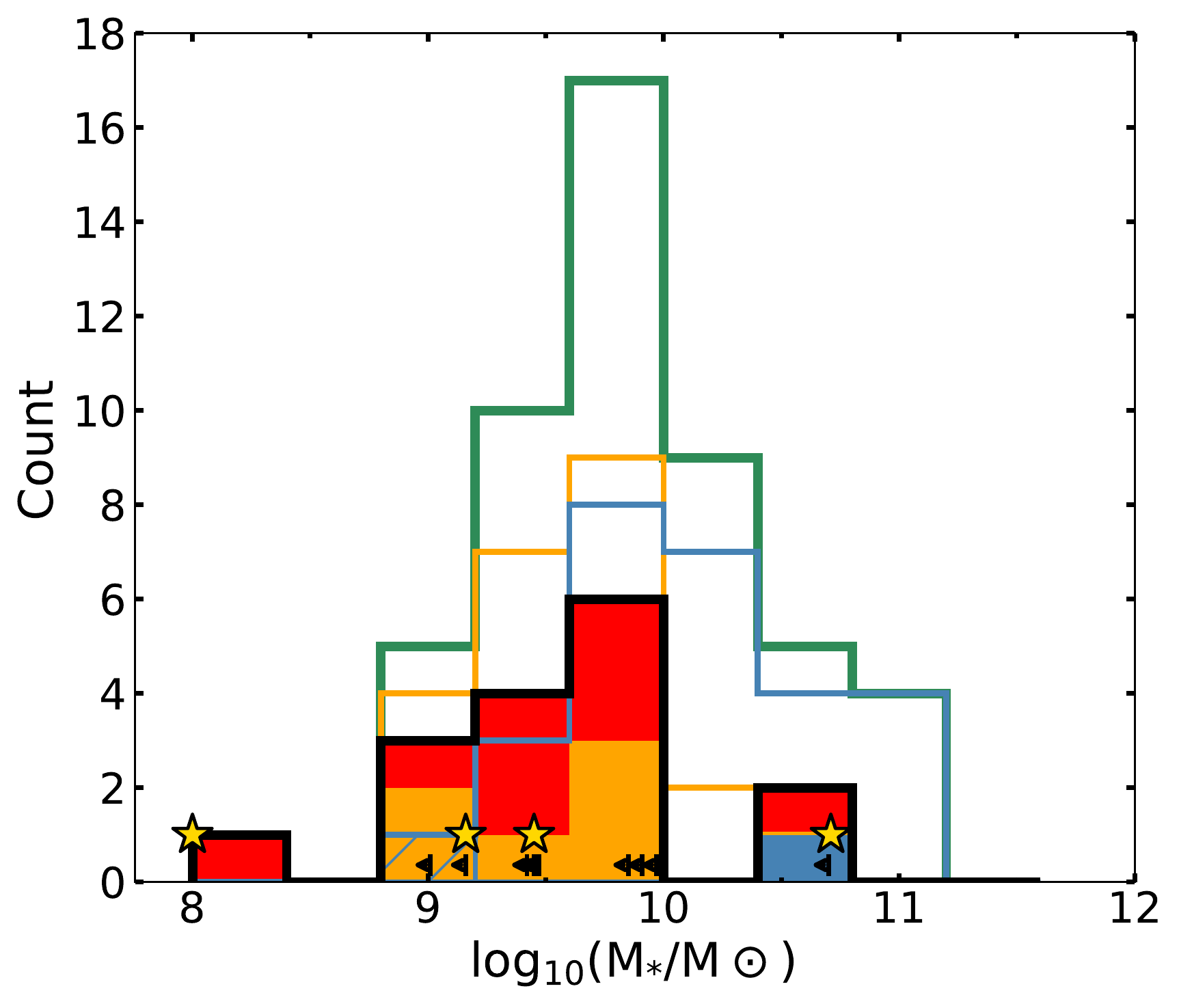}
    \includegraphics[width=6cm]{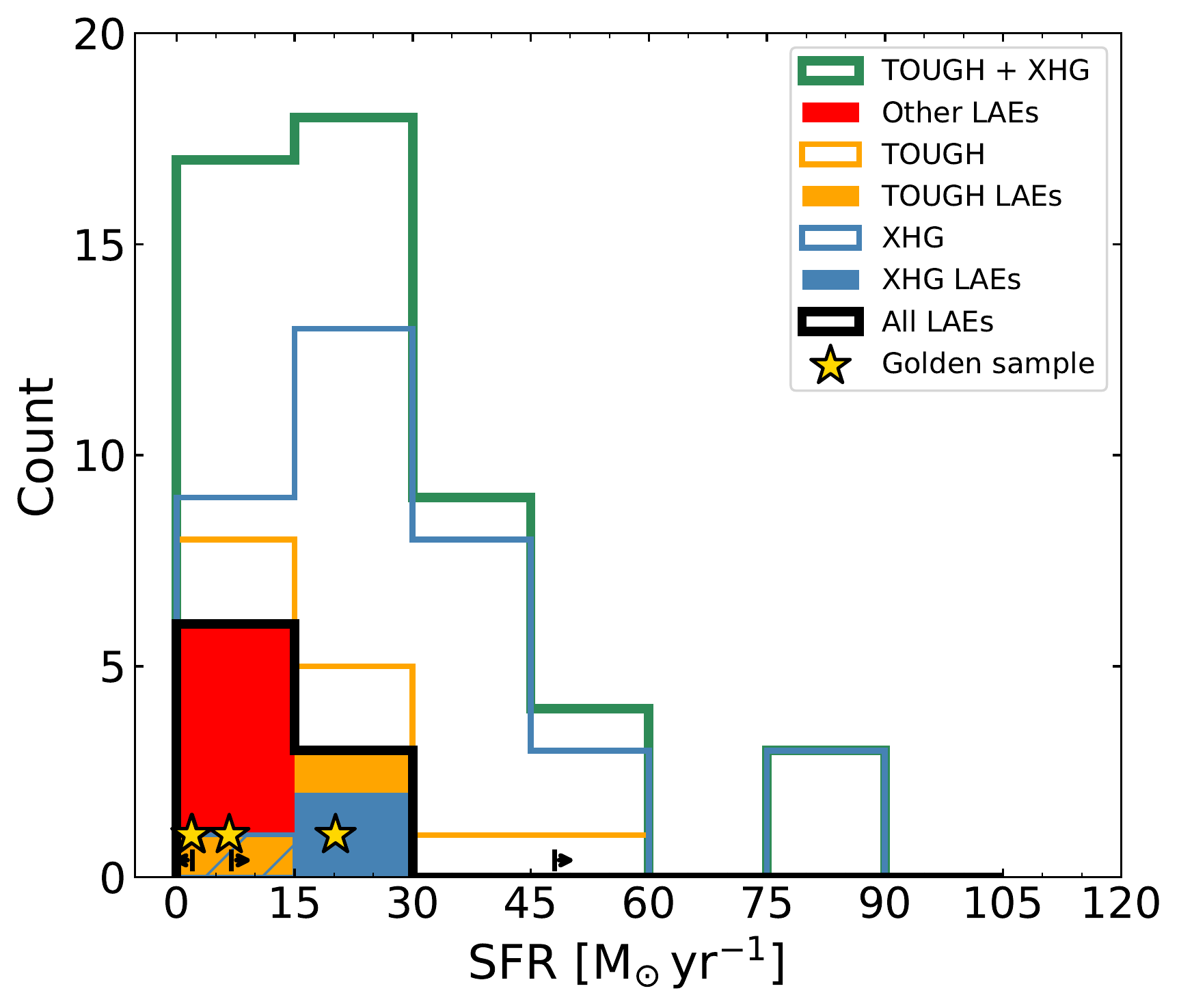}
    \caption{Distribution of the M$_{UV}$ (left panel), stellar masses (middle panel) and SFR (right panel) of the overall LAE-LGRBs (black solid line). The colored areas inside the black histogram correspond to the LAEs detected in the TOUGH (yellow), XHG (blue) and other LAE-LGRBs outside these two samples (red). The dashed blue area is for GRB\,070110 which is detected in both TOUGH and XHG samples.
    The colored areas are cumulative and not superimposed.
    The overall TOUGH and XHG distribution are represented by the yellow and blue empty histograms, respectively. The green histogram is the sum of the TOUGH and XHG samples.
    The arrows represent limits and the yellow stars are for the {\it golden sample} presented in Sect. \ref{GoldSample}.
  }
    \label{Lvs}
\end{figure*}

The results of the three samples presented above are detailed in the tables of Appendix \ref{tablerecap}. By merging them and removing the overlapping cases, we find 14 LAEs out of 84 LGRB host and afterglow spectra. 
The statistics on the LAE-LGRBs (restricted to the host galaxy spectra only) are summarized in Table \ref{Tab_fesc}.

To complete the census, we should add to this merged sample the complementary LAE-LGRBs individually reported in the literature. 
To this purpose, we selected from two GRB databases\footnote{ \url{http://www.astro.caltech.edu/grbox/grbox.php} and \url{http://www.mpe.mpg.de/~jcg/grbgen.html}}, maintained by D. Perley and J. Greiner, all LGRBs detected until December 1, 2020, with a spectroscopic redshift $z>1.6$.
This results in 234 LGRBs (see Fig. \ref{histo}) with 193 having a spectrum covering the 
\Lya line and for which it is possible for us to verify whether the \Lya emission is detected.
For those not included in the sample presented in the previous sections, we searched the literature for articles or Gamma-ray Coordinates Network (GCN\footnote{\url{https://gcn.gsfc.nasa.gov/}}) Circulars claiming a detection of \Lya line. If unpublished spectra were available in the ESO archive, we retrieved and reduced them.

Overall we found thirteen additional LAE-LGRBs from the literature, and two LAE-LGRBs from unpublished data: GRB\,081121 at $z=2.5134$ and 081222 $z=2.770$. Both objects have X-shooter spectra (Prog. ID: 097.D-0672; PI: S.D. Vergani). We reduced the data following the procedure described in Sect. \ref{Data_reduction}. Their \Lya emission is shown in Fig. \ref{2DLya}.
The measured \Lya fluxes of GRB\,081121 and 081222 corrected for Galactic extinction and slit loss are $F_{\rm Ly\alpha}=\rm (4.9 \pm 0.5) \times 10^{-17}\ erg\ s^{-1}\ cm^{-2}$ and $F_{\rm Ly\alpha}=\rm (1.6 \pm 0.4) \times 10^{-17}\ erg\ s^{-1}\ cm^{-2}$, respectively. Additionally to the \Lya line, GRB\,081121 shows the presence of the [OII] doublet, [OIII]$\lambda$5007 and \Hb lines. Residual sky lines contaminate strongly the [OIII]$\lambda$4959 line and partially \Hb. We determine $F_{\rm [OII]}=\rm (1.4 \pm 0.4) \times 10^{-17}$\ergscm, $F_{\rm [OIII]\lambda5007}=\rm (9.0 \pm 0.4) \times 10^{-17}$\ergscm~and $F_{\rm H\beta} \leq 1.2 \times 10^{-17}$\ergscm~(all corrected for Galactic extinction and slit loss).  

In total, to date, there are 29 confirmed LAE-LGRBs (see Fig. \ref{FluxLimit} and \ref{histo}). Eleven of them have $\rm EW_{0}$ $>20$ \AA, five are below this threshold and the continuum flux measurement is lacking for the rest of them. Table \ref{Tab_recap} of the Appendix gives a summary of the LAE-LGRBs and their \Lya fluxes. 
These fluxes should be considered as lower limits. It is well known (\citealt[e.g.,][]{Fynbo1999, Steidel2011, Wisotzki2016, Leclercq2017}) that  
a significant fraction of the \Lya emission of high-redshift galaxies takes place in \Lya halos extending several kpc away of the galaxy. In our measurements we consider only the flux falling in the slit, without any correction taking into account the spatial distribution of the \Lya emission of our objects (unknown for almost all of them).

\subsection{Comparison of LAE-LGRB properties to LGRBs that are not LAEs and LAEs that are not LGRB hosts} \label{prop_LAEs}

LGRB host galaxies are selected in principle only based on the fact that they host a LGRB explosion. Therefore, they are star-forming systems, with a young star formation, and preferentially with sub-solar metallicity (see e.g., \citealt{Lyman2017, Palmerio2019}). In the following, we look at LAE-LGRB statistics and compare it to those of LAEs in galaxy surveys at a similar redshift range to see if these systems have an enhanced or suppressed \Lya emission compared to the general population of star-forming galaxies. We also compare the properties of the LAE-LGRBs to those of LGRB host galaxies not showing \Lya emission (Fig.\,\ref{Lvs}) to try to identify possible driving factors of \Lya escape.
We stress that in some cases only a small fraction of objects in our samples have available information to be used for such comparison. We refer the reader to Tables in Appendix \ref{tablerecap} for the values and references of properties discussed in the following and shown in Fig.\,\ref{Lvs}.

In the survey of galaxies at $z\sim2$ presented by \cite{Steidel2004}, 40\% of galaxies with available spectra show the presence of \Lya emission (are LAEs) and 10\% have $\rm EW_{0}$(\Lya) $> 20$ \AA\, \citep{Erb2014}. \cite{Reddy2008} found 65\% LAEs for LBGs at $z\sim3$, and 23\% with $\rm EW_{0}$(\Lya) $> 20$ \AA.  \cite{Pentericci2010} found a fraction of 50\% LAEs in their selected GOOD-MUSIC LBG sample at $z\sim3$, and 18\% with $\rm EW_{0}$(\Lya) $> 20$ \AA.
The LAE statistics of the TOUGH and XHG samples presented in previous sub-sections of Sect.\,\ref{LAEdetections} is lower than that of those studies (see Table \ref{Tab_fesc}, column f$_{tot}$). However, the selection criteria of the parent sample are not the same and may contribute to this difference.
Most of the above-mentioned studies are based on galaxies with magnitude brighter than $R=25.5$ mag. This corresponds to the median value of the TOUGH sample, and about a half of TOUGH LAEs are fainter galaxies.
If we apply this selection to the TOUGH sample (hence reduced to 17 objects) we obtain a fraction of $22\% \pm 15\%$ LAEs and $7\% \pm 7\%$ LAEs having $\rm EW_{0}$(\Lya) $> 20$ \AA. 
Those numbers should be considered as lower limits as the \Lya flux limit reached by the \cite{Steidel2003, Steidel2004} spectroscopic survey is deeper than ours (typically $\sim 10^{-18}$\ergscm\, instead of $\sim 10^{-17}$\ergscm). 
The fraction of LAEs in the TOUGH sample could be as high as 47\%, assuming that all the objects with flux limit values above that of the weakest detected \Lya are all LAEs. 
Therefore, the lower statistics found for LGRB samples than for galaxy surveys could be explained by the different \Lya flux limits.
The magnitude selection of \cite{Pentericci2010} reaches $R=26$ mag. Applying the same cut to the TOUGH sample would not change so much the results compared to the $R=25.5$ mag cut.

We do not find any particular correlation of the LGRB host galaxy properties (stellar mass, SFR, metallicity, UV magnitude, dust extinction or \HI column density along LGRB line of sight) with \Lya luminosity or $\rm EW_{0}$, but our results are limited by the poor statistics. 
As shown in Fig. \ref{Lvs}, whenever the information on the SFR of LAE-LGRBs is available, its value is below 30\,$\rm M_{\odot}\ yr^{-1}$, except for GRB\,080602 which has a lower 
limit of 48\,$\rm M_{\odot}\ yr^{-1}$ \citep{Palmerio2019}. 
If we focus on stellar masses, we see that LAE-LGRBs are associated preferentially with stellar masses $\rm M_{\star}<10^{10}\ M_\odot$, as also found by \cite{Pentericci2010} 
for LBG samples at $z \sim 3$. 
However, due to the poor statistics we cannot conclude if this is an intrinsic properties of LAEs or a consequence of the characteristics of the parent samples.

The low fraction of LAEs in the XHG sample is likely due to the difference in the M$_{\rm UV}$ and stellar mass range covered by the TOUGH and XHG samples. 
Indeed, by construction, the latter is biased towards LGRBs with highly extinguished afterglows \citep{Kruhler2015}, that on average may originate in more massive and dusty galaxies. Over the 30 objects of the XHG sample having information on E(B-V) (even if with large errors), 70\% have E(B-V) values higher than the most extinguished LAE-LGRB in the sample.
This is consistent with the fact that dust is considered having a significant impact on \Lya suppression 
\citep{Neufeld1990, Laursen2009} and that LAEs are preferentially found in systems with low extinction \citep[e.g.,][]{Shapley2003, Pentericci2007, Pentericci2010, Matthee2016}.
The metallicities of the LAE-LGRBs also support this trend as their values are among the lowest values of the TOUGH and XHG sample (see Appendix \ref{tablerecap}).

\begin{figure}[!ht]
\centering
\includegraphics[width=\hsize]{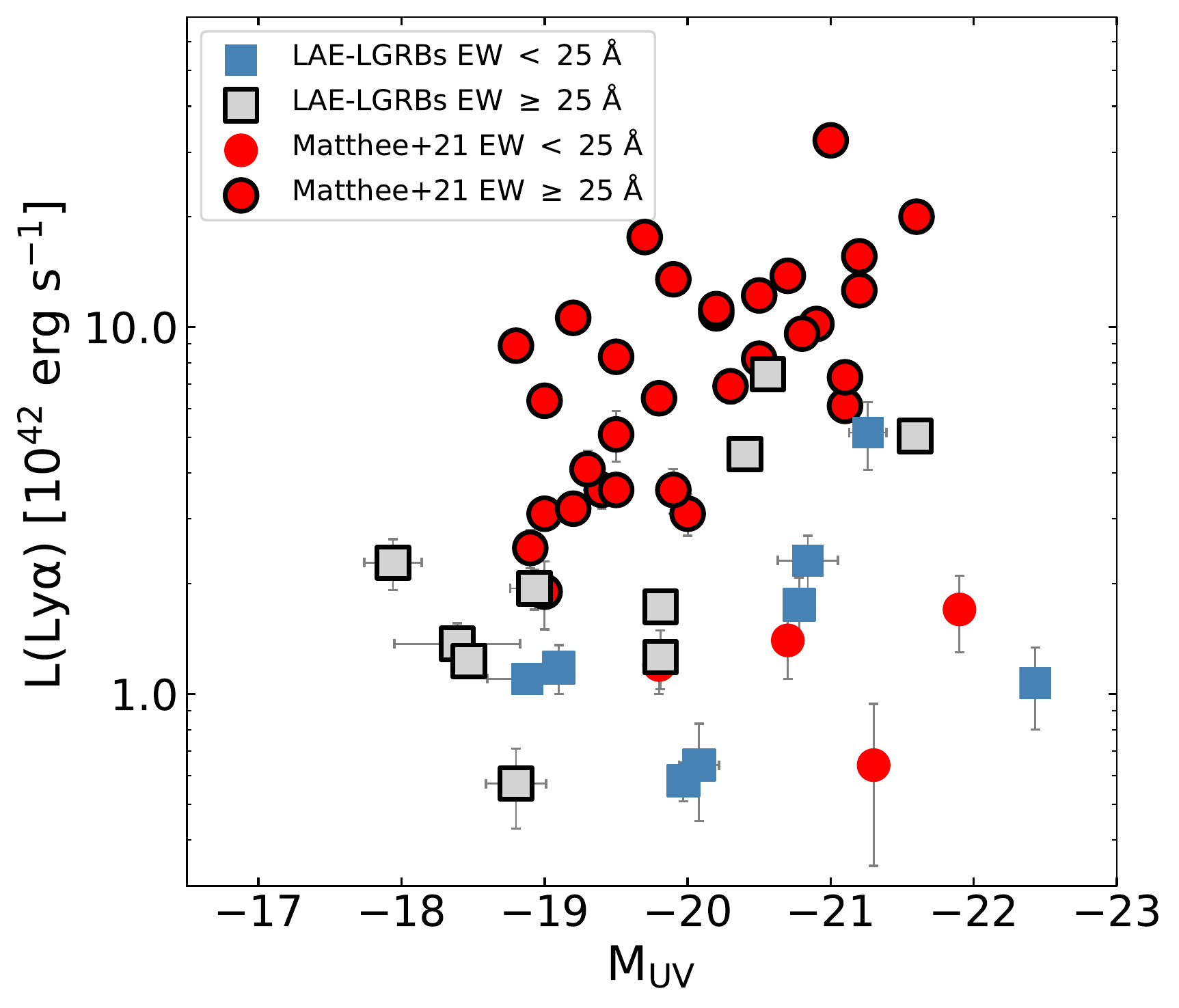}
 \caption{Distribution of luminosities of LAE-LGRBs and LAEs of the XLS-$z2$ sample \citep{Matthee2021} as a function of the galaxy UV absolute magnitude. Systems having $\rm EW_{0}$ $>25\AA$ are shown with black bold contours. The remaining LAEs have fainter or undetermined $\rm EW_{0}$.
}
 \label{Jorryt}
\end{figure}

%% FCs
\begin{figure*}[]
    \centering
    \includegraphics[width=0.75\textwidth]{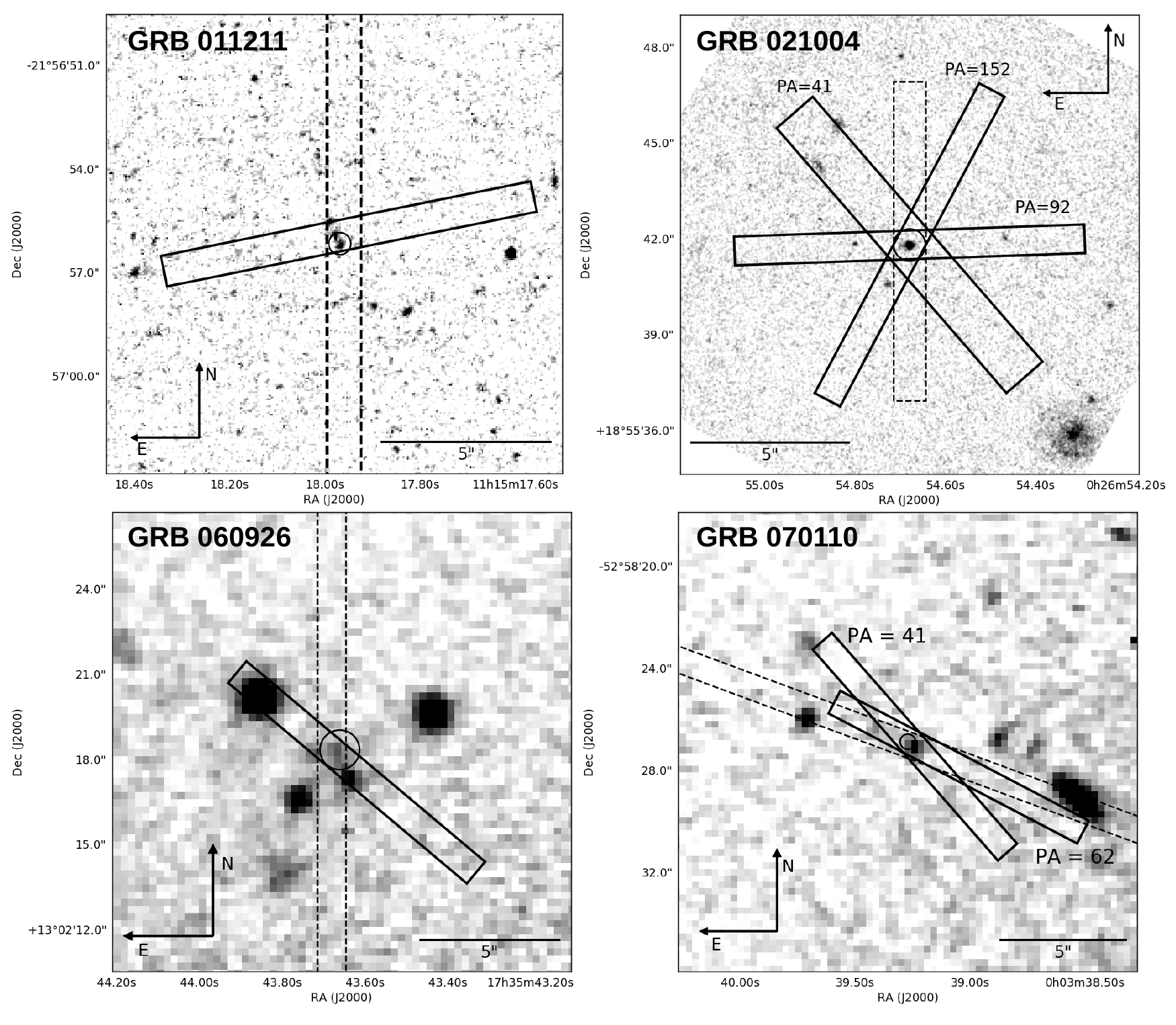}
    \caption{Images of the fields of the four LAE-LGRBs of the {\it golden sample}. The slit position of the afterglow (dashed lines) and host galaxy spectra (solid line) are superimposed. The afterglow position is represented by a circle (\citealt{Jakobsson2003,Fox2002,Holland2006,Malesani2007}, for GRB\,011211, 021004, 060926 and 070110, respectively). GRB\,011211 field: HST/STIS image (Prog. ID: 8867; P.I.: S.R. Kulkarni); GRB\,021004 field: HST/ACS WFC2 F814 image (Prog. ID: 9405; P.I.: A. Fruchter); GRB\,060926A field: VLT/FORS1 image (Prog. ID: 079.A-0253(A); P.I.: P. Jakobsson); GRB\,070110 field: VLT/FORS2 image (Prog. I.D.: 098.D-0416(A); PI: S. Schulze). }
    \label{FCs}
\end{figure*}

\begin{table*}[!h]
    \caption{Observing log of the observations of the four GRB host galaxy spectra of the {\it golden sample} analysed in this paper.}      
    \centering
    \small
    \begin{tabular}{l*{12}{c}r}
    	\hline\hline
     	\rule[0.2cm]{0cm}{0.2cm}GRB host & Redshift & \multicolumn{3}{c}{Exposure time (s)} & \multicolumn{3}{c}{Slit width} & PA & Obs. date & Seeing   & Airmass \\

     	\rule[-0.2cm]{0cm}{0.2cm}  
     	&  	     & UVB & VIS & NIR 					  & UVB & VIS & NIR 			  &  	($\deg$)	  &  			           & (") &  				 \\
     	\hline
     	\rule[0.2cm]{0cm}{0.2cm}GRB 011211   & 2.1434 & $4 \times 1800$ & $4 \times 1800$ & $4 \times 900$	& 1.0 & 0.9 & 0.9  & 101 & 2010-03-18  & 0.5   & 1.4 \\

     	GRB 021004                           & 2.3298 & $4 \times 1200$ & $4 \times 1200$ & $8 \times 650$	& 1.6 & 1.5 & 1.5  & 41  & 2009-11-21  & 1.3   & 1.4 \\

     	GRB 060926                           & 3.2090 & $2 \times 1800$ & $2 \times 1800$ & $6 \times 600$  & 1.0 & 0.9 & 0.9  & 47  & 2010-04-18  & 0.8   & 1.3 \\

     	\rule[-0.2cm]{0cm}{0.2cm}GRB 070110  & 2.3523 & $4 \times 1680$  & $4 \times 1680$ & $8 \times 840$ & 1.0 & 0.9 & 0.9  & 42 & 2010-10-29	 & 0.8   & 1.4 \\
    	\hline
    \end{tabular}
    \tablefoot{The columns indicate the GRB name, redshift determined from emission lines (see Sect.\,\ref{GoldSample}), exposure time, slit width and position angle (PA) used in this study, as well as the observing date, average seeing and airmass of the observations.}
    \label{Tlog}
\end{table*}

\cite{Matthee2021} predicted that the bulk of \Lya population of UV galaxies with $\rm M_{1500}\sim-20\pm1$ at $z=2$ should be below $\sim 2\times10^{42}$erg\,s$^{-1}$, and the fraction of LAEs with $\rm EW_{0}$ $>25$ \AA~should be of $10-30$\%. We can see in Fig. \ref{Jorryt} that most of the THOUGH unbiased sample of LGRB hosts is below this limit. This is due to the fact that LGRB samples have no pre-selection except for the LGRB explosions. Therefore, LAE-LGRBs are representative of \Lya emission from the bulk of UV-selected galaxies at $z\sim2$. Furthermore, the fraction of LAEs with $\rm EW_{0}$ $>25$ \AA~among the TOUGH sample is of $\sim25\%$, in agreement with \cite{Matthee2021} expectations for UV-selected galaxies.

\section{The golden sample} \label{GoldSample}

\begin{table*}[!ht]
\caption{Fluxes of Balmer and oxygen lines and derived properties.}
\centering
\small
\begin{tabular}{c | c c c c c c }
 \hline\hline
 \rule[0.2cm]{0cm}{0.2cm}GRB host & H$\beta$ & H$\alpha$ & [OII]$\lambda3726,3729$ & [OIII]$\lambda4959,5007$ & [OIII]/[OII] & SFR \\
 & [$\rm 10^{-17} erg\ s^{-1}\ cm^{-2}$] & [$\rm 10^{-17} erg\ s^{-1}\ cm^{-2}$] & [$\rm 10^{-17} erg\ s^{-1}\ cm^{-2}$] & [$\rm 10^{-17} erg\ s^{-1}\ cm^{-2}$] & & [$\rm M_{\odot}\ yr^{-1}$] \\
 \hline
 \rule[0.2cm]{0cm}{0.2cm}GRB 011211 & < 0.4*  		  & <1.2           & < 1.6 & $2.7 \pm 0.3$ & > 1.7 & < 2.0          \\
 						 GRB 021004 & $1.2 \pm 0.2$* & $3.3 \pm 0.4$  & < 1.7 & $16.9 \pm 0.5$ & > 10 & $6.7 \pm 0.9$  \\
						 GRB 060926 & $1.6 \pm 0.3$  & $4.6 \pm 0.9$* & $2.3 \pm 0.4^{\#}$ & $3.5 \pm 0.4^{\dagger}$ & $1.5 \pm 0.4$ & $20.2 \pm 4.0$ \\ 
\rule[-0.2cm]{0cm}{0.2cm}GRB 070110 & $0.3 \pm 0.2$* & $0.9 \pm 0.4$  & $1.8 \pm 0.3^{\#}$ & $4.0 \pm 0.3$ & $2.2 \pm 0.4$ & $1.9 \pm 0.6$  \\
 \hline
 \end{tabular}
 \tablefoot{Fluxes are corrected for Galactic extinction and slit loss. No correction for host-intrinsic extinction is applied. 
Upper limits are at 3$\sigma$. 
$^*$: flux estimated using $\rm H\alpha\ = 2.86 \times H\beta$. 
$^{\dagger}$: [OIII]$\lambda4959$ line not detected. We fixed its flux as being 1/3 of the [OIII]$\lambda5007$ flux.
$^{\#}$: [OII]$\rm \lambda3729$ line falling on a sky line.
We fixed its flux as being 3/2 of the [OII]$\lambda3726$ flux. 
}
 \label{TH}
\end{table*}

\begin{table*}[!ht]
\caption{Fluxes and properties of the observed \Lya emission line.}             
\centering
\small
\begin{tabular}{c | c c c c c c}
 \hline\hline
 \rule[0.2cm]{0cm}{0.2cm}GRB host &	F(\Lya) & $\rm EW_0(Ly\alpha)$ & $\rm FWHM_0(Ly\alpha)$ & \Lya red-peak shift & $\rm f_{esc}(Ly\alpha)$ & $\rm F_i(Ly\alpha)$	 \\
 								  & [$\rm 10^{-17} erg\ s^{-1}\ cm^{-2}$]    & [\AA]	  & [$\rm km\ s^{-1}$]	   & [$\rm km\ s^{-1}$]	         &  & [$\rm 10^{-17} erg\ s^{-1}\ cm^{-2}$]                       \\
 \hline
 \rule[0.2cm]{0cm}{0.2cm}GRB 011211  & $1.6 \pm 0.2$  & $14 \pm 3$	 & $249 \pm 60$ & $280 \pm 30$   &  > 0.16 &  < 10.4      \\  
GRB 021004                           & $16.9 \pm 0.3$ & $105 \pm 10$ & $228 \pm 20$ & $180 \pm 20$   & $0.59 \pm 0.08$ & $28.8 \pm 3.5$\\ 

GRB 060926  						 & $5.3 \pm 0.4$  & $37 \pm 7$	 & $417 \pm 40$ & $250\pm 20$    & $0.13 \pm 0.03$ & $40.0 \pm 7.9$\\

\rule[-0.2cm]{0cm}{0.2cm}GRB 070110  & $2.8 \pm 0.5$  & $33 \pm 8$	 & $412 \pm 80$ & $285 \pm 40$   & $0.36 \pm 0.09$ & $7.8 \pm 2.7$\\ 
 \hline
\end{tabular}
 \tablefoot{The \Lya fluxes are corrected for Galactic extinction and slit loss. No correction for host-intrinsic extinction is applied. The $\rm EW_0(Ly\alpha)$ and $\rm FWHM_0(Ly\alpha)$ are respectively the rest-frame equivalent width and FWHM of the observed \Lya line. The \Lya red-peak shift is the velocity shift of the \Lya peak redward the theoretical \Lya line center, as calculated from the systemic redshift of the galaxy (see Table \ref{Tlog}). The escape fraction of \Lya photons $\rm (f_{esc}$(\Lya)) and the intrinsic flux of the \Lya line ($\rm F_i(Ly\alpha)$) are determined from Balmer lines as described in Sect. \ref{GoldSample}. We stress that these escape fractions are not corrected for dust extinction as explained in Sect. \ref{GoldSample}.}
 \label{TLya}
\end{table*}

One of the strengths of GRBs is that they give us the possibility to combine the information on the host galaxy gas retrieved from the afterglow spectra with that obtained through direct host galaxy observations, once the afterglow faded. This advantage can be useful to characterize LAE-LGRBs and to test \Lya models.

For this purpose, we selected from the census presented in Sect. \ref{LAEdetections} a {\it golden sample} of LAE-LGRBs fulfilling the following criteria:

    \begin{enumerate}[label=(\roman*)]
    \item Availability of both afterglow and host galaxy spectra;
    \item \Lya detection in the afterglow or host galaxy spectra with spectral resolution R>1000, as lower resolution can lead to a misinterpretation of the line profile \citep[e.g.,][]{Verhamme2015, Gronke2015};
    \item Detection of other host galaxy emission lines in order to determine the systemic redshift of the galaxy, the intrinsic \Lya and galaxy properties.
    \end{enumerate} 

We stress the importance of X-shooter observations to fulfill point (iii), thanks to its wide spectral coverage from $\sim300$\,nm to 2500\,nm. 

We find only four objects fulfilling the above-mentioned criteria: GRBs 011211, 021004 (reported independently in the literature), 060926 (part of the XHG sample), and 070110 (part of both the XHG and TOUGH sample). 
The images of their fields, with the slits used to obtain the afterglow and host-galaxy spectra, are shown in Fig. \ref{FCs}.
We reduced the X-shooter spectra of the four GRB hosts (see log of the observations in Table \ref{Tlog}) and measured the emission line fluxes (reported in Table \ref{TH} and \ref{TLya}) accordingly to the procedure described in Sect.\,\ref{Data_reduction}. 
These fluxes only represent the \Lya photons falling in the slit, without applying any correction to take into account the two-dimensional spatial extension of the \Lya emission. If we consider a 2D-Gaussian approximation, the resulting fluxes would be on average a factor $\sim$\,2 higher.
The \Lya fluxes of these host galaxies have previously been reported in \citet{MilvangJensen2012}. 
Our measurements using X-shooter spectra provide lower values but consistent within the uncertainties (at 2 or 3 $\sigma$). This is a good agreement, taking into account that different observing techniques and slit position angles may cover different parts of the diffuse \Lya emission.

For each host galaxy we determined the star-formation rate (SFR) (or put limits on it), and the \Lya properties by the following method. 
We used the H$\alpha$ flux (measured or converted from the H$\beta$ flux, assuming H$\alpha=2.86\times \rm H\beta$; \citealt{Osterbrock1989}) to derive the H$\alpha$ luminosity corrected for Milky Way extinction ($\rm L(H\alpha^{MWcor}_{obs})$). 
We converted the H$\alpha$ luminosity into SFR using the relation from \citet{Kennicutt1998}, scaled to the \citet{Chabrier2003} initial mass function as:
\begin{equation}
 \rm SFR(H\alpha) = 4.6 \times 10^{-42} ~L(H\alpha^{MWcor}_{obs}) ~M_{\odot}\ yr^{-1} \, .
 \label{eqSFR}
\end{equation} 

\noindent We cannot retrieve information on the dust extinction of the four host galaxies from the emission lines, due to the detection of only one of the Balmer lines (H$\alpha$ or H$\beta$) in the spectra.
Therefore, the SFR values are determined without applying any host-galaxy dust correction and should be considered as lower limits. 

Under the assumption that the \Lya and the Balmer lines originate from the same regions and are produced by the same recombination process, it is possible to infer the intrinsic properties of the \Lya emission line from the H$\alpha$ or H$\beta$ lines.
We determined the $\rm FWHM_i$(\Lya) from a Gaussian fit of the H$\alpha$ (or H$\beta$) line profile that we corrected for instrumental dispersion.
In case B recombination, the theoretical ratio between the \Lya and the H$\alpha$ lines is $\sim$\,8.7 \citep{Brocklehurst1971}. Using this value and assuming no extinction, 
we converted the Balmer flux to intrinsic \Lya flux ($\rm F_i(Ly\alpha)$). 
We determined the \fescLya as the ratio of the observed \Lya flux to the intrinsic one. Following the consideration on \Lya fluxes reported above, the \fescLya values should be considered as lower limits.
The rest-frame \Lya equivalent width ($\rm EW_{0}(Ly\alpha$)) is determined by dividing the rest-frame \Lya flux by the \Lya continuum level. 

We will detail more on the analysis of the data of each GRB in the following sections. The fluxes of the identified lines, the SFR, and the \Lya emission characteristics are reported in Tables \ref{TH} and \ref{TLya}.

In general, we remark that the \Lya emission line profile and velocity spread of our sample (see next sections) are typical of LAEs at $z\sim2$. 
Similarly to those of the XLS-$z2$ sample of LAEs presented in \cite{Matthee2021}, they show an asymmetric profile with a prominent peak redshifted from the systemic redshift of the host by $\gtrsim200$\,km\,s$^{-1}$, and extending over $\gtrsim500$\,km\,s$^{-1}$. 
However, the velocity shift of the ISM absorption lines detected in the afterglow spectra with respect to the systemic velocity of the galaxies ($\sim-100$\,km\,s$^{-1}$) are not as high as the the XLS-$z2$ sample ($\sim-260$\,km\,s$^{-1}$), likely testifying of a more static environment.

    %--------------------------------------------------------------------
    \subsection{GRB\,011211} \label{GRB011211}

The photometry of the afterglow and host galaxy of GRB\,011211 has been published in \citet{Jakobsson2003}. The detection of \Lya emission through narrow-band filters was reported by \citet{Fynbo2003}. The host galaxy of GRB\,011211 has a multi-component morphology \citep{Jakobsson2003,Fynbo2003}. The GRB site is in the southeast part of the system, while the \Lya emission peaks in the central-north part of the system and extends over the entire system. 

We present here the X-shooter observation of the host galaxy (Prog. ID: 084.A-0631; PI: S. Piranomonte, see Table \ref{Tlog}), not previously published. 
At the afterglow position, we clearly identify the [OIII]$\lambda\lambda4959,5007$ doublet and \Lya line (see Fig.\,\ref{LyaLines}). 
From the [OIII]$\lambda\lambda4959,5007$, we derive a redshift of $2.1434 \pm 0.0001$. 
The \Lya peak is redshifted by $\rm 280 \pm 30\ km\ s^{-1}$ compared to the host galaxy redshift. 
The observed \Lya line properties are reported in Table \ref{TLya}.
The spatial extension of the \Lya line in the 2D spectrum is 2\farc5, compared to 1\farc8 for the brightest nebular emission line [OIII]$\lambda5007$.
From the $\rm 3\sigma$ upper limit on the H$\alpha$ flux we determine a $\rm SFR < 2.0\ M_{\odot}\ yr^{-1}$.
\citet{Perley2013} obtained an average dust attenuation of $\rm A_V = 0.19^{+0.70}_{-0.00}$ mag from the host galaxy 
Spectral Energy Distribution (SED) fitting.

The afterglow spectrum shows the detection of the GRB Damped \Lya system (DLA) with $\rm log(N_{HI}/cm^{-2}) = 20.4 \pm 0.2 $ (\citealt{Vreeswijk2006}), with many associated absorption lines. From the low-ionization state lines (LIS) redshift reported by \cite{Vreeswijk2006} we determine the velocity of the ISM with respect to the systemic redshift of the host galaxy, $\rm V_{LIS}~=~-160 \pm 180\ km\ s^{-1}$, with important uncertainties due to the low resolution of VLT/FORS2 spectra. The \Lya emission peak would fall in the \Lya absorption trough but is not detected in the afterglow spectra, likely due to a combination of line faintness, spectral resolution and noisy spectral region. From the power-law fitting of the afterglow, \citet{Jakobsson2003} determined $\rm A_V = 0.08 \pm 0.08$ mag along the line of sight.

    %--------------------------------------------------------------------
    \subsection{GRB\,021004} \label{GRB021004}
 
The afterglow and host galaxy of GRB\,021004 have been intensively observed (see \citealt{Fynbo2005} and references therein).
\citet{Moller2002}, showed the presence of \Lya emission in the optical spectrum of the
GRB afterglow, which was later confirmed by many other works \citep[e.g.,][]{Mirabal2003,Starling2005a}.
The host galaxy of GRB\,021004 has a compact core with a faint second component and the GRB site is located at the center of its host \citep{Fynbo2005}. 
The X-shooter observation of the host galaxy (Prog. ID: 084.A-0631; PI: S. Piranomonte, see Table \ref{Tlog}) has previously been published in \citet{Vergani2011a}. In Fig. \ref{FCs}, we show the 
image of GRB\,021004 field with superimposed the slits used to observe the host galaxy and the afterglow. 
\cite{Vergani2011a} report the detection of [OIII]$\lambda5007$ emission line, at the same redshift as the GRB, from the galaxy eastward of the GRB host in the spectrum obtained with the slit position angle PA=92$^{\circ}$.  
Its proximity to the GRB host (projected distance of 14\,kpc) may indicate a possible interaction between this galaxy and the GRB host. 

\begin{figure}[!ht]
    \centering
    \includegraphics[width=\hsize]{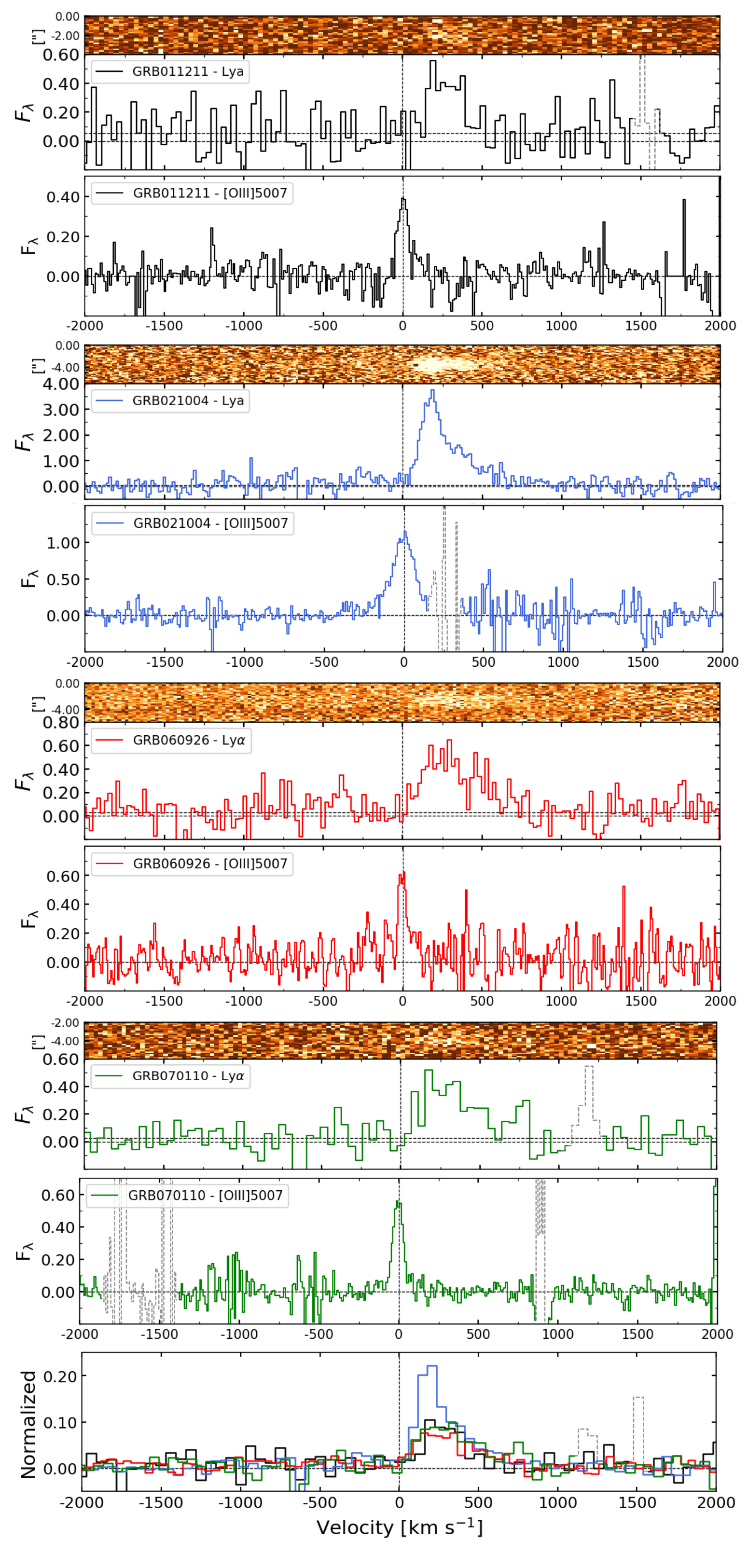}
    \caption{\Lya and [OIII]$\lambda5007$ emission lines from the X-shooter spectra of the four GRB host galaxies GRB\,011211, 021004, 060926 and 070110 from top to bottom respectively. 
    The lines are centered at the systemic redshift of the galaxy determined from the [OIII] emissions lines.
    The flux density $\rm F_{\lambda}$ is in units of $\rm 10^{-17} erg\ s^{-1}\ cm^{-2}$ \AA$^{-1}$. The bottom panel shows the superposition of the four \Lya lines rebinned to $\rm 0.8$ \AA~and normalized to the \Lya line flux.}
    \label{LyaLines}
\end{figure}
%----------------------------------------

For our analysis we used the X-shooter spectrum obtained with slit PA of 41$^{\circ}$ that, due to its larger slit width, minimizes the  loss of \Lya flux. We reduced the spectrum following the procedure described in Sect. \ref{LAEdetections}, and identify the [OIII]$\lambda\lambda4959,5007$ doublet, H$\alpha$ and \Lya emission lines (see Fig.\,\ref{LyaLines}). H$\beta$ falls exactly on a sky line and the [OII]$\lambda3727$ doublet is not detected ($3\sigma$ upper limit of $\rm 1.7 \times 10^{-17} erg\ s^{-1}\ cm^{-2}$). We derive a redshift of $z=2.3298 \pm 0.0001$ for the host galaxy.
We estimate the line fluxes and SFR following the procedure described in Sect. \ref{GoldSample}. 
We find a $\rm SFR = 6.7 \pm 0.8$ \Msunyr (without host extinction correction).
The \Lya emission from the host galaxy is an asymmetric line redshifted by $\rm 180 \pm 20\ km\ s^{-1}$ with an $\rm EW_0 = 105 \pm 10$ \AA. 
The spatial extension of the \Lya line in the 2D spectrum is 4\farc7, compared to 2\farc0 for the brightest nebular emission line [OIII]$\lambda5007$.
The SED fitting performed by \citet{DeUgartePostigo2005} determined $\rm A_V = 0.06^{+0.08}_{-0.06}$ mag, whereas \citet{Perley2013} report $\rm A_V = 0.42^{+0.09}_{-0.07}$ mag. 

The \Lya emission is also detected in the afterglow spectrum at the same velocity and with a similar shape profile as that detected in the host galaxy spectrum.
The afterglow spectrum shows also the \Lya absorption of the GRB sub-DLA system, with $\rm log(N_{HI}/cm^{-2}) = 19.5 \pm 0.5 $ \citep{Fynbo2005}. 

The VLT/UVES afterglow spectrum of GRB\,021004 presents a plethora of absorption lines with complex velocity structures \citep{Fiore2005,Chen2007,CastroTirado2010}, spanning thousands of km\,s$^{-1}$. Although a progenitor-star wind origin was claimed, this was firmly excluded (at least for some of these absorbing systems) by the detection of low-ionization transitions (see \citealt{Chen2007}). Here we focus only on the lowest velocity component, associated with the GRB sub-DLA, therefore likely representing the cold and warm ISM gas of the GRB host galaxy.
Using the shift of the AlII$\lambda1670$ and CII$\lambda1334$ LIS from the systemic redshift of the galaxy, we determine a ISM velocity $\rm V_{LIS} = -80 \pm 20\ km\ s^{-1}$. 
From the spectral flux distribution of the afterglow at several epochs, \citet{DeUgartePostigo2005} determined an average $\rm A_V = 0.20 \pm 0.08$ mag along the GRB line of sight.

    %--------------------------------------------------------------------
    \subsection{GRB\,060926} \label{GRB060926}
    
In the VLT/FORS1 image (Prog. ID: 079.A-0253(A); P.I.: P.~Jakobsson) shown in Fig. \ref{FCs}, we see an offset between the GRB position and the brightest part of the system. Both are covered by the slit used for the host spectroscopic observations. 
In the X-shooter host galaxy observations obtained under program 085.A-0795(A) (presented here for the first time; PI: H. Flores; see Table \ref{Tlog}), we identify the [OIII]$\lambda5007$, [OII]$\lambda3726$, H$\beta$ and \Lya lines (see Fig.\,\ref{LyaLines}). These lines, except \Lya, are used to derive a redshift of $z=3.2090 \pm 0.0001$ for the host galaxy which is consistent with the redshift derived in \citet{Kruhler2015} with a different data set. 
Even with seeing conditions of $\sim$0\farc8, from the emission lines it is not possible to separate the different parts of the host galaxy systems. 
The spatial extension of the \Lya line in the 2D spectrum is 2\farc7, compared to 1\farc4 for the brightest nebular emission line [OIII]$\lambda5007$.

Due to the high redshift of the host, the wavelength coverage of X-shooter does not allow us to observe the H$\alpha$ line. 
We determine a SFR $= \rm 20.2 \pm 4.0\ M_{\odot}\ yr^{-1}$, assuming $\rm H\alpha = 2.86 \times H\beta$ \citep{Osterbrock1989}. As we do not have any information on the host extinction, the value inferred above should be considered as a lower limit of the SFR.
The \Lya emission from the host galaxy is an asymmetric line redshifted by $\rm 250 \pm 20\ km\ s^{-1}$ from the systemic redshift.
The observed \Lya equivalent width and intrinsic properties are reported in Table \ref{TLya}. 

The afterglow spectrum of GRB\,060926 has been published in \citet{Fynbo2009}. The \Lya line is clearly detected in the trough of the DLA. From the  SiII$\lambda1526$, AlII$\lambda1670$ and CII$\lambda1334$ LIS redshift, we determine a ISM velocity of $\rm V_{LIS} = -30 \pm 180\ km\ s^{-1}$.
The \HI~column density derived from the fit of the \Lya absorption \citep{Fynbo2009} is among the highest values probed by GRB afterglow, $\rm log(N_{HI}/cm^{-2}) = 22.6 \pm 0.15 $.

%% PLOT FIT UNCONSTRAINED MODELS
 \begin{figure*}
    \centering
    \includegraphics[width=0.7\textwidth]{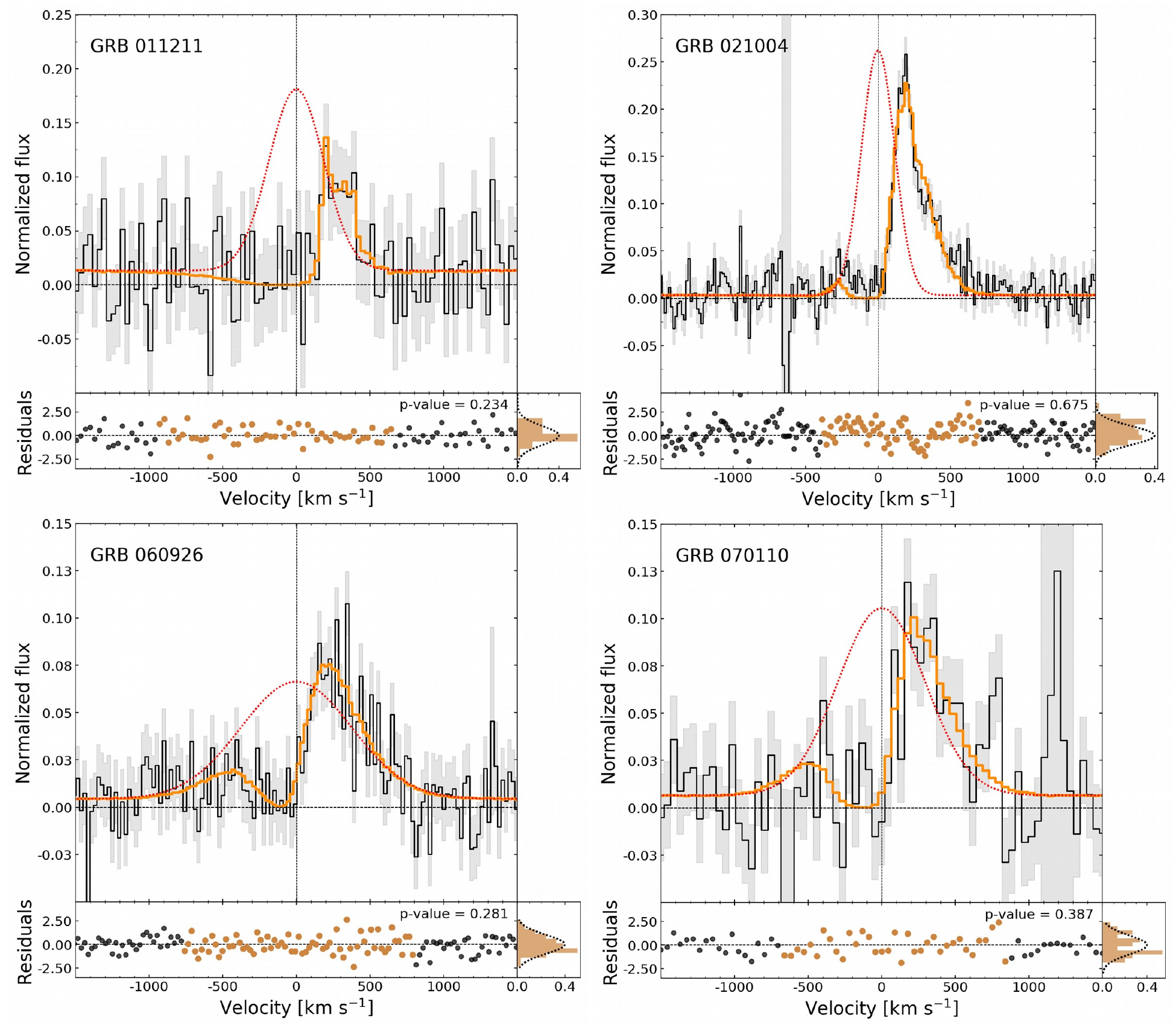}
    \caption{Best fit of the \Lya line obtained by the unconstrained shell model. For each LGRB host: in the top panels, the solid black line corresponds to the data with their error bars (grey), the dashed orange line is the best fit, and the dotted red line corresponds to the intrinsic \Lya emission predicted by the models; in the bottom panels the dots correspond to the normalized residuals between the observation and the model, and those in brown are the residuals covering the fitted \Lya line. The distribution of the residuals is projected in the right bottom panels.
    We also superimpose a Gaussian distribution with $\rm \mu = 0$ and $\sigma = 1$ (dotted black line) as a reference for comparison. We test the deviation of the normalized residual distribution from this Gaussian distribution by using the Shapiro-Wilk test and we provide the p-value of the test in the bottom panels. 
    A p-value above 0.05 indicates that it is no longer possible to reject the null hypothesis that the normalized residual distribution is drawn from a Gaussian distribution at a 95\% confidence level.
    }
    \label{unconst}
\end{figure*}

    %--------------------------------------------------------------------
    \subsection{GRB\,070110} \label{GRB070110}

The near-infrared X-shooter spectrum of GRB\,070110 host galaxy
has been studied in \citet{Kruhler2015}, but the UVB spectrum has not been previously published. The [OIII]$\lambda\lambda4959,5007$, [OII]$\lambda3726$, H$\alpha$ and \Lya lines are identified in the host galaxy spectrum (see Fig.\,\ref{LyaLines}), allowing a redshift determination of $z=2.3523 \pm 0.0001$.
The measured H$\alpha$ flux corresponds to a SFR $=\rm 1.9 \pm 0.6\ M_{\odot}\ yr^{-1}$.
The \Lya emission from the host galaxy shows an asymmetric profile and is redshifted by $\rm 285 \pm 40\ km\ s^{-1}$ from the systemic redshift (see also Table \ref{TLya} for other \Lya properties). 
The spatial extension of the \Lya line in the 2D spectrum is 2\farc1, compared to 2\farc0 for the brightest nebular emission line [OIII]$\lambda5007$.

The afterglow spectrum of GRB\,070110 has been published in \citet{Fynbo2009}. They report the detection of the \Lya emission line in the trough of the GRB-DLA.
From the SiII$\rm \lambda1526$, AlII$\rm \lambda1670$ and CII$\rm \lambda1334$ LIS redshift, we determine an ISM velocity $\rm V_{LIS} = -20 \pm 50\ km\ s^{-1}$. 
The \HI~column density derived from the fit of the \Lya absorption \citep{Fynbo2009} is $\rm log(N_{HI}/cm^{-2}) = 21.7 \pm 0.10$. 
\citet{Troja2007} determined $\rm A_V = 0.08 \pm 0.08$ mag along the GRB afterglow line of sight.

%--------------------------------------------------------------------
\section{Modelling of the Lyman-alpha line} 
\label{LyaLine}

    %% TABLE PARAMETERS UNCONSTRAINED MODELS
\begin{table*}[!ht]
    \caption{Best-fit results of the unconstrained shell model}             
    \centering
    \resizebox{0.7\textwidth}{!}{%
     \begin{tabular}{c | c c c c c c c }
         \hline\hline
         \rule[0.2cm]{0cm}{0.2cm} & \multicolumn{7}{c}{Shell model results from unconstrained fitting} \\
         \hline
          \rule[0.2cm]{0cm}{0.2cm}GRB host 	& $\rm \Delta z$ 	& $\rm log(N_{HI}/cm^{-2})$        & $\rm V_{exp}$     & $\log\rm (T/K)$             & $\rm \tau_d$ & $\rm FWHM_i$(\Lya) & $\rm EW_i$(\Lya)  \\
         			& 			&  & [$\rm km\ s^{-1}$] &  &                & [$\rm km\ s^{-1}$]      & [\AA]               \\
         \hline
         \rule[0.2cm]{0cm}{0.2cm}GRB 011211 & $30^{+45}_{-45}$  & $19.9^{+0.3}_{-0.3}$ & $112^{+33}_{-27}$   &  $4.52^{+0.66}_{-1.08}$   & $2.8^{+1.6}_{-1.7}$  & $440^{+130}_{-150}$ & $63^{+38}_{-31}$ \\

        GRB 021004  						& $5^{+35}_{-35}$  & $19.6^{+0.1}_{-0.1}$  & $122^{+6}_{-6}$     &  $4.99^{+0.15}_{-0.15}$   & $0.365^{+0.104}_{-0.081}$  & $270^{+22}_{-24}$ & $157^{+18}_{-13}$ \\

        GRB 060926  						& $0^{+45}_{-45}$   & $19.5^{+0.2}_{-0.1}$ & $164^{+35}_{-29}$   &  $3.59^{+1.04}_{-0.57}$   & $0.52^{+0.99}_{-0.40}$  & $890^{+115}_{-115}$ & $63^{+35}_{-20}$  \\

        \rule[-0.2cm]{0cm}{0.2cm}GRB 070110 & $0^{+45}_{-45}$   & $19.7^{+0.3}_{-0.2}$ & $182^{+58}_{-45}$   &  $4.37^{+0.94}_{-1.07}$   & $1.14^{+2.00}_{-0.91}$  & $700^{+145}_{-160}$ & $53^{+32}_{-25}$ \\
        \hline
    \end{tabular}}
     \tablefoot{$\rm \Delta z=(z_{fit}-z_{host})\times c$, where $\rm z_{fit}$ is the \Lya redshift of the best fit, $\rm z_{host}$ is the redshift derived from the X-shooter emission lines (Table \ref{prop_model}), and c is the speed of light; $\log(\rm N_{\rm HI}/{\rm cm^{-2}})$ is the \HI column density of the shell; $\rm V_{exp}$ is the shell expansion velocity; $\log\rm (T/K)$ is the temperature of the gas; $\rm \tau_d$ is the dust optical depth; $\rm FWHM_i$(\Lya) and $\rm EW_i$(\Lya) are the intrinsic FWHM and equivalent width of the \Lya line, respectively.}
    \label{tableComp_unconst} 
\end{table*}

 \begin{table*}[ht!]
\caption{Observationally determined values of the parameters of the shell model}             
\label{prop_model}      
\centering
\resizebox{0.8\textwidth}{!}{%
 \begin{tabular}{c | c c c c c c c c}
 \hline\hline
 \rule[0.2cm]{0cm}{0.2cm}GRB host 	& Redshift$\rm ^{HG}$ 	& $\log(\rm N_{\rm HI}^{\rm OA}/{\rm cm^{-2}})$        & $\rm V_{LIS}^{OA}$             & $\rm A_v^{OA}$ & $\rm \tau_d^{OA}$ & $\rm FWHM_i^{HG}$(\Lya) & $\rm EW_i^{HG}$(\Lya)  & Refs \\
 			& 			&  & [$\rm km\ s^{-1}$]  &  [mag]  &        & [$\rm km\ s^{-1}$]      & [\AA]  &             \\
 \hline
 \rule[0.2cm]{0cm}{0.2cm}GRB 011211 & 2.1434  & $20.4 \pm 0.20$ & $-160\pm 180$ & $0.08 \pm 0.08$ & $0.52 \pm 0.52$ & $50^{+10}_{-10}$* & < 64  & (1)  \\

GRB 021004  						& 2.3298  & $19.5 \pm 0.50$  & $-80\pm 20$ 	   & $0.20 \pm 0.08$ & $1.29 \pm 0.52$ & $200^{+20}_{-20}$ & $160 \pm 40$ & (2)  \\

GRB 060926  						& 3.2090  & $22.6 \pm 0.15$ & $-30\pm 180$ 	& -               & -       & $50^{+20}_{-50}$  & $275 \pm 85$ &      \\

\rule[-0.2cm]{0cm}{0.2cm}GRB 070110 & 2.3523  & $21.7 \pm 0.10$ & $-20\pm 50$  	   & $0.08 \pm 0.08$ & $0.52 \pm 0.52$ & $30^{+10}_{-30}$  & $75 \pm 35$ & (3)  \\
\hline
\end{tabular}}
 \tablefoot{We refer to Sect.\,\ref{GoldSample} for the value determinations, except for the dust optical depth,
 $\rm \tau_d$, that is 
 converted from $\rm A_V$ by considering a dust albedo of A $= 0.5$ (see \citealt{Verhamme2006, Gronke2015}). 
 "*" is for value determined from the FWHM of the [OIII]$\lambda5007$ line. "$^{\rm OA}$" is for the parameters related to optical afterglow observations and "$^{\rm HG}$" for the host galaxy ones. \\
 \textbf{References:} The $\rm A_V$ values are reported in (1): \citet{Jakobsson2003}; (2): \citet{DeUgartePostigo2005}; (3): \citet{Troja2007}, see Sect. \ref{GoldSample} for more details. 
 }
\end{table*}

\subsection{Description of the model} 
\label{Model}

Following \cite{Vielfaure2020}, we model the observed \Lya line with the "shell-model"
fitting pipeline described in \citet{Gronke2015}.
This simplistic model consists of an expanding, homogeneous, spherical shell composed of uniformly mixed neutral hydrogen and dust \citep{Ahn2003,Verhamme2006}.
The expanding geometry was motivated by the \HI~outflows which seem to be ubiquitous at both low- and high-redshifts \citep[e.g.,][]{Shapley2003,Wofford2013,RiveraThorsen2015,Chisholm2015}. 
The \Lya emitting source is placed at the centre of the sphere filled with ionized gas and surrounded by a neutral and dusty shell. 
This source would correspond to the star-forming region hosting the LGRB. Such assumption is reasonable considering that LGRBs are produced by massive stars 
and are found in the central star-forming region of their host (typically at $\sim$\,1 kpc from the center, \citealt[e.g.,][]{Lyman2017}).
The photons are collected in all directions after their scattering through the neutral shell.

The shell model is defined by seven parameters that we briefly detail here.
Four parameters describe the shell properties: 
\begin{itemize}
    \item the \textit{radial expansion velocity} ($\rm V_{exp}$).
     This parameter is physically linked to the broadening and the peak shift of the \Lya line. For outflowing shell, the velocity expansion regulates the blue peak intensity by scattering the photons in the red wing of the line. A higher velocity of the gas increases these effects but suppresses the interaction of the photons with the neutral gas. See for example \citet{Verhamme2006} for more details on this effect. The $\rm V_{exp}$ of the neutral gas can be linked to the
    velocity shift of the LIS lines ($\rm V_{LIS}$) when available.
    \item The \textit{\HI~column density} ($\rm N_{HI}$). The effect of increasing $\rm N_{HI}$ is to broaden the line by shifting the \Lya photons from the line center.
    \item The \textit{dust optical depth} ($\rm \tau_d$) at wavelengths in the vicinity of \Lya. The dust grains absorb and scatter the \Lya photons. An increasing dust content will favor this process and affect the intensity of the line but can also affect its shape.
    An increasing $\rm N_{HI}$ boosts the interaction of the photons with dust so both parameters are linked. 
    \item The \textit{effective temperature} of the gas (T). Its effect is complex and its impact on the width of the line and the position of the peak depends on the other parameter values. 
\end{itemize}

We consider a Gaussian profile for the intrinsic \Lya emission and an adjacent flat UV continuum, therefore three additional parameters are used:

\begin{itemize}
    \item the \textit{redshift} of the emitter (z) corresponding to the center of the intrinsic \Lya line;
    \item the \textit{intrinsic equivalent width} of the \Lya line ($\rm \rm EW_i(Ly\alpha)$);
    \item the \textit{intrinsic FWHM} of the \Lya line ($\rm \rm FWHM_i(Ly\alpha)$).
\end{itemize}
        
We refer the reader to \citet{Verhamme2006} and \citet{Gronke2015} for a more detailed description on the model and parameters.     
The observation of LGRB host galaxy and afterglow allows us to constrain these parameters even for faint star-forming galaxies.
However, while host galaxy observations provide information for the overall galaxy, the afterglow constrains parameters along the LGRB line of sight.
Therefore, by using the shell model with afterglow observations we assume that the LGRB region is the main source of \Lya emission in the host, and that the LGRB line of sight probes gas properties characteristic of the environment surrounding the LGRB region. 
In the following, to differentiate between the parameters constrained by these two types of observation, we annotate with X$^{\rm OA}$ the parameters related to optical afterglow observations and X$^{\rm HG}$ the host galaxy ones. 
As a first step, we fit the observed \Lya lines without constraints, as it would typically be done for the majority of high-z LAE observations, where most of the parameters can hardly be constrained observationally. 
Then, we fit the \Lya profiles using the observational constraints, discussing as well the effects of constraining the parameters with only the host galaxy or afterglow observations.

   \subsection{Unconstrained \Lya profile fitting} \label{UnconFitting} 

To evaluate the predictions of the shell model, we fit the observed \Lya lines of the four LGRBs of the {\it golden sample} using an improved model grid and process originally described in \citet{Gronke2015}, without considering any prior.
The automatic fitting succeeds in reproducing the line profile satisfactory in all four cases.  
In Fig. \ref{unconst}, we present for each spectrum the best fit corresponding to the lowest $\rm \chi^2$. In all cases but GRB\,011211 the models fit a blue peak which is not clearly seen in the spectra but consistent with the uncertainties. 
The distributions of normalized residuals presented at the bottom of each spectrum are consistent with a unit Gaussian distribution, indicating a good agreement between the models and the observations.

While the unconstrained shell model succeeds in reproducing the observed profiles, the comparison between the best shell parameters returned by the fitting (Table\,\ref{tableComp_unconst}) and the corresponding values determined from the observations (Table \,\ref{prop_model}) reveals important discrepancies for GRB\,060926 and GRB\,070110. Their observed \HI~column density is higher than the fit results, and the $\rm FWHM_i(Ly\alpha)$ fitting values are largely overestimated. 

The \Lya escape fraction (\fescLya; see Table\,\ref{tableComp_con}) is also a by-product of the shell model and is found to be consistent (within the uncertainties) with the value determined from the observations (see Table\,\ref{TLya}) only in the case of GRB\,070110. 
Particularly, \fescLya is discrepant for GRB\,011211 and 021004 
whereas the fit and best-fitting parameters are in agreement with the observations. 
The disagreement on \fescLya implies also that the H$\alpha$ flux and subsequent SFR obtained by the fit are inconsistent with the values determined from the observations (see Tables\,\ref{TH} and \ref{tableComp_con}).
For GRB\,011211 the H$\alpha$ flux is less constrained (only $3\sigma$ upper limit) but a lower H$\alpha$ flux would imply an increased difference. 
We caution, however, that the \Lya escape fractions derived from the spectral fitting procedure is highly uncertain as it depends on the fitted value of $\tau_{\rm d}$ which mostly has only minor effects on the spectral shape \citep[see discussion in][]{Gronke2015}.
The disagreement could also come from the lack of extinction correction of the observed \Ha flux in our calculation of \fescLya and SFR (see Sect.\ref{GoldSample}). 
An $A_V \approx 0.6$ mag for both GRB\,011211 and 021004 (considering an SMC extinction curve; \citealt{Japelj2015}) would result in consistent H$\alpha$ and \fescLya values within $1\sigma$.

%% TABLE PROPERTIES UNCONSTRAINED MODELS
\begin{table}[ht!]
    \caption{Properties derived from the best fit results (unconstrained model)}             
    \label{tableComp_con}      
    \centering
    \small
     \begin{tabular}{c | c c c c}
         \hline\hline
         \rule[0.2cm]{0cm}{0.2cm} & \multicolumn{4}{|c}{Values obtained from the unconstrained fitting results} \\
         \hline
         \rule[0.2cm]{0cm}{0.2cm}GRB host & \fescLya  & $\rm F_i(Ly\alpha)$ & $\rm F(H\alpha)$               & SFR(H$\alpha$)           \\
         & (1)      & (2)                  & (3)     & (4)                         \\
         \hline
        \rule[0.2cm]{0cm}{0.2cm}011211 & $0.03^{+0.07}_{-0.02}$ & $53.1^{+124.5}_{-37.5}$ & $6.1^{+14.3}_{-4.3}$ & $10.1^{+24.0}_{-7.2}$ \\

         021004                         & $0.27^{+0.07}_{-0.05}$ & $62.6^{+16.3}_{-11.6}$ & $7.2^{+1.9}_{-1.4}$  &  $14.7^{+3.9}_{-2.9}$ \\

          060926                         & $0.69^{+0.19}_{-0.22}$ & $7.8^{+7.5}_{-7.5}$ & $0.9^{+0.4}_{-0.4}$  &  $3.9^{+1.8}_{-1.8}$  \\

        \rule[-0.2cm]{0cm}{0.2cm} 070110 & $0.43^{+0.32}_{-0.24}$ & $7.0^{+5.3}_{-4.4}$ & $0.8^{+0.6}_{-0.5}$  & $1.6^{+1.3}_{-1.1}$   \\
        \hline
    \end{tabular}
     \tablefoot{ Flux units are $\rm 10^{-17} erg\ s^{-1}cm^{-2}$. (1): Escape fraction of \Lya photons calculated by the models according to the method described in \citet{Gronke2015}.
     (2): Intrinsic \Lya flux determined from the flux of the observed \Lya line and the \Lya escape fraction calculated by the models.
     (3): H$\alpha$ flux inferred from the flux of the observed \Lya line and the \fescLya~provided by the best fit results, calculated using F(H$\alpha$) = F(\Lya)/8.7 (case B of the theory of the recombination \citep{Brocklehurst1971}).
     (4): SFR (in $\rm M_{\odot}\ yr^{-1}$) derived using the H$\alpha$ flux of column (3) and following the method described in Sect. \ref{GoldSample}. }
\end{table}

%% TABLE FIT PARAMETERS CONSTRAINED MODELS
\begin{table*}[ht]
    \caption{Results of the shell model fit with constrained parameters}
    \label{tableComp_const}      
    \centering
    \footnotesize{
     \begin{tabular}{c | c c c c c c c }
         \hline\hline
         \rule[0.2cm]{0cm}{0.2cm} & \multicolumn{7}{c}{Shell model results from constrained fitting} \\
         \hline
          \rule[0.2cm]{0cm}{0.2cm}GRB host 	& $\rm \Delta z$ 	& $\log(\rm N_{\rm HI}/{\rm cm^{-2}})$        & $\rm V_{exp}$     & log(T/K)             & $\rm \tau_d$ & $\rm FWHM_i$(\Lya) & $\rm EW_i$(\Lya)  \\
         			& 			    &  & [km s$^{-1}$] &   &                & [km s$^{-1}$]               &  [\AA]               \\
         \hline
         \rule[0.2cm]{0cm}{0.2cm}GRB 011211 & $20^{+45}_{-45}$ & $20.1^{+0.2}_{-0.1}$ & $74^{+27}_{-24}$   &  $3.91^{+0.8}_{-0.75}$   & $0.50^{+0.29}_{-0.24}$  & $30^{+50}_{-20}$ & $45^{+14}_{-18}$  \\

        GRB 021004  						& $5^{+45}_{-45}$  & $19.6^{+0.1}_{-0.1}$  & $118^{+5}_{-6}$   &  $4.99^{+0.16}_{-0.15}$   & $0.402^{+0.103}_{-0.093}$  & $227^{+11}_{-12}$ & $161^{+13}_{-11}$  \\

        GRB 060926  						& $70^{+45}_{-45}$     & $21.5^{+0.1}_{-0.1}$        & $3^{+1}_{-1}$     &  $2.94^{+0.16}_{-0.11}$          & $0.003^{+0.004}_{-0.002}$  & $70^{+20}_{-20}$ & $23^{+9}_{-6}$  \\

        \rule[-0.2cm]{0cm}{0.2cm}GRB 070110 & $35^{+45}_{-45}$  & $21.5^{+0.1}_{-0.1}$ & $3^{+1}_{-1}$     &  $3.04^{+0.12}_{-0.16}$   &$0.008^{+0.010}_{-0.005}$  & $53^{+20}_{-20}$ & $33^{+25}_{-13}$  \\
        \hline
    \end{tabular}
    }
    \begin{center}
     \footnotesize{
     \textbf{Notes.} Same as Table \ref{tableComp_unconst} for models constrained with z, $\rm N_{HI}$, $\rm V_{exp}$, $\rm \tau_d$, $\rm FWHM_i(Ly\alpha)$ and $\rm EW_i(Ly\alpha)$ 
     }
    %  }
    \end{center}
\end{table*}

    \subsection{Constrained \Lya profile fitting} \label{ConFitting}

%% PLOT FIT CONSTRAINED MODELS
\begin{figure*}[]
    \centering
    \includegraphics[width=0.8\textwidth]{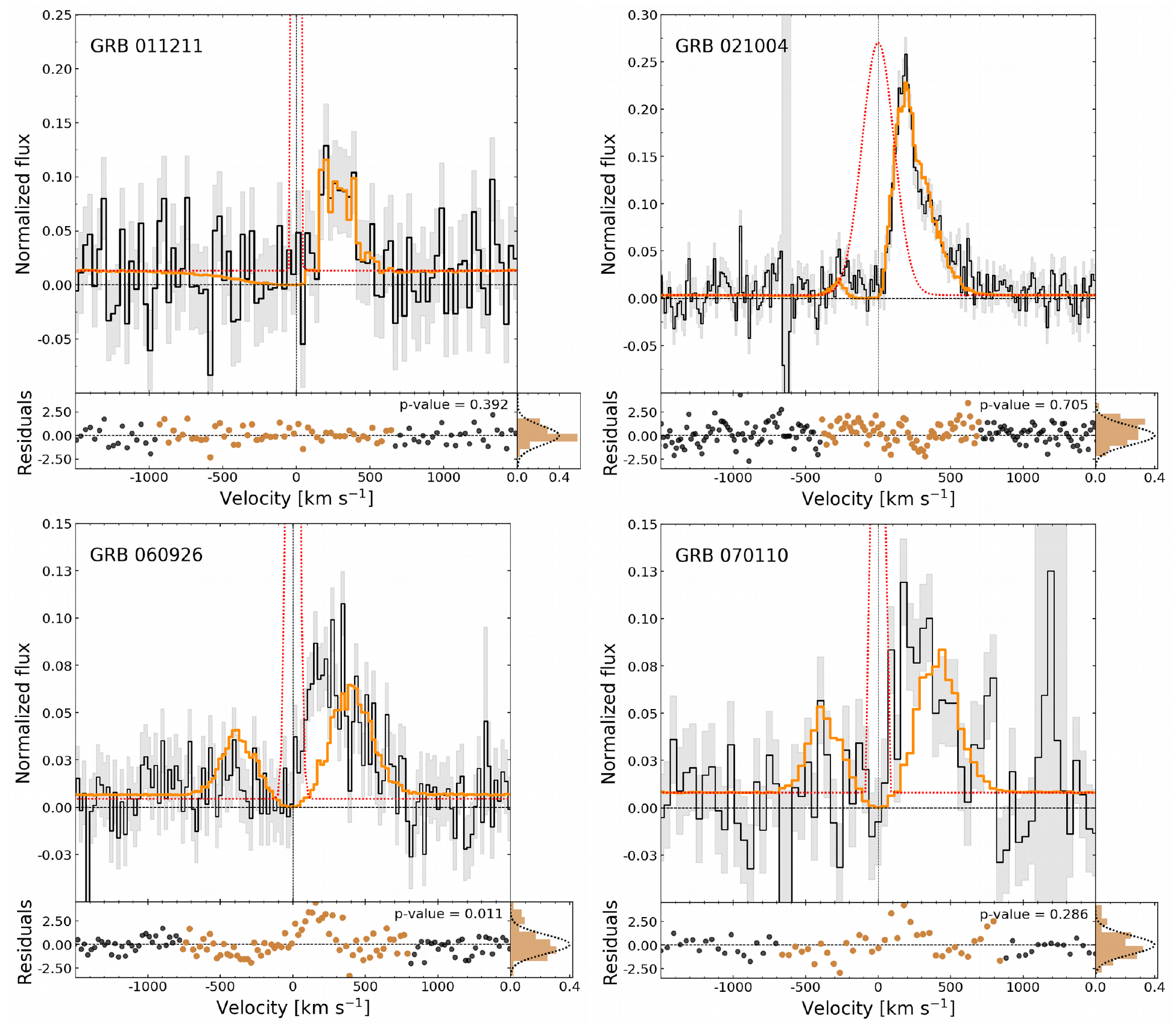}
    \caption{Same as Fig. \ref{unconst} but with the shell model parameters $z$, $\rm N_{HI}$, $\rm V_{exp}$, $\rm \tau_d$, $\rm FWHM_i(Ly\alpha)$ and $\rm EW_i(Ly\alpha)$ constrained by the values determined from the observations.}
    \label{const}
\end{figure*}

We then proceeded to fit the \Lya using the observational constraints described in Sect. \ref{GoldSample} and listed in Table\,\ref{prop_model}, as priors.
Using the afterglow and host galaxy spectra, as described in Sect. \ref{GoldSample}, we are able to constrain up to 6 out of the 7 parameters taken as input by the model. 
The host galaxy observations provide the information on the observed \Lya line and the Balmer lines used to constrain the intrinsic \Lya emission ($\rm FWHM_i^{HG}$(\Lya) and $\rm EW_i^{HG}$(\Lya)). 
They also provide the nebular emission lines used to constrain the redshift ($\rm z^{HG}$). 
The afterglow observations constrain the gas properties: the hydrogen column density (\NHIOA), the velocity of the gas ($\rm V_{LIS}^{OA}$) and the information on the dust content ($\rm \tau_d^{OA}$).
It was not possible to constrain the \HI temperature. The effect of this parameter on the final profile is complex and usually it is degenerated with the other parameters \citep{Gronke2015}. We let this parameter free for the 4 cases.
For GRB\,060926 we lack information about the dust extinction, and for GRB\,011211 we have only upper limits for \Ha. 
To be consistent with the constraints on the \NHIOA values derived from the afterglow, the dust extinctions are taken from the afterglow SED fitting reported in Sect. \ref{GoldSample}.
The grid we use does not extend up to $\log(\rm N_{\rm HI}/{\rm cm^{-2}}) = 22.6$, so this value is not firmly constrained for GRB\,060926 but constrained to the highest value of the grid which is 21.8. 
We report the results of the fitting in Table \ref{tableComp_const}. In Fig. \ref{const}, the intrinsic \Lya derived from the model and the best fit obtained with the lowest $\rm \chi^2$ are superimposed to the data.

Not surprisingly, we find a good agreement between the fitting and the observations of the two hosts with lower \NHIOA (GRB\,011211 and 021004). As shown by the residuals at the bottom of Fig. \ref{const}, the fit is worse than in the unconstrained case for GRB\,070110 and fails to reproduce the \Lya profile and parameters for GRB\,060926, which has an extreme $\log(\rm N_{\rm HI}^{\rm OA}/{\rm cm^{-2}})$ value of 22.6.
In these two cases the models tend to fit a double peak profile with higher red peak velocity shift ($\rm \sim400\ km\ s^{-1}$) than observed ($\rm \sim280\ km\ s^{-1}$) and quasi-static \HI~shell ($\rm 3 \pm 1\ km\ s^{-1}$) which results in a double peak profile. 
By relaxing the constraints, without considering priors for $\rm FWHM_i(Ly\alpha)^{HG}$, we manage to fit the profile of GRB\,070110 in a better way but the best-fitting $\rm FWHM_i(Ly\alpha)$ becomes extremely narrow with an unrealistic value of $\rm 8^{+6}_{-3}\ km\ s^{-1}$. 

We also performed two sets of fitting constraining separately the parameter values determined from the host galaxy and the afterglow observations. 
We find that the profiles of the fitting constrained with the host galaxy observations ($\rm z^{HG}$, $\rm FWHM_i^{HG}$(\Lya) and $\rm EW_i^{HG}$(\Lya)) are very similar to the unconstrained cases, well reproducing the data (see Appendix \ref{shell_host}). Differently from the unconstrained case, the blue peak is not present, which is due to the constraint on $\rm FWHM_i^{HG}$(\Lya).
The predictions for the properties of the gas are also similar to the unconstrained cases. Especially, the \NHI values are lower than observed through LGRB line of sight. The fitting constrained with the afterglow observations ($\rm z^{HG}$, \NHIOA, $\rm V_{LIS}^{OA}$ and $\rm \tau_d^{OA}$) only gives very similar results to the fully constrained case (see Appendix \ref{shell_OA}). 
The models successfully reproduce the low \NHI cases but fail to reproduce the two cases with high \NHIOA values.

These results are in agreement with previous finding that the most constraining parameters are \NHI and $\rm V_{exp}$ \citep[e.g.,][]{Gronke2015}.

\section{Discussion} \label{Discussion}

    \subsection{Implications of the shell-model fits} \label{DiscShell}

The unconstrained fitting succeeds in reproducing all \Lya profiles of the {\it golden sample} (and GRB\,191004B; see \citealt{Vielfaure2020}). However, if we compare the parameter values resulting from the fitting with those observed, we find that for GRB\,060926 and GRB\,070110 there are important discrepancies. A fit performed constraining the parameters values with the observed ones does not succeed in reproducing the observed line profile of these two cases. 

The $\rm FWHM_i(Ly\alpha)$ provided by the unconstrained models for GRB\,060926 and 070110 are very large and disagree with the $\rm FWHM_i(Ly\alpha)^{HG}$ we determined assuming that the Balmer photons and \Lya photons are produced by recombination in the same regions. Such high values are usually returned by the shell model fitting to reproduce the line profile of the emitters having \Lya peaks and wings substantially shifted from the systemic redshift. 
A similar discrepancy has also been found by \cite{Hashimoto2015} in the study of $z\sim2$ LAEs and in the \cite{Yang2016} and \cite{Orlitova2018} studies on low-redshift GP galaxies. 
The latter also reports a discrepancy between the systemic redshift derived from the models and the observed ones, whereas they agree in our case.
\cite{Hashimoto2015} propose that other sources of \Lya emission than star formation, such as gravitational cooling, could account for the large intrinsic width of the \Lya emission line.
\cite{Orlitova2018} try to explain the FWHM discrepancy considering complex kinematic structures (as testified by 
broad H$\alpha$ and H$\beta$ wings observed in their GP spectra) but with negative results. They conclude that the discrepancy is a model failure and not a physical effect. From our data we cannot recover possible kinematic structures. 
A likely cause for the mismatch is a widening of the "intrinsic" \Lya spectrum due to radiative transfer effects \citep[as discussed, e.g., in][]{Yang2016}.
In fact, the study of \citet{Gronke2018a} supports such a scenario \citep[also see the recent study by ][where $\gg 100$\kms observed spectra were modeled with narrow intrinsic spectra]{Li2020}. There, the authors post-process a hydrodynamic simulation of a galactic disk using \Lya radiative transfer, and show that the spectrum is indeed widened compared to H$\alpha$ due to ISM turbulence (cf. their figure 1 where the intrinsic spectrum of width $\sim$100\kms is widened to $\sim$400\kms due to radiative transfer effects within the gaseous disk)\footnote{In \citet{Gronke2018a}, the subsequent effect of the ``shell'' can be taken quite literally as they consider cosmic-ray driven outflows which are likely smooth and cold \citep[e.g.,][]{Girichidis2018} -- like a shell. However, in general this can also be due to CGM effects, or, a lack of numerical resolution leading to a ``shell'' by smoothing out otherwise multiphase structure \citep{Gronke2018a}.}. This could in principle explain the effect seen here where the intrinsic width obtained from the shell-model fitting is much wider than the one inferred from the observations.

\begin{figure*}[h!]
\centering
\includegraphics[width=0.90\textwidth]{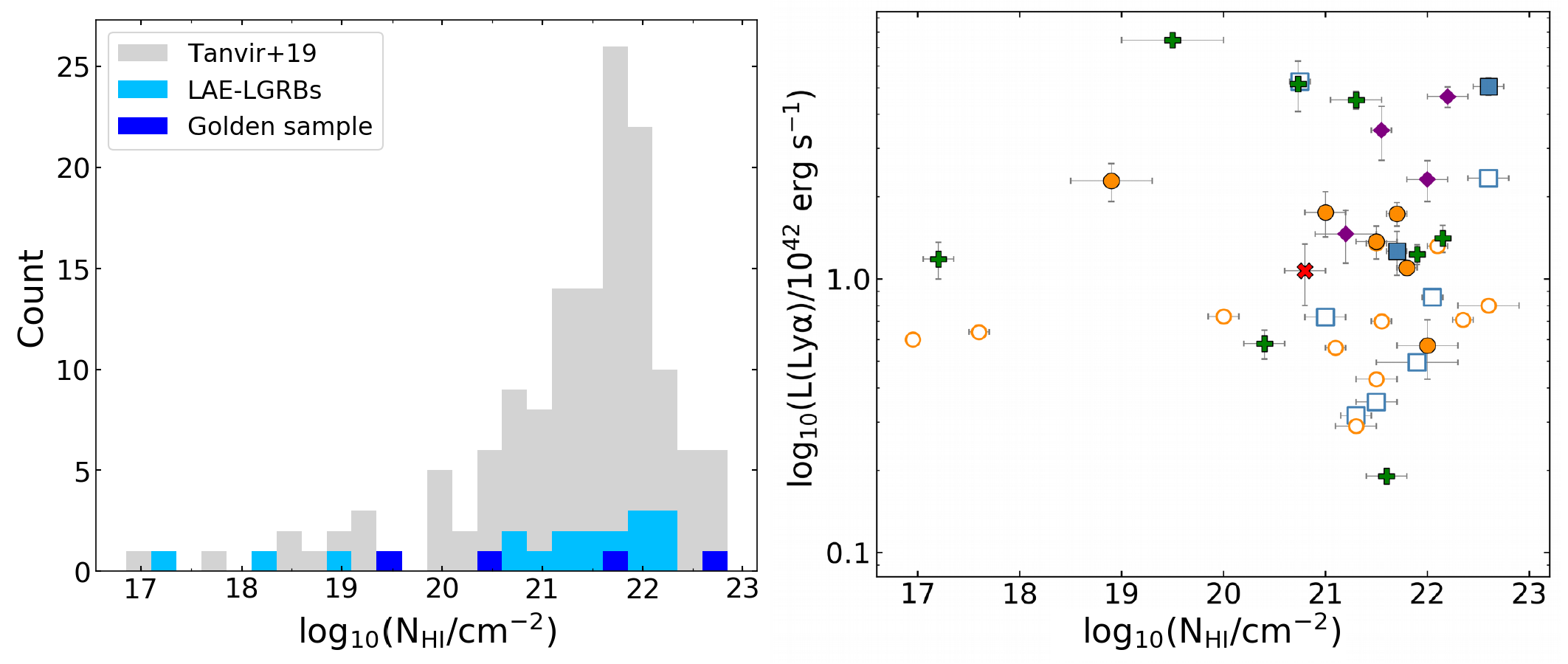}
 \caption{{\it Left Panel}: $\rm N_{HI}$ distribution of the 140 LGRBs reported in \citet{Tanvir2019} (grey). The 
 22/29 LAE-LGRBs having this information are shown in light blue, except for the GRB in the {\it golden sample} that are represented in dark blue. The different histograms are superimposed and not cumulative.
 {\it Right Panel}: \Lya luminosity distribution as function of $\rm N_{HI}$ column density. The symbols are the same as in Fig. \ref{FluxLimit}.
}
 \label{NHI_z}
\end{figure*} 

Another important discrepant value is $\rm N_{HI}$. The fit returns lower values than those determined from the afterglow spectroscopy in half of our sample. 
The two GRBs of the discrepant cases have much higher \NHIOA than the other GRBs in the {\it golden sample}, 
with $\log(\rm N_{\rm HI}^{\rm OA}/{\rm cm^{-2}}) > 21.5$ as opposed to values of $\log(\rm N_{\rm HI}^{\rm OA}/{\rm cm^{-2}})\sim20$.
The \Lya emission corresponds to a resonant transition making its escape unfavorable when the photons pass through a static high-\HI~density medium.
In the shell model, $\rm N_{HI}$ impacts the \Lya peak shift, with low $\rm N_{HI}$ values corresponding to small shifts and vice-versa. Indeed, when we constrain its value, the model is not able to reproduce the line profile for GRB\,060926 and 070110, and results in a peak shifted to larger velocities ($\sim$400\,\kms).

When considering all the LGRBs with a measured \NHIOA (reported in \citealt{Tanvir2019}), we see (Fig. \ref{NHI_z}) that such high \NHIOA values are common among LAE-LGRBs. Even if
the fraction of LAE-LGRBs from sub-DLA/Lyman-limit systems ($\log(\rm N_{\rm HI}^{\rm OA}/{\rm cm^{-2}})<20.3$; $22\% \pm 11\%$) seems to be higher (considering also the poor statistics) than that from DLA ($\log(\rm N_{\rm HI}^{\rm OA}/{\rm cm^{-2}})>20.3$; $15\% \pm 4\%$), there is no strong evidence for \Lya suppression associated with high \NHIOA.
Also, the \Lya luminosity seems not to depend on \NHIOA (see Fig. \ref{NHI_z}, right panel).

The discrepancy among the observed \NHIOA values and those found by the fitting, as well as the high \NHIOA values found among LAE-LGRBs, could be explained by the difference between the medium probed by \Lya photons and the afterglow emission.
LGRB lines of sight may go through regions with high \NHIOA, while \Lya photons may have escaped through lower $\rm N_{HI}$ lines of sight from the GRB young star-forming region. This would imply an anisotropic environment.
The theoretical studies of resonant line transfer through simplified anisotropic geometries \citep[e.g.,][]{Dijkstra2012,2014A&A...563A..77B,Eide2018} as well as turbulent medium with ionized channels \citep[e.g.,][]{Kimm2019, Kakiichi2019} show that \Lya spectra are shaped by the lowest density pathways. This predicts that in general, the column density probed by \Lya is less or equal to the one observed along the line of sight, in agreement to what we find for the LAE-LGRB modelling presented here.
To put constraints on the 
\HI anisotropy is important also to the consequences this can have on the ionizing escape fraction of galaxies \citep[e.g.,][]{Vielfaure2020}.
Already from this work where we find $\log (\rm N_{\rm HI}^{OA}({\rm LGRB}) / N_{\rm HI}(Ly\alpha))$\,$\sim$\,2-3 in two of the cases, we can speculate that it is difficult to explain this large anisotropy by purely turbulent driving. For reasonable Mach numbers $\mathcal{M}\lesssim 5$ studies show only $\sim$\,1\,dex difference (for $\sim 10\%$ of the sightlines; \citealp{Safarzadeh2016}) which suggests the need of radiative or mechanical feedback causing larger anisotropies \citep{Kimm2019, Kakiichi2019, Cen2020}.

The study of LGRB lines of sight shows that, from a statistical point of view, low-density channels surrounding star-forming regions should be rare \citep[][]{Tanvir2019}.  
This is consistent with LGRB progenitors being massive stars typically formed in dense molecular clouds and residing in gas-rich star-forming galaxies \citep[e.g.,][]{Jakobsson2006}.
The observation of the time-variability of fine-structure transitions, in LGRB afterglow spectra, has shown that the distance of the dominant absorbing clouds ranges from $\sim$50 pc to 1 kpc \citep[e.g.,][]{Vreeswijk2013}. 
The neutral gas probed by LGRB afterglows should therefore be connected to the star-forming regions where LGRBs explode. Since these regions are also expected to be the main birthplaces of \Lya photons, the properties of LGRB lines of sight could represent the average conditions of the environment surrounding \Lya emission regions.
If we consider the proportion of sub-DLA LGRBs as representative on average of the low-density channels through which the gas has favourable conditions for the \Lya radiation to escape, they would correspond to 12\% of all the possible lines of sight\footnote{Interestingly, previous studies of anisotropic \Lya escape show that such low opening angles can in principle be sufficient to set the properties of the isotropically emergent spectrum \citep{DijkstraGronkeSobral2016,Eide2018,Kakiichi2018}.}.

Even if LGRBs sites pinpoint regions with high star-formation activity (and therefore a significant production of \Lya photons), another possibility to explain the $\rm N_{HI}$ discrepancy is that most of the \Lya photons that succeed to escape from the galaxy may not originate from the young star-forming regions probed by GRBs
(as also proposed by \citealt{Vreeswijk2004} for GRB\,030323).
One way to investigate both scenarios would be through galaxy simulations, studying the global, the line of sight and the \Lya emission properties and comparing them to those determined by LGRB afterglow and host observations. This is beyond the scope of this paper and will be the object of a further paper.

%--------------------------------------------------------------------
\begin{figure*}[]
\centering
\includegraphics[width=\textwidth]{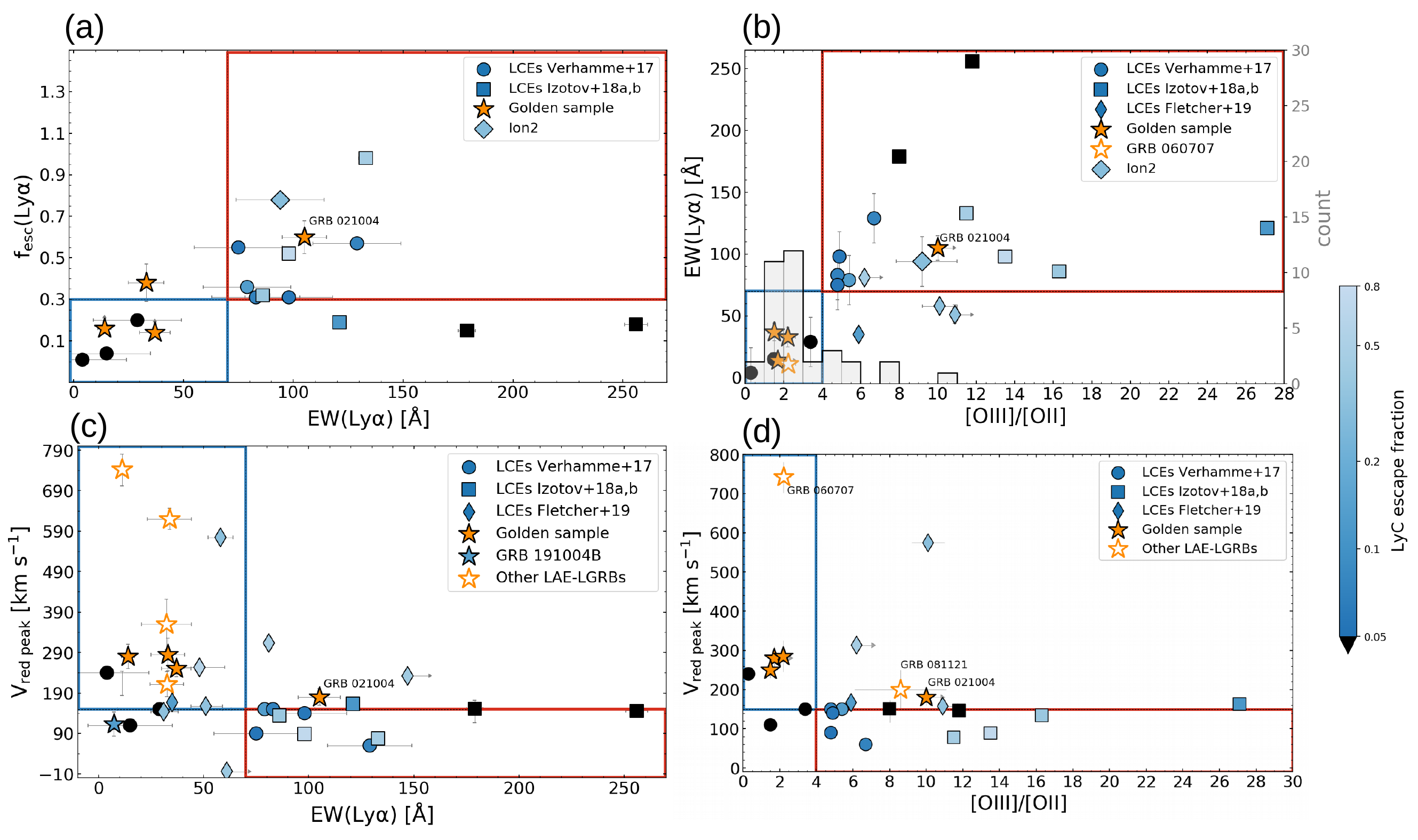}
 \caption{Comparison of the properties of the four LAE-LGRBs from the {\it golden sample} (orange stars) to other LAE-LGRBs and Lyman continuum emitters (LCEs) reported in \citet{Verhamme2017}, \citet{Izotov2018a, Izotov2018b} and \citet{Fletcher2019}. The white stars with orange contours represent LAE-LGRBs for which information on indirect indicators is available. The colorbar codes the \fescLyC of the LCEs. In the four panels the blue rectangle represents the space parameter of low LyC leakage (\fescLyC<5\%) while the red rectangle marks the space parameter of high LyC leakage (\fescLyC>5\%) as defined by \citet{Verhamme2017}.
 Panel (a) shows the escape fraction of \Lya photons as a function of the rest-frame \Lya EW.  
 Panel (b) shows the correlation between the rest-frame \Lya EW and the [OIII]/[OII] ratio. The distribution of [OIII]/[OII] ratio for the LGRB host galaxies from \citet{Kruhler2015} is also superimposed.
 Panel (c) shows the relation between the \Lya red peak velocity shift and the rest-frame \Lya EW. 
 Panel (d) shows is the correlation between the \Lya red peak velocity shift and the [OIII]/[OII] ratio.
 }
 \label{prop_comp}
\end{figure*}

The situation is even more complex if we look at the morphological properties of the host galaxies in more detail.   
The images of the host galaxies reveal a patchy and irregular morphology (see Fig.~\ref{FCs}).  
The HST images obtained for GRB\,021004 \citep{Fynbo2005} shows that the GRB position is superimposed to the strongest UV emitting region. This is not the case for GRB\,011211 \citep{Fynbo2003}, where the \Lya emission envelops the entire galaxy but the peak of emission is clearly separated from the GRB position.
Interestingly, this does not prevent the match between the shell model and the observations. 
A complex morphology is also found for other LAE-LGRB hosts (GRB\,000926 \citealt{Fynbo2002}, GRB\,050315 and GRB\,061222A \citealt{Blanchard2016}).
Such irregular and clumpy systems may represent mergers \citep[e.g.,][]{Conselice2003} or just clumps of star formation. The presence of a companion galaxy $\sim$15\,kpc away from the GRB\,021004 host, and the high velocity absorption found in its afterglow spectra, may suggest interaction of the two galaxies and the presence of outflowing gas.
In this light, the success of a very simple configuration as the one of the shell model in reproducing the \Lya profile and the observed parameters is maybe surprising.

    \subsection{Comparison to LyC leakers} \label{PropLAEs}

In Fig. \ref{prop_comp}, we compare properties of LAE-LGRBs to those of known LyC leakers in the literature.
For this comparison, we consider the LyC leakers at $z\sim3.1$ reported by \citet{Fletcher2019} and \textit{Ion2}  \citep[$z\sim3.2$,][]{Barros2016, Vanzella2016}.
We also consider those detected in the local Universe analysed in \citet{Verhamme2017} as well as the green peas studied in \citet{Izotov2018a, Izotov2018b}.
The properties compared are the \Lya escape fraction, the rest-frame \Lya equivalent width, the \Lya red peak velocity ($\rm v_{peak}$) and also the [OIII]/[OII] ratio (O$_{32}$). The last three properties have been proposed in the literature to correlate with ionizing and \Lya photons leakage. \citet{Verhamme2017} have pointed out rest-frame EW(\Lya) $> 70$ \AA, \fescLya $>0.3$, $\rm v_{peak}<150$\,\kms and O$_{32}>4$ as good indirect indicators of high LyC leakage (\fescLyC>5\%). The O$_{32}$ ratio is a proxy of the ionization state of the gas in the star forming regions and has been proposed as a marker of density-bounded \HII regions \citep{Jaskot2014,Nakajima2013,Nakajima2014,Stasinska2015}.
Its positive correlation with the escape fraction of ionizing and \Lya photons leakage have been shown in several studies for local GPs or high-redshift LAEs \citep[e.g.,][]{Nakajima2014, Izotov2016b, Izotov2018a, Izotov2018b, Barros2016, Verhamme2017, Fletcher2019}. 

From the comparisons we see that, except for GRB\,060707 and GRB\,060605 that show very high $\rm v_{peak}$, the LAE-LGRBs fall in the same parameter space as LyC leakers and follow the correlations between the indirect indicators found by \citet{Verhamme2017}. 
Following the study of \citet{Verhamme2017} on low-redshift LyC leakers, we can cut these plots in two regions 
corresponding to strong LyC leakers (\fescLyC$>5$\%, red rectangle) and weak LyC leakers (\fescLyC$<5$\%, blue rectangle). 
All LAE-LGRBs fall in the category of the weak LyC leakers except GRB\,021004 which appears systematically in good agreement with the region of the strong leakers. 
In panel (b) of Fig. \ref{prop_comp}, we also superimposed the distribution of [OIII]/[OII] ratio for the GRB sample of \citet{Kruhler2015} and GRB\,081121 reported here.
We see that for the majority of the GRBs this ratio is around two, with seven cases at $\rm O_{32} > 4$. GRB\,021004 has a high value of $\rm O_{32} > 10$ and is the strongest LAE of our {\it golden sample} with a substantial \fescLya~of 60\%, in agreement with potentially high \fescLyC.
Nevertheless, the \Lya profile of GRB\,021004 is a single peak with no residual flux at the \Lya line center. This is not the typical line shape observed for confirmed LyC emitters which have the tendency to show double- or triple-peak profile \citep{Verhamme2017,Rivera2017,Vanzella2020}.
Also our shell-model fitting suggests the column densities to be too large to allow LyC photons to escape.
However, this model already predicted similar \HI column density (\lognhicm$ \approx 19-20$) for four Green Pea galaxies out of the five with detected LyC emission reported in \citet{Yang2017}, suggesting that LyC emission can escape through holes in the ISM even if the \Lya photons probe denser neutral gas. 
The only LAE-LGRB for which LyC leakage has been detected along the LGRB line of sight (\fescLyC=$0.35^{+0.10}_{-0.11}$) is GRB\,191004B \citep{Vielfaure2020}. However, in Fig. \ref{prop_comp} (panel (c)) we can see that it does not fall in the high escape fraction region but in the lower left area. The reasons could be that (i) \fescLyC~is lower at the scale of the galaxy than along the LGRB line of sight, and (ii) the indicators of strong LyC leakage evolves with redshift.
As a comparison, the LyC emitters from \citet{Fletcher2019} are found out of the high escape region (red rectangle). They show lower rest-frame EW(\Lya) and higher $\rm v_{peak}$ than the local LyC emitters whereas their escape fraction of ionising photons is significantly higher (\fescLyC$ = 15-60$\%). 
This could also suggest that strong LyC leakers span a wider parameters space than predicted by the study of local LyC emitters. Overall it is clear that this kind of studies are still limited by the poor statistic and the current results show the difficulty of characterizing LyC leakage based only on these properties.

%--------------------------------------------------------------------
\section{Conclusions} \label{Conclusions}

In this paper we have studied the \Lya emission of LGRB host galaxies.
First, we provide a new census of LAEs among LGRB host galaxies. To date, there are 29 LAE-LGRBs. 
The fraction of LAEs among LGRB hosts varies from $\sim$10\% to 40\% depending on the sample and threshold considered.
Such statistics are lower than those found for LBG samples at similar redshift range, but they become comparable when taking into account only LAEs with rest-frame EW(\Lya) $> 20$ \AA. These results can be explained by the different selection criteria of the parent samples and by the shallower spectral observations of LGRB samples compared to LBG ones. 
We compared the properties of LAE-LGRBs to those of LGRB hosts in general and find 
evidences of \Lya emission suppression in dusty hosts.
We showed that, since LGRB hosts are not selected following usual criteria and techniques used for \Lya emission searches in star forming galaxies, they probe the regime of \Lya emission of the bulk of the UV-selected galaxy population at intermediate (and possibly high) redshifts.

We then selected a sub-sample of four LGRBs that allowed the combination of the emission properties of the host galaxies with the information on the ISM probed by the afterglow. We fit the \Lya emission of these galaxies using the shell model and we use the ancillary observational properties to constrain and test the model. 
We find that, without priors on the parameters, the shell model succeeds in reproducing the four profiles. However, the properties predicted by the model differ from the observed ones for the two cases (GRB\,060926 and 070110) having a high \NHIOA.
Similarly, constraining the model parameters using the values determined from the observations, the shell model succeeds in reproducing the two lower \NHIOA cases but fails for the two highest \NHIOA ones.
These results may be due to the fact that (i) the $\rm N_{HI}$ of the gas probed by the LGRB afterglow ($\rm N_{\rm HI}^{\rm OA}({\rm LGRB})$) is not representative of that encountered by the \Lya photons ($\rm N_{\rm HI}(\rm Ly\alpha)$); specifically we find $\rm N_{\rm HI}(\rm Ly\alpha) \le N_{\rm HI}^{\rm OA}({\rm LGRB})$, consistent with the picture that \Lya photons escape preferentially through low-column density channels and (ii) as known, the shell model relies on a too simplistic gas configuration. 
A consistent scenario would be provided by a turbulent medium widening the intrinsic \Lya spectrum, and an anisotropic \HI distribution allowing the escape of \Lya photons through lower density channels than probed by the LGRB sightline. 
Such a description is consistent with mechanical and radiative feedback mechanisms produced by starburst and supernova that clear out channels from the active regions within the galaxies.
Nevertheless, we stress that the afterglows of most LGRBs and LAE-LGRBs show $\log(\rm N_{\rm HI}^{\rm OA}/{\rm cm^{-2}}) >20.3$, implying that likely the \Lya photons produced by massive stars will predominantly be surrounded by such high-density gas, and low-density channels should be rare ($\sim$\,10\%).
Another possibility is that, even if dense star-forming regions produce a high fraction of \Lya photons, most of \Lya escaping photons come from less dense regions overall the galaxy. We will investigate further both scenarios in the future taking advantage of galaxy simulations.

Finally, we compared the properties of the LAE-LGRBs with those of LyC leakers in the literature. Specifically, we considered the commonly used \Lya indirect indicators of LyC leakage at low redshift and find that only one LAE-LGRB (GRB\,021004) has the corresponding values for being a strong LyC leaker. However (strong) LyC reported in the literature at $z\sim3$, span a wider range of parameter values than those commonly considered to point towards strong LyC leakage, questioning the use of such indicators at least at high redshift.   

The analysis presented in this paper needs larger statistics. 
To this purpose, we have been awarded X-shooter time to observe eleven more LAE-LGRBs reported in the census.
In the future it will be possible to extend these studies to higher redshift. Indeed, new space GRB missions like {\it Gamow Explorer} \citep{White2020} and {\it THESEUS}\footnote{\url{https://www.isdc.unige.ch/theseus/}} (\citealt{amati2018}; pre-selected as M5 European Space Agency mission), in synergy with ELT and JWST observations, will enable us to perform similar studies at $z>3.5$, and with better statistics.

% -------------------------------------------------------------
\begin{acknowledgements}
We thanks A. Verhamme for very useful discussions. 
This work is part of the {\it BEaPro} project (P.I.: S.D. Vergani) funded by the French {\it Agence Nationale de la Recherche} (ANR-16-CE31-0003) and supporting JBV PhD thesis.
JBV is now supported by LabEx UnivEarthS (ANR-10-LABX-0023 and ANR-18-IDEX-0001).
MG is supported by NASA through the NASA Hubble Fellowship grant HST-HF2-51409 and acknowledges support from HST grants HST-GO-15643.017-A, HST-AR-15039.003-A, and XSEDE grant TG-AST180036.
The Cosmic Dawn Center is funded by the Danish National Research Foundation under grant no.\ 140. 
BMJ is supported in part by Independent Research Fund Denmark grant DFF - 7014-00017.
Based on observations collected at the European Southern Observatory under ESO programmes: 079.A-0253, 084.A-0631, 084.A-0260, 084.A-0303, 084.D-0265,085.A-0009, 086.A-0073, 086.B-0954, 086.A-0533, 086.A-0874,
087.A-0055, 087.A-0451, 087.B-0737, 088.A-0051, 088.A-0644,
089.A-0067, 089.A-0120, 089.D-0256, 089.A-0868, 090.A-0088,
090.A-0760, 090.A-0825, 091.A-0342, 091.A-0703, 091.A-0877,
091.C-0934, 0091.C-0934, 092.D-0056,
092.D-0633, 092.A-0076, 092.A-0124, 092.A-0231, 093.A-0069,
094.A-0593, 094.A-0134, 095.A-0045, 095.B-0811,
096.A-0079, 097.A-0036, 097.D-0672, 098.D-0416, 098.A-0136, and 098.A-0055.
\end{acknowledgements}

\bibliographystyle{aa} % style aa.bst
\bibliography{refs} % your references Yourfile.bib

\begin{thebibliography}{148}
\expandafter\ifx\csname natexlab\endcsname\relax\def\natexlab#1{#1}\fi

\bibitem[{Ahn(2000)}]{Ahn2000}
Ahn, S.-H. 2000, \apj, 530, L9

\bibitem[{{Ahn}(2004)}]{Ahn2004}
{Ahn}, S.-H. 2004, \apjl, 601, L25

\bibitem[{{Ahn} {et~al.}(2003){Ahn}, {Lee}, \& {Lee}}]{Ahn2003}
{Ahn}, S.-H., {Lee}, H.-W., \& {Lee}, H.~M. 2003, \mnras, 340, 863

\bibitem[{{Amati} {et~al.}(2018){Amati}, {O'Brien}, {G{\"o}tz}, {Bozzo},
  {Tenzer}, {Frontera}, {Ghirlanda}, {Labanti}, {Osborne}, {Stratta}, {Tanvir},
  {Willingale}, {Attina}, {Campana}, {Castro-Tirado}, {Contini}, {Fuschino},
  {Gomboc}, {Hudec}, {Orleanski}, {Renotte}, {Rodic}, {Bagoly}, {Blain},
  {Callanan}, {Covino}, {Ferrara}, {Le Floch}, {Marisaldi}, {Mereghetti},
  {Rosati}, {Vacchi}, {D'Avanzo}, {Giommi}, {Piranomonte}, {Piro}, {Reglero},
  {Rossi}, {Santangelo}, {Salvaterra}, {Tagliaferri}, {Vergani}, {Vinciguerra},
  {Briggs}, {Campolongo}, {Ciolfi}, {Connaughton}, {Cordier}, {Morelli},
  {Orland ini}, {Adami}, {Argan}, {Atteia}, {Auricchio}, {Balazs}, {Baldazzi},
  {Basa}, {Basak}, {Bellutti}, {Bernardini}, {Bertuccio}, {Braga}, {Branchesi},
  {Brandt}, {Brocato}, {Budtz-Jorgensen}, {Bulgarelli}, {Burderi}, {Camp},
  {Capozziello}, {Caruana}, {Casella}, {Cenko}, {Chardonnet}, {Ciardi},
  {Colafrancesco}, {Dainotti}, {D'Elia}, {De Martino}, {De Pasquale}, {Del
  Monte}, {Della Valle}, {Drago}, {Evangelista}, {Feroci}, {Finelli},
  {Fiorini}, {Fynbo}, {Gal-Yam}, {Gendre}, {Ghisellini}, {Grado}, {Guidorzi},
  {Hafizi}, {Hanlon}, {Hjorth}, {Izzo}, {Kiss}, {Kumar}, {Kuvvetli}, {Lavagna},
  {Li}, {Longo}, {Lyutikov}, {Maio}, {Maiorano}, {Malcovati}, {Malesani},
  {Margutti}, {Martin-Carrillo}, {Masetti}, {McBreen}, {Mignani}, {Morgante},
  {Mundell}, {Nargaard-Nielsen}, {Nicastro}, {Palazzi}, {Paltani}, {Panessa},
  {Pareschi}, {Pe'er}, {Penacchioni}, {Pian}, {Piedipalumbo}, {Piran}, {Rauw},
  {Razzano}, {Read}, {Rezzolla}, {Romano}, {Ruffini}, {Savaglio}, {Sguera},
  {Schady}, {Skidmore}, {Song}, {Stanway}, {Starling}, {Topinka}, {Troja}, {van
  Putten}, {Vanzella}, {Vercellone}, {Wilson-Hodge}, {Yonetoku}, {Zampa},
  {Zampa}, {Zhang}, {Zhang}, {Zhang}, {Zhang}, {Antonelli}, {Bianco}, {Boci},
  {Boer}, {Botticella}, {Boulade}, {Butler}, {Campana}, {Capitanio}, {Celotti},
  {Chen}, {Colpi}, {Comastri}, {Cuby}, {Dadina}, {De Luca}, {Dong}, {Ettori},
  {Gandhi}, {Geza}, {Greiner}, {Guiriec}, {Harms}, {Hernanz}, {Hornstrup},
  {Hutchinson}, {Israel}, {Jonker}, {Kaneko}, {Kawai}, {Wiersema}, {Korpela},
  {Lebrun}, {Lu}, {MacFadyen}, {Malaguti}, {Maraschi}, {Meland ri}, {Modjaz},
  {Morris}, {Omodei}, {Paizis}, {P{\'a}ta}, {Petrosian}, {Rachevski}, {Rhoads},
  {Ryde}, {Sabau-Graziati}, {Shigehiro}, {Sims}, {Soomin}, {Sz{\'e}csi},
  {Urata}, {Uslenghi}, {Valenziano}, {Vianello}, {Vojtech}, {Watson}, \&
  {Zicha}}]{amati2018}
{Amati}, L., {O'Brien}, P., {G{\"o}tz}, D., {et~al.} 2018, Advances in Space
  Research, 62, 191

\bibitem[{{Appenzeller} {et~al.}(1998){Appenzeller}, {Fricke}, {F{\"u}rtig},
  {G{\"a}ssler}, {H{\"a}fner}, {Harke}, {Hess}, {Hummel}, {J{\"u}rgens},
  {Kudritzki}, {Mantel}, {Meisl}, {Muschielok}, {Nicklas}, {Rupprecht},
  {Seifert}, {Stahl}, {Szeifert}, \& {Tarantik}}]{Appenzeller1998}
{Appenzeller}, I., {Fricke}, K., {F{\"u}rtig}, W., {et~al.} 1998, The
  Messenger, 94, 1

\bibitem[{Arabsalmani {et~al.}(2018)Arabsalmani, Le~Floc'h, Dannerbauer,
  Feruglio, Daddi, Ciesla, Charmandaris, Japelj, Vergani, Duc, Basa, Bournaud,
  \& Elbaz}]{Arabsalmani2018b}
Arabsalmani, M., Le~Floc'h, E., Dannerbauer, H., {et~al.} 2018, \mnras, 476,
  2332

\bibitem[{{Bagley} {et~al.}(2017){Bagley}, {Scarlata}, {Henry}, {Rafelski},
  {Malkan}, {Teplitz}, {Dai}, {Baronchelli}, {Colbert}, {Rutkowski}, {Mehta},
  {Dressler}, {McCarthy}, {Bunker}, {Atek}, {Garel}, {Martin}, {Hathi}, \&
  {Siana}}]{Bagley2017}
{Bagley}, M.~B., {Scarlata}, C., {Henry}, A., {et~al.} 2017, \apj, 837, 11

\bibitem[{{Behrens} {et~al.}(2014){Behrens}, {Dijkstra}, \&
  {Niemeyer}}]{2014A&A...563A..77B}
{Behrens}, C., {Dijkstra}, M., \& {Niemeyer}, J.~C. 2014, \aap, 563, A77

\bibitem[{{Berger} \& {Rauch}(2008)}]{Berger2008}
{Berger}, E. \& {Rauch}, M. 2008, GCN, 8542, 1

\bibitem[{Blanchard {et~al.}(2016)Blanchard, Berger, \& fai
  Fong}]{Blanchard2016}
Blanchard, P.~K., Berger, E., \& fai Fong, W. 2016, \apj, 817, 144

\bibitem[{{Bolmer} {et~al.}(2019){Bolmer}, {Ledoux}, {Wiseman}, {De Cia},
  {Selsing}, {Schady}, {Greiner}, {Savaglio}, {Burgess}, {D'Elia}, {Fynbo},
  {Goldoni}, {Hartmann}, {Heintz}, {Jakobsson}, {Japelj}, {Kaper}, {Tanvir},
  {Vreeswijk}, \& {Zafar}}]{Bolmer2019}
{Bolmer}, J., {Ledoux}, C., {Wiseman}, P., {et~al.} 2019, \aap, 623, A43

\bibitem[{Brocklehurst(1971)}]{Brocklehurst1971}
Brocklehurst, M. 1971, \mnras, 153, 471

\bibitem[{Castro-Tirado {et~al.}(2010)Castro-Tirado, Møller, García-Segura,
  Gorosabel, Pérez, de~Ugarte~Postigo, Solano, Barrado, Klose, Kann,
  Castro~Cerón, Kouveliotou, Fynbo, Hjorth, Pedersen, Pian, Rol, Palazzi,
  Masetti, Tanvir, Vreeswijk, Andersen, Fruchter, Greiner, Wijers, \& van~den
  Heuvel}]{CastroTirado2010}
Castro-Tirado, A.~J., Møller, P., García-Segura, G., {et~al.} 2010, \aap,
  517, A61

\bibitem[{{Cen}(2020)}]{Cen2020}
{Cen}, R. 2020, \apjl, 889, L22

\bibitem[{Chabrier(2003)}]{Chabrier2003}
Chabrier, G. 2003, \pasp, 115, 763

\bibitem[{{Chen} {et~al.}(2007){Chen}, Prochaska, Ramirez‐Ruiz, Bloom,
  Dessauges‐Zavadsky, \& Foley}]{Chen2007}
{Chen}, H., Prochaska, J.~X., Ramirez‐Ruiz, E., {et~al.} 2007, \apj, 663, 420

\bibitem[{Chen(2012)}]{Chen2012}
Chen, H.-W. 2012, \mnras, 419, 3039

\bibitem[{Chisholm {et~al.}(2015)Chisholm, Tremonti, Leitherer, Chen, Wofford,
  \& Lundgren}]{Chisholm2015}
Chisholm, J., Tremonti, C.~A., Leitherer, C., {et~al.} 2015, \apj, 811, 149

\bibitem[{Conselice {et~al.}(2003)Conselice, Bershady, Dickinson, \&
  Papovich}]{Conselice2003}
Conselice, C.~J., Bershady, M.~A., Dickinson, M., \& Papovich, C. 2003, \aj,
  126, 1183–1207

\bibitem[{Cucchiara {et~al.}(2015)Cucchiara, Fumagalli, Rafelski, Kocevski,
  Prochaska, Cooke, \& Becker}]{Cucchiara2015}
Cucchiara, A., Fumagalli, M., Rafelski, M., {et~al.} 2015, \apj, 804, 51

\bibitem[{{de Barros} {et~al.}(2016){de Barros}, {Vanzella}, {Amor{\'\i}n},
  {Castellano}, {Siana}, {Grazian}, {Suh}, {Balestra}, {Vignali}, {Verhamme},
  {Zamorani}, {Mignoli}, {Hasinger}, {Comastri}, {Pentericci},
  {P{\'e}rez-Montero}, {Fontana}, {Giavalisco}, \& {Gilli}}]{Barros2016}
{de Barros}, S., {Vanzella}, E., {Amor{\'\i}n}, R., {et~al.} 2016, \aap, 585,
  A51

\bibitem[{{de Ugarte Postigo} {et~al.}(2005){de Ugarte Postigo}, Castro-Tirado,
  Gorosabel, Jóhannesson, Björnsson, Gudmundsson, Bremer, Pak, Tanvir,
  Castro~Cerón, Guzyi, Jelínek, Klose, Pérez-Ramírez, Aceituno,
  Campo~Bagatín, Covino, Cardiel, Fathkullin, Henden, Huferath, Kurata,
  Malesani, Mannucci, Ruiz-Lapuente, Sokolov, Thiele, Wisotzki, Antonelli,
  Bartolini, Boattini, Guarnieri, Piccioni, Pizzichini, {M. del Principe},
  di~Paola, Fugazza, Ghisellini, Hunt, Konstantinova, Masetti, Palazzi, Pian,
  Stefanon, Testa, \& Tristram}]{DeUgartePostigo2005}
{de Ugarte Postigo}, Castro-Tirado, A.~J., Gorosabel, J., {et~al.} 2005, \aap,
  443, 841

\bibitem[{{D'Elia} {et~al.}(2019){D'Elia}, {Fynbo}, {Izzo}, {Malesani},
  {Heintz}, {Tanvir}, {de Ugarte Postigo}, {Vergani}, \&
  {collabaration}}]{DElia2019}
{D'Elia}, V., {Fynbo}, J.~P.~U., {Izzo}, L., {et~al.} 2019, GCN, 25956, 1

\bibitem[{Dessauges-Zavadsky {et~al.}(2010)Dessauges-Zavadsky, D'Odorico,
  Schaerer, Modigliani, Tapken, \& Vernet}]{DessaugesZavadsky2010}
Dessauges-Zavadsky, M., D'Odorico, S., Schaerer, D., {et~al.} 2010, \aap, 510,
  A26

\bibitem[{{Dijkstra} {et~al.}(2016){Dijkstra}, {Gronke}, \&
  {Sobral}}]{DijkstraGronkeSobral2016}
{Dijkstra}, M., {Gronke}, M., \& {Sobral}, D. 2016, \apj, 823, 74

\bibitem[{Dijkstra \& Kramer(2012)}]{Dijkstra2012}
Dijkstra, M. \& Kramer, R. 2012, \mnras, 424, 1672

\bibitem[{D’Avanzo {et~al.}(2010)D’Avanzo, Perri, Fugazza, Salvaterra,
  Chincarini, Margutti, Wu, Thöne, Fernández-Soto, Ukwatta, Burrows, Gehrels,
  Meszaros, Toma, Zhang, Covino, Campana, D’Elia, Della~Valle, \&
  Piranomonte}]{DAvanzo2010}
D’Avanzo, P., Perri, M., Fugazza, D., {et~al.} 2010, \aap, 522, A20

\bibitem[{Eide {et~al.}(2018)Eide, Gronke, Dijkstra, \& Hayes}]{Eide2018}
Eide, M.~B., Gronke, M., Dijkstra, M., \& Hayes, M. 2018, \apj, 856, 156

\bibitem[{{El{\'\i}asd{\'o}ttir} {et~al.}(2009){El{\'\i}asd{\'o}ttir}, {Fynbo},
  {Hjorth}, {Ledoux}, {Watson}, {Andersen}, {Malesani}, {Vreeswijk},
  {Prochaska}, {Sollerman}, \& {Jaunsen}}]{Eliasdottir2009}
{El{\'\i}asd{\'o}ttir}, {\'A}., {Fynbo}, J.~P.~U., {Hjorth}, J., {et~al.} 2009,
  \apj, 697, 1725

\bibitem[{Erb {et~al.}(2014)Erb, Steidel, Trainor, Bogosavljevi{\'{c}},
  Shapley, Nestor, Kulas, Law, Strom, Rudie, Reddy, Pettini, Konidaris, Mace,
  Matthews, \& McLean}]{Erb2014}
Erb, D.~K., Steidel, C.~C., Trainor, R.~F., {et~al.} 2014, \apj, 795, 33

\bibitem[{Fiore {et~al.}(2005)Fiore, D’Elia, Lazzati, Perna, Sbordone,
  Stratta, Meurs, Ward, Antonelli, Chincarini, Covino, Di~Paola, Fontana,
  Ghisellini, Israel, Frontera, Marconi, Stella, Vietri, \& Zerbi}]{Fiore2005}
Fiore, F., D’Elia, V., Lazzati, D., {et~al.} 2005, \apj, 624, 853

\bibitem[{{Fletcher} {et~al.}(2019){Fletcher}, {Tang}, {Robertson}, {Nakajima},
  {Ellis}, {Stark}, \& {Inoue}}]{Fletcher2019}
{Fletcher}, T.~J., {Tang}, M., {Robertson}, B.~E., {et~al.} 2019, \apj, 878, 87

\bibitem[{{Fox}(2002)}]{Fox2002}
{Fox}, D.~W. 2002, GCN, 1564, 1

\bibitem[{Friis {et~al.}(2015)Friis, De~Cia, Krühler, Fynbo, Ledoux,
  Vreeswijk, Watson, Malesani, Gorosabel, Starling, Jakobsson, Varela,
  Wiersema, Drachmann, Trotter, Thöne, de~Ugarte~Postigo, D'Elia, Elliott,
  Maturi, Goldoni, Greiner, Haislip, Kaper, Knust, LaCluyze, Milvang-Jensen,
  Reichart, Schulze, Sudilovsky, Tanvir, \& Vergani}]{Friis2015}
Friis, M., De~Cia, A., Krühler, T., {et~al.} 2015, \mnras, 451, 167

\bibitem[{Fynbo {et~al.}(2005)Fynbo, Gorosabel, Smette, Fruchter, Hjorth,
  Pedersen, Levan, Burud, Sahu, Vreeswijk, Bergeron, Kouveliotou, Tanvir,
  Thorsett, Wijers, Castro~Ceron, Castro‐Tirado, Garnavich, Holland,
  Jakobsson, Moller, Nugent, Pian, Rhoads, Thomsen, Watson, \&
  Woosley}]{Fynbo2005}
Fynbo, J. P.~U., Gorosabel, J., Smette, A., {et~al.} 2005, \apj, 633, 317

\bibitem[{Fynbo {et~al.}(2003)Fynbo, Jakobsson, Møller, Hjorth, Thomsen,
  Andersen, Fruchter, Gorosabel, Holland, Ledoux, Pedersen, Rhoads, Weidinger,
  \& Wijers}]{Fynbo2003}
Fynbo, J. P.~U., Jakobsson, P., Møller, P., {et~al.} 2003, \aap, 406, L63

\bibitem[{Fynbo {et~al.}(2009)Fynbo, Jakobsson, Prochaska, Malesani, Ledoux,
  de~Ugarte~Postigo, Nardini, Vreeswijk, Wiersema, Hjorth, Sollerman, Chen,
  Thöne, Björnsson, Bloom, Castro-Tirado, Christensen, De~Cia, Fruchter,
  Gorosabel, Graham, Jaunsen, Jensen, Kann, Kouveliotou, Levan, Maund, Masetti,
  Milvang-Jensen, Palazzi, Perley, Pian, Rol, Schady, Starling, Tanvir, Watson,
  Xu, Augusteijn, Grundahl, Telting, \& Quirion}]{Fynbo2009}
Fynbo, J. P.~U., Jakobsson, P., Prochaska, J.~X., {et~al.} 2009, \apj, 185, 526

\bibitem[{Fynbo {et~al.}(2002)Fynbo, Møller, Thomsen, Hjorth, Gorosabel,
  Andersen, Egholm, Holland, Jensen, Pedersen, \& Weidinger}]{Fynbo2002}
Fynbo, J. P.~U., Møller, P., Thomsen, B., {et~al.} 2002, \aap, 388, 425

\bibitem[{{Fynbo} {et~al.}(1999){Fynbo}, Møller, \& Warren}]{Fynbo1999}
{Fynbo}, J.~U., Møller, P., \& Warren, S.~J. 1999, \mnras, 305, 849

\bibitem[{Gehrels \& Razzaque(2013)}]{Gehrels2013a}
Gehrels, N. \& Razzaque, S. 2013, Front. Phys., 8, 661

\bibitem[{{Girichidis} {et~al.}(2018){Girichidis}, {Naab}, {Hanasz}, \&
  {Walch}}]{Girichidis2018}
{Girichidis}, P., {Naab}, T., {Hanasz}, M., \& {Walch}, S. 2018, \mnras, 479,
  3042

\bibitem[{Graham \& Fruchter(2017)}]{Graham2017}
Graham, J.~F. \& Fruchter, A.~S. 2017, \apj, 834, 170

\bibitem[{{Gronke}(2017)}]{Gronke2017}
{Gronke}, M. 2017, \aap, 608, A139

\bibitem[{{Gronke} {et~al.}(2015){Gronke}, Bull, \& Dijkstra}]{Gronke2015}
{Gronke}, M., Bull, P., \& Dijkstra, M. 2015, \apj, 812, 123

\bibitem[{{Gronke} {et~al.}(2018){Gronke}, {Girichidis}, {Naab}, \&
  {Walch}}]{Gronke2018a}
{Gronke}, M., {Girichidis}, P., {Naab}, T., \& {Walch}, S. 2018, \apjl, 862, L7

\bibitem[{Hartoog {et~al.}(2015)Hartoog, Malesani, Fynbo, Goto, Krühler,
  Vreeswijk, De~Cia, Xu, Møller, Covino, D’Elia, Flores, Goldoni, Hjorth,
  Jakobsson, Krogager, Kaper, Ledoux, Levan, Milvang-Jensen, Sollerman, Sparre,
  Tagliaferri, Tanvir, de~Ugarte~Postigo, Vergani, Wiersema, Datson, Salinas,
  Mikkelsen, \& Aghanim}]{Hartoog2015}
Hartoog, O.~E., Malesani, D., Fynbo, J. P.~U., {et~al.} 2015, \aap, 580, A139

\bibitem[{Hashimoto {et~al.}(2015)Hashimoto, Verhamme, Ouchi, Shimasaku,
  Schaerer, Nakajima, Shibuya, Rauch, Ono, \& Goto}]{Hashimoto2015}
Hashimoto, T., Verhamme, A., Ouchi, M., {et~al.} 2015, \apj, 812, 157

\bibitem[{Heintz {et~al.}(2019)Heintz, Ledoux, Fynbo, Jakobsson, Noterdaeme,
  Krogager, Bolmer, Møller, Vergani, Watson, Zafar, De~Cia, Tanvir, Malesani,
  Japelj, Covino, \& Kaper}]{Heintz2019a}
Heintz, K.~E., Ledoux, C., Fynbo, J. P.~U., {et~al.} 2019, \aap, 621, A20

\bibitem[{{Henry} {et~al.}(2015){Henry}, {Scarlata}, {Martin}, \&
  {Erb}}]{Henry2015}
{Henry}, A., {Scarlata}, C., {Martin}, C.~L., \& {Erb}, D. 2015, \apj, 809, 19

\bibitem[{Hjorth {et~al.}(2012)Hjorth, Malesani, Jakobsson, Jaunsen, Fynbo,
  Gorosabel, Krühler, Levan, Michałowski, Milvang-Jensen, Møller, Schulze,
  Tanvir, \& Watson}]{Hjorth2012}
Hjorth, J., Malesani, D., Jakobsson, P., {et~al.} 2012, \apj, 756, 187

\bibitem[{Hjorth {et~al.}(2003)Hjorth, Sollerman, Møller, Fynbo, Woosley,
  Kouveliotou, Tanvir, Greiner, Andersen, Castro-Tirado, Cerón, Fruchter,
  Gorosabel, Jakobsson, Kaper, Klose, Masetti, Pedersen, Pedersen, Pian,
  Palazzi, Rhoads, Rol, Heuvel, Vreeswijk, Watson, \& Wijers}]{Hjorth2003a}
Hjorth, J., Sollerman, J., Møller, P., {et~al.} 2003, Nature, 423, 847

\bibitem[{{Holland} {et~al.}(2006){Holland}, {Barthelmy}, {Burrows}, {Capalbi},
  {Conciatore}, {Cummings}, {Gehrels}, {Guidorzi}, {Kennea}, {Krimm},
  {Mangano}, {Markwardt}, {Marshall}, {Osborne}, {Page}, {Palmer}, {Romano},
  {Sato}, {Stamatikos}, {vanden Berk}, \& {Ziaeepour}}]{Holland2006}
{Holland}, S.~T., {Barthelmy}, S.~D., {Burrows}, D.~N., {et~al.} 2006, GCN,
  5612, 1

\bibitem[{{Izotov} {et~al.}(2016){Izotov}, {Schaerer}, {Thuan}, {Worseck},
  {Guseva}, {Orlitov{\'a}}, \& {Verhamme}}]{Izotov2016b}
{Izotov}, Y.~I., {Schaerer}, D., {Thuan}, T.~X., {et~al.} 2016, \mnras, 461,
  3683

\bibitem[{{Izotov} {et~al.}(2018a){Izotov}, {Schaerer}, {Worseck}, {Guseva},
  {Thuan}, {Verhamme}, {Orlitov{\'a}}, \& {Fricke}}]{Izotov2018a}
{Izotov}, Y.~I., {Schaerer}, D., {Worseck}, G., {et~al.} 2018a, \mnras, 474,
  4514

\bibitem[{{Izotov} {et~al.}(2018b){Izotov}, {Worseck}, {Schaerer}, {Guseva},
  {Thuan}, {Fricke}, \& {Orlitov{\'a}}}]{Izotov2018b}
{Izotov}, Y.~I., {Worseck}, G., {Schaerer}, D., {et~al.} 2018b, \mnras, 478,
  4851

\bibitem[{Jakobsson {et~al.}(2005)Jakobsson, Björnsson, Fynbo, Jóhannesson,
  Hjorth, Thomsen, Møller, Watson, Jensen, Östlin, Gorosabel, \&
  Gudmundsson}]{Jakobsson2005}
Jakobsson, P., Björnsson, G., Fynbo, J. P.~U., {et~al.} 2005, \mnras, 362, 245

\bibitem[{Jakobsson {et~al.}(2006)Jakobsson, Fynbo, Ledoux, Vreeswijk, Kann,
  Hjorth, Priddey, Tanvir, Reichart, Gorosabel, Klose, Watson, Sollerman,
  Fruchter, de~Ugarte~Postigo, Wiersema, Björnsson, Chapman, Thöne, Pedersen,
  \& Jensen}]{Jakobsson2006}
Jakobsson, P., Fynbo, J. P.~U., Ledoux, C., {et~al.} 2006, \aap, 460, L13

\bibitem[{Jakobsson {et~al.}(2003)Jakobsson, Hjorth, Fynbo, Gorosabel,
  Pedersen, Burud, Levan, Kouveliotou, Tanvir, Fruchter, Rhoads, Grav, Hansen,
  Michelsen, Andersen, Jensen, Pedersen, Thomsen, Weidinger, Bhargavi, Cowsik,
  \& Pandey}]{Jakobsson2003}
Jakobsson, P., Hjorth, J., Fynbo, J. P.~U., {et~al.} 2003, \aap, 408, 941

\bibitem[{Jakobsson {et~al.}(2004)Jakobsson, Hjorth, Fynbo, Weidinger,
  Gorosabel, Ledoux, Watson, Björnsson, Gudmundsson, Wijers, Møller,
  Pedersen, Sollerman, Henden, Jensen, Gilmore, Kilmartin, Levan,
  Castro~Cerón, Castro-Tirado, Fruchter, Kouveliotou, Masetti, \&
  Tanvir}]{Jakobsson2004b}
Jakobsson, P., Hjorth, J., Fynbo, J. P.~U., {et~al.} 2004, \aap, 427, 785

\bibitem[{{Jakobsson} {et~al.}(2012){Jakobsson}, {Hjorth}, {Malesani},
  {Chapman}, {Fynbo}, {Tanvir}, {Milvang-Jensen}, {Vreeswijk}, {Letawe}, \&
  {Starling}}]{Jakobsson2012}
{Jakobsson}, P., {Hjorth}, J., {Malesani}, D., {et~al.} 2012, \apj, 752, 62

\bibitem[{{Japelj} {et~al.}(2015){Japelj}, {Covino}, {Gomboc}, {Vergani},
  {Goldoni}, {Selsing}, {Cano}, {D'Elia}, {Flores}, {Fynbo}, {Hammer},
  {Hjorth}, {Jakobsson}, {Kaper}, {Kopa{\v{c}}}, {Kr{\"u}hler}, {Melandri},
  {Piranomonte}, {S{\'a}nchez-Ram{\'\i}rez}, {Tagliaferri}, {Tanvir}, {de
  Ugarte Postigo}, {Watson}, \& {Wijers}}]{Japelj2015}
{Japelj}, J., {Covino}, S., {Gomboc}, A., {et~al.} 2015, \aap, 579, A74

\bibitem[{Japelj {et~al.}(2016)Japelj, Vergani, Salvaterra, D’Avanzo,
  Mannucci, Fernandez-Soto, Boissier, Hunt, Atek, Rodríguez-Muñoz, Scodeggio,
  Cristiani, Le~Floc’h, Flores, Gallego, Ghirlanda, Gomboc, Hammer, Perley,
  Pescalli, Petitjean, Puech, Rafelski, \& Tagliaferri}]{Japelj2016a}
Japelj, J., Vergani, S.~D., Salvaterra, R., {et~al.} 2016, \aap, 590, A129

\bibitem[{{Jaskot} \& {Oey}(2014)}]{Jaskot2014}
{Jaskot}, A.~E. \& {Oey}, M.~S. 2014, \apjl, 791, L19

\bibitem[{{Kakiichi} {et~al.}(2018){Kakiichi}, {Ellis}, {Laporte}, {Zitrin},
  {Eilers}, {Ryan-Weber}, {Meyer}, {Robertson}, {Stark}, \&
  {Bosman}}]{Kakiichi2018}
{Kakiichi}, K., {Ellis}, R.~S., {Laporte}, N., {et~al.} 2018, \mnras, 479, 43

\bibitem[{{Kakiichi} \& {Gronke}(2019)}]{Kakiichi2019}
{Kakiichi}, K. \& {Gronke}, M. 2019, arXiv e-prints, arXiv:1905.02480

\bibitem[{{Kennicutt}(1998)}]{Kennicutt1998}
{Kennicutt}, Robert~C., J. 1998, \araa, 36, 189

\bibitem[{{Kimm} {et~al.}(2019){Kimm}, {Blaizot}, {Garel}, {Michel-Dansac},
  {Katz}, {Rosdahl}, {Verhamme}, \& {Haehnelt}}]{Kimm2019}
{Kimm}, T., {Blaizot}, J., {Garel}, T., {et~al.} 2019, \mnras, 486, 2215

\bibitem[{{Kruehler} \& Schady(2017)}]{Kruehler2017}
{Kruehler}, T. \& Schady, P. 2017, Figshare,
  https://doi.org/10.6084/m9.figshare.\-4776886.v3

\bibitem[{Krühler {et~al.}(2015)Krühler, Malesani, Fynbo, Hartoog, Hjorth,
  Jakobsson, Perley, Rossi, Schady, Schulze, Tanvir, Vergani, Wiersema, Afonso,
  Bolmer, Cano, Covino, D’Elia, de~Ugarte~Postigo, Filgas, Friis, Graham,
  Greiner, Goldoni, Gomboc, Hammer, Japelj, Kann, Kaper, Klose, Levan,
  Leloudas, Milvang-Jensen, Nicuesa~Guelbenzu, Palazzi, Pian, Piranomonte,
  Sánchez-Ramírez, Savaglio, Selsing, Tagliaferri, Vreeswijk, Watson, \&
  Xu}]{Kruhler2015}
Krühler, T., Malesani, D., Fynbo, J. P.~U., {et~al.} 2015, \aap, 581, A125

\bibitem[{Krühler {et~al.}(2012)Krühler, Malesani, Milvang-Jensen, Fynbo,
  Hjorth, Jakobsson, Levan, Sparre, Tanvir, \& Watson}]{Kruhler2012}
Krühler, T., Malesani, D., Milvang-Jensen, B., {et~al.} 2012, \apj, 758, 46

\bibitem[{Kulkarni {et~al.}(1998)Kulkarni, Djorgovski, Ramaprakash, Goodrich,
  Bloom, Adelberger, Kundic, Lubin, Frail, Frontera, Feroci, Nicastro, Barth,
  Davis, Filippenko, \& Newman}]{Kulkarni1998}
Kulkarni, S.~R., Djorgovski, S.~G., Ramaprakash, A.~N., {et~al.} 1998, Nature,
  393, 35

\bibitem[{{Kunth} {et~al.}(1998){Kunth}, {Mas-Hesse}, {Terlevich}, {Terlevich},
  {Lequeux}, \& {Fall}}]{Kunth1998}
{Kunth}, D., {Mas-Hesse}, J.~M., {Terlevich}, E., {et~al.} 1998, \aap, 334, 11

\bibitem[{{Laursen} {et~al.}(2009){Laursen}, {Sommer-Larsen}, \&
  {Andersen}}]{Laursen2009}
{Laursen}, P., {Sommer-Larsen}, J., \& {Andersen}, A.~C. 2009, \apj, 704, 1640

\bibitem[{{Leclercq} {et~al.}(2017){Leclercq}, {Bacon}, {Wisotzki}, {Mitchell},
  {Garel}, {Verhamme}, {Blaizot}, {Hashimoto}, {Herenz}, {Conseil},
  {Cantalupo}, {Inami}, {Contini}, {Richard}, {Maseda}, {Schaye}, {Marino},
  {Akhlaghi}, {Brinchmann}, \& {Carollo}}]{Leclercq2017}
{Leclercq}, F., {Bacon}, R., {Wisotzki}, L., {et~al.} 2017, \aap, 608, A8

\bibitem[{{Li} {et~al.}(2020){Li}, {Steidel}, {Gronke}, \& {Chen}}]{Li2020}
{Li}, Z., {Steidel}, C.~C., {Gronke}, M., \& {Chen}, Y. 2020, arXiv e-prints,
  arXiv:2008.09130

\bibitem[{Lidman {et~al.}(2012)Lidman, Hayes, Jones, Schaerer, Westra, Tapken,
  Meisenheimer, \& Verhamme}]{Lidman2012}
Lidman, C., Hayes, M., Jones, D.~H., {et~al.} 2012, \mnras, 420, 1946

\bibitem[{{Lyman} {et~al.}(2017){Lyman}, {Levan}, {Tanvir}, {Fynbo}, {McGuire},
  {Perley}, {Angus}, {Bloom}, {Conselice}, {Fruchter}, {Hjorth}, {Jakobsson},
  \& {Starling}}]{Lyman2017}
{Lyman}, J.~D., {Levan}, A.~J., {Tanvir}, N.~R., {et~al.} 2017, \mnras, 467,
  1795

\bibitem[{{Malesani} {et~al.}(2007){Malesani}, {Jaunsen}, \&
  {Vreeswijk}}]{Malesani2007}
{Malesani}, D., {Jaunsen}, A.~O., \& {Vreeswijk}, P.~M. 2007, GCN, 6015, 1

\bibitem[{Malesani {et~al.}(2013)Malesani, Xu, Fynbo, Kruehler, Perley,
  Vergani, \& Goldoni}]{Malesani2013}
Malesani, D., Xu, D., Fynbo, J. P.~U., {et~al.} 2013, GCN, 14291, 1

\bibitem[{{Matthee} {et~al.}(2021){Matthee}, {Sobral}, {Hayes}, {Pezzulli},
  {Gronke}, {Schaerer}, {Naidu}, {R{\"o}ttgering}, {Calhau}, {Paulino-Afonso},
  {Santos}, \& {Amor{\'\i}n}}]{Matthee2021}
{Matthee}, J., {Sobral}, D., {Hayes}, M., {et~al.} 2021, arXiv e-prints,
  arXiv:2102.07779

\bibitem[{{Matthee} {et~al.}(2016){Matthee}, {Sobral}, {Oteo}, {Best}, {Smail},
  {R{\"o}ttgering}, \& {Paulino-Afonso}}]{Matthee2016}
{Matthee}, J., {Sobral}, D., {Oteo}, I., {et~al.} 2016, \mnras, 458, 449

\bibitem[{Milvang-Jensen {et~al.}(2012)Milvang-Jensen, Fynbo, Malesani, Hjorth,
  Jakobsson, \& Møller}]{MilvangJensen2012}
Milvang-Jensen, B., Fynbo, J. P.~U., Malesani, D., {et~al.} 2012, \apj, 756, 25

\bibitem[{Mirabal {et~al.}(2003)Mirabal, Halpern, Chornock, Filippenko,
  Terndrup, Armstrong, Kemp, Thorstensen, Tavarez, \& Espaillat}]{Mirabal2003}
Mirabal, N., Halpern, J.~P., Chornock, R., {et~al.} 2003, 595, 15

\bibitem[{{Modigliani} {et~al.}(2010){Modigliani}, {Goldoni}, {Royer},
  {Haigron}, {Guglielmi}, {Fran{\c{c}}ois}, {Horrobin}, {Bristow}, {Vernet},
  {Moehler}, {Kerber}, {Ballester}, {Mason}, \& {Christensen}}]{Modigliani2010}
{Modigliani}, A., {Goldoni}, P., {Royer}, F., {et~al.} 2010, SPIE Conference
  Series, Vol. 7737, {The X-shooter pipeline}, 773728

\bibitem[{Møller {et~al.}(2002)Møller, Fynbo, Hjorth, Thomsen, Egholm,
  Andersen, Gorosabel, Holland, Jakobsson, Jensen, Pedersen, Pedersen, \&
  Weidinger}]{Moller2002}
Møller, P., Fynbo, J. P.~U., Hjorth, J., {et~al.} 2002, \aap, 396, L21

\bibitem[{Nakajima \& Ouchi(2014)}]{Nakajima2014}
Nakajima, K. \& Ouchi, M. 2014, \mnras, 442, 900

\bibitem[{Nakajima {et~al.}(2013)Nakajima, Ouchi, Shimasaku, Hashimoto, Ono, \&
  Lee}]{Nakajima2013}
Nakajima, K., Ouchi, M., Shimasaku, K., {et~al.} 2013, \apj, 769, 3

\bibitem[{{Neufeld}(1990)}]{Neufeld1990}
{Neufeld}, D.~A. 1990, \apj, 350, 216

\bibitem[{{Orlitov{\'a}} {et~al.}(2018){Orlitov{\'a}}, {Verhamme}, {Henry},
  {Scarlata}, {Jaskot}, {Oey}, \& {Schaerer}}]{Orlitova2018}
{Orlitov{\'a}}, I., {Verhamme}, A., {Henry}, A., {et~al.} 2018, \aap, 616, A60

\bibitem[{{Osterbrock}(1989)}]{Osterbrock1989}
{Osterbrock}, D.~E. 1989, {Astrophysics of gaseous nebulae and active galactic
  nuclei} (\apj)

\bibitem[{Ouchi {et~al.}(2009)Ouchi, Ono, Egami, Saito, Oguri, McCarthy,
  Farrah, Kashikawa, Momcheva, Shimasaku, Nakanishi, Furusawa, Akiyama, Dunlop,
  Mortier, Okamura, Hayashi, Cirasuolo, Dressler, Iye, Jarvis, Kodama, Martin,
  McLure, Ohta, Yamada, \& Yoshida}]{Ouchi2009}
Ouchi, M., Ono, Y., Egami, E., {et~al.} 2009, \apj, 696, 1164

\bibitem[{Palmerio {et~al.}(2019)Palmerio, Vergani, Salvaterra, Sanders,
  Japelj, Vidal-García, D’Avanzo, Corre, Perley, Shapley, Boissier, Greiner,
  Le~Floc’h, \& Wiseman}]{Palmerio2019}
Palmerio, J.~T., Vergani, S.~D., Salvaterra, R., {et~al.} 2019, \aap, 623, A26

\bibitem[{Pei(1992)}]{Pei1992}
Pei, Y.~C. 1992, \apj, 395, 130

\bibitem[{{Pentericci} {et~al.}(2007){Pentericci}, {Grazian}, {Fontana},
  {Salimbeni}, {Santini}, {de Santis}, {Gallozzi}, \&
  {Giallongo}}]{Pentericci2007}
{Pentericci}, L., {Grazian}, A., {Fontana}, A., {et~al.} 2007, \aap, 471, 433

\bibitem[{{Pentericci} {et~al.}(2010){Pentericci}, {Grazian}, {Scarlata},
  {Fontana}, {Castellano}, {Giallongo}, \& {Vanzella}}]{Pentericci2010}
{Pentericci}, L., {Grazian}, A., {Scarlata}, C., {et~al.} 2010, \aap, 514, A64

\bibitem[{{Perley} {et~al.}(2009){Perley}, {Cenko}, {Bloom}, {Chen}, {Butler},
  {Kocevski}, {Prochaska}, {Brodwin}, {Glazebrook}, {Kasliwal}, {Kulkarni},
  {Lopez}, {Ofek}, {Pettini}, {Soderberg}, \& {Starr}}]{Perley2009}
{Perley}, D.~A., {Cenko}, S.~B., {Bloom}, J.~S., {et~al.} 2009, \aj, 138, 1690

\bibitem[{{Perley} {et~al.}(2016a){Perley}, {Kr{\"u}hler}, {Schulze}, {de
  Ugarte Postigo}, {Hjorth}, {Berger}, {Cenko}, {Chary}, {Cucchiara}, {Ellis},
  {Fong}, {Fynbo}, {Gorosabel}, {Greiner}, {Jakobsson}, {Kim}, {Laskar},
  {Levan}, {Micha{\l}owski}, {Milvang-Jensen}, {Tanvir}, {Th{\"o}ne}, \&
  {Wiersema}}]{Perley2016a}
{Perley}, D.~A., {Kr{\"u}hler}, T., {Schulze}, S., {et~al.} 2016a, \apj, 817, 7

\bibitem[{{Perley} {et~al.}(2013){Perley}, {Levan}, {Tanvir}, {Cenko}, {Bloom},
  {Hjorth}, {Kr{\"u}hler}, {Filippenko}, {Fruchter}, {Fynbo}, {Jakobsson},
  {Kalirai}, {Milvang-Jensen}, {Morgan}, {Prochaska}, \&
  {Silverman}}]{Perley2013}
{Perley}, D.~A., {Levan}, A.~J., {Tanvir}, N.~R., {et~al.} 2013, \apj, 778, 128

\bibitem[{{Perley} {et~al.}(2016b){Perley}, {Tanvir}, {Hjorth}, {Laskar},
  {Berger}, {Chary}, {de Ugarte Postigo}, {Fynbo}, {Kr{\"u}hler}, {Levan},
  {Micha{\l}owski}, \& {Schulze}}]{Perley2016b}
{Perley}, D.~A., {Tanvir}, N.~R., {Hjorth}, J., {et~al.} 2016b, \apj, 817, 8

\bibitem[{{Planck Collaboration} {et~al.}(2016){Planck Collaboration}, {Ade},
  {Aghanim}, {Arnaud}, {Ashdown}, {Aumont}, {Baccigalupi}, {Banday},
  {Barreiro}, \& {Bartlett}}]{Planck2016}
{Planck Collaboration}, {Ade}, P.~A.~R., {Aghanim}, N., {et~al.} 2016, \aap,
  594, A13

\bibitem[{Prochaska {et~al.}(2007)Prochaska, Chen, Dessauges-Zavadsky, \&
  Bloom}]{Prochaska2007}
Prochaska, J.~X., Chen, H.-W., Dessauges-Zavadsky, M., \& Bloom, J.~S. 2007,
  \apj, 666, 267

\bibitem[{{Reddy} {et~al.}(2008){Reddy}, {Steidel}, {Pettini}, {Adelberger},
  {Shapley}, {Erb}, \& {Dickinson}}]{Reddy2008}
{Reddy}, N.~A., {Steidel}, C.~C., {Pettini}, M., {et~al.} 2008, \apjs, 175, 48

\bibitem[{{Rivera-Thorsen} {et~al.}(2017){Rivera-Thorsen}, {Dahle}, {Gronke},
  {Bayliss, M.}, {Rigby, J. R.}, {Simcoe, R.}, {Bordoloi, R.}, {Turner, M.}, \&
  {Furesz, G.}}]{Rivera2017}
{Rivera-Thorsen}, T.~E., {Dahle}, H., {Gronke}, M., {et~al.} 2017, A\&A, 608,
  L4

\bibitem[{{Rivera-Thorsen} {et~al.}(2015){Rivera-Thorsen}, {Hayes},
  {{\"O}stlin}, {Duval}, {Orlitov{\'a}}, {Verhamme}, {Mas-Hesse}, {Schaerer},
  {Cannon}, {Ot{\'\i}-Floranes}, {Sand berg}, {Guaita}, {Adamo}, {Atek},
  {Herenz}, {Kunth}, {Laursen}, \& {Melinder}}]{RiveraThorsen2015}
{Rivera-Thorsen}, T.~E., {Hayes}, M., {{\"O}stlin}, G., {et~al.} 2015, \apj,
  805, 14

\bibitem[{{Safarzadeh} \& {Scannapieco}(2016)}]{Safarzadeh2016}
{Safarzadeh}, M. \& {Scannapieco}, E. 2016, \apjl, 832, L9

\bibitem[{{Salvaterra} {et~al.}(2011){Salvaterra}, Ferrara, \&
  Dayal}]{Salvaterra2011}
{Salvaterra}, R., Ferrara, A., \& Dayal, P. 2011, \mnras, 414, 847

\bibitem[{{Salvaterra} {et~al.}(2013){Salvaterra}, {Maio}, {Ciardi}, \&
  {Campisi}}]{Salvaterra2013}
{Salvaterra}, R., {Maio}, U., {Ciardi}, B., \& {Campisi}, M.~A. 2013, \mnras,
  429, 2718

\bibitem[{{Salvaterra} {et~al.}(2009){Salvaterra}, Valle, Campana, Chincarini,
  Covino, D’Avanzo, Fernández-Soto, Guidorzi, Mannucci, Margutti, Thöne,
  Antonelli, Barthelmy, De~Pasquale, D’Elia, Fiore, Fugazza, Hunt, Maiorano,
  Marinoni, Marshall, Molinari, Nousek, Pian, Racusin, Stella, Amati,
  Andreuzzi, Cusumano, Fenimore, Ferrero, Giommi, Guetta, Holland, Hurley,
  Israel, Mao, Markwardt, Masetti, Pagani, Palazzi, Palmer, Piranomonte,
  Tagliaferri, \& Testa}]{Salvaterra2009}
{Salvaterra}, R., Valle, M.~D., Campana, S., {et~al.} 2009, Nature, 461, 1258

\bibitem[{{Savaglio}(2006)}]{Savaglio2006}
{Savaglio}, S. 2006, New Journal of Physics, 8, 195

\bibitem[{Schaerer {et~al.}(2011)Schaerer, Hayes, Verhamme, \&
  Teyssier}]{Schaerer2011}
Schaerer, D., Hayes, M., Verhamme, A., \& Teyssier, R. 2011, \aap, 531, A12

\bibitem[{Schaerer \& Verhamme(2008)}]{Schaerer2008}
Schaerer, D. \& Verhamme, A. 2008, \aap, 480, 369

\bibitem[{Schlafly \& Finkbeiner(2011)}]{Schlafly2011}
Schlafly, E.~F. \& Finkbeiner, D.~P. 2011, \apj, 737, 103

\bibitem[{{Schulze} {et~al.}(2015){Schulze}, {Chapman}, {Hjorth}, {Levan},
  {Jakobsson}, {Bj{\"o}rnsson}, {Perley}, {Kr{\"u}hler}, {Gorosabel}, {Tanvir},
  {de Ugarte Postigo}, {Fynbo}, {Milvang-Jensen}, {M{\o}ller}, \&
  {Watson}}]{Schulze2015}
{Schulze}, S., {Chapman}, R., {Hjorth}, J., {et~al.} 2015, \apj, 808, 73

\bibitem[{Selsing {et~al.}(2019)Selsing, Malesani, Goldoni, Fynbo, Krühler,
  Antonelli, Arabsalmani, Bolmer, Cano, Christensen, Covino, D’Avanzo,
  D’Elia, De~Cia, de~Ugarte~Postigo, Flores, Friis, Gomboc, Greiner, Groot,
  Hammer, Hartoog, Heintz, Hjorth, Jakobsson, Japelj, Kann, Kaper, Ledoux,
  Leloudas, Levan, Maiorano, Melandri, Milvang-Jensen, Palazzi, Palmerio,
  Perley, Pian, Piranomonte, Pugliese, Sánchez-Ramírez, Savaglio, Schady,
  Schulze, Sollerman, Sparre, Tagliaferri, Tanvir, Thöne, Vergani, Vreeswijk,
  Watson, Wiersema, Wijers, Xu, \& Zafar}]{Selsing2019}
Selsing, J., Malesani, D., Goldoni, P., {et~al.} 2019, \aap, 623, A92

\bibitem[{{Shapley} {et~al.}(2003){Shapley}, {Steidel}, {Pettini}, \&
  {Adelberger}}]{Shapley2003}
{Shapley}, A.~E., {Steidel}, C.~C., {Pettini}, M., \& {Adelberger}, K.~L. 2003,
  \apj, 588, 65

\bibitem[{Sobral {et~al.}(2015)Sobral, Matthee, Darvish, Schaerer, Mobasher,
  Röttgering, Santos, \& Hemmati}]{Sobral2015}
Sobral, D., Matthee, J., Darvish, B., {et~al.} 2015, \apj, 808, 139

\bibitem[{Starling {et~al.}(2005)Starling, Wijers, Hughes, Tanvir, Vreeswijk,
  Rol, \& Salamanca}]{Starling2005a}
Starling, R. L.~C., Wijers, R. a. M.~J., Hughes, M.~A., {et~al.} 2005, \mnras,
  360, 305

\bibitem[{Stasińska {et~al.}(2015)Stasińska, Izotov, Morisset, \&
  Guseva}]{Stasinska2015}
Stasińska, G., Izotov, Y., Morisset, C., \& Guseva, N. 2015, \aap, 576, A83

\bibitem[{{Steidel} {et~al.}(2003){Steidel}, {Adelberger}, {Shapley},
  {Pettini}, {Dickinson}, \& {Giavalisco}}]{Steidel2003}
{Steidel}, C.~C., {Adelberger}, K.~L., {Shapley}, A.~E., {et~al.} 2003, \apj,
  592, 728

\bibitem[{Steidel {et~al.}(2011)Steidel, Bogosavljevi{\'{c}}, Shapley,
  Kollmeier, Reddy, Erb, \& Pettini}]{Steidel2011}
Steidel, C.~C., Bogosavljevi{\'{c}}, M., Shapley, A.~E., {et~al.} 2011, The
  Astrophysical Journal, 736, 160

\bibitem[{{Steidel} {et~al.}(2004){Steidel}, {Shapley}, {Pettini},
  {Adelberger}, {Erb}, {Reddy}, \& {Hunt}}]{Steidel2004}
{Steidel}, C.~C., {Shapley}, A.~E., {Pettini}, M., {et~al.} 2004, \apj, 604,
  534

\bibitem[{Tanvir {et~al.}(2009)Tanvir, Fox, Levan, Berger, Wiersema, Fynbo,
  Cucchiara, Krühler, Gehrels, Bloom, Greiner, Evans, Rol, Olivares, Hjorth,
  Jakobsson, Farihi, Willingale, Starling, Cenko, Perley, Maund, Duke, Wijers,
  Adamson, Allan, Bremer, Burrows, Castro-Tirado, Cavanagh, de~Ugarte~Postigo,
  Dopita, Fatkhullin, Fruchter, Foley, Gorosabel, Kennea, Kerr, Klose, Krimm,
  Komarova, Kulkarni, Moskvitin, Mundell, Naylor, Page, Penprase, Perri,
  Podsiadlowski, Roth, Rutledge, Sakamoto, Schady, Schmidt, Soderberg,
  Sollerman, Stephens, Stratta, Ukwatta, Watson, Westra, Wold, \&
  Wolf}]{Tanvir2009}
Tanvir, N.~R., Fox, D.~B., Levan, A.~J., {et~al.} 2009, Nature, 461, 1254

\bibitem[{Tanvir {et~al.}(2019)Tanvir, Fynbo, de~Ugarte~Postigo, Japelj,
  Wiersema, Malesani, Perley, Levan, Selsing, Cenko, Kann, Milvang-Jensen,
  Berger, Cano, Chornock, Covino, Cucchiara, D’Elia, Gargiulo, Goldoni,
  Gomboc, Heintz, Hjorth, Izzo, Jakobsson, Kaper, Krühler, Laskar, Myers,
  Piranomonte, Pugliese, Rossi, Sánchez-Ramírez, Schulze, Sparre, Stanway,
  Tagliaferri, Thöne, Vergani, Vreeswijk, Wijers, Watson, \& Xu}]{Tanvir2019}
Tanvir, N.~R., Fynbo, J. P.~U., de~Ugarte~Postigo, A., {et~al.} 2019, \mnras,
  483, 5380

\bibitem[{Thöne {et~al.}(2011)Thöne, Campana, Lazzati, de~Ugarte~Postigo,
  Fynbo, Christensen, Levan, Aloy, Hjorth, Jakobsson, Levesque, Malesani,
  Milvang-Jensen, Roming, Tanvir, Wiersema, Gladders, Wuyts, \&
  Dahle}]{Thone2011}
Thöne, C.~C., Campana, S., Lazzati, D., {et~al.} 2011, \mnras, 414, 479

\bibitem[{{Totani} {et~al.}(2006){Totani}, {Kawai}, {Kosugi}, {Aoki}, {Yamada},
  {Iye}, {Ohta}, \& {Hattori}}]{Totani2006}
{Totani}, T., {Kawai}, N., {Kosugi}, G., {et~al.} 2006, \pasj, 58, 485

\bibitem[{Troja {et~al.}(2007)Troja, Cusumano, O’Brien, Zhang, Sbarufatti,
  Mangano, Willingale, Chincarini, Osborne, Marshall, Burrows, Campana,
  Gehrels, Guidorzi, Krimm, La~Parola, Liang, Mineo, Moretti, Page, Romano,
  Tagliaferri, Zhang, Page, \& Schady}]{Troja2007}
Troja, E., Cusumano, G., O’Brien, P.~T., {et~al.} 2007, \apj, 665, 599

\bibitem[{{van Dokkum}(2001)}]{VanDokkum2001}
{van Dokkum}, P.~G. 2001, \pasp, 113, 1420

\bibitem[{{Vanzella} {et~al.}(2020){Vanzella}, {Caminha}, {Calura}, {Cupani},
  {Meneghetti}, {Castellano}, {Rosati}, {Mercurio}, {Sani}, {Grillo}, {Gilli},
  {Mignoli}, {Comastri}, {Nonino}, {Cristiani}, {Giavalisco}, \&
  {Caputi}}]{Vanzella2020}
{Vanzella}, E., {Caminha}, G.~B., {Calura}, F., {et~al.} 2020, \mnras, 491,
  1093

\bibitem[{{Vanzella} {et~al.}(2016){Vanzella}, de~Barros, Vasei, Alavi,
  Giavalisco, Siana, Grazian, Hasinger, Suh, Cappelluti, Vito, Amorin,
  Balestra, Brusa, Calura, Castellano, Comastri, Fontana, Gilli, Mignoli,
  Pentericci, Vignali, \& Zamorani}]{Vanzella2016}
{Vanzella}, E., de~Barros, S., Vasei, K., {et~al.} 2016, \apj, 825, 41

\bibitem[{Vergani {et~al.}(2011)Vergani, Piranomonte, Petitjean, Flores,
  Goldoni, Rodrigues, Hammer, \& Covino}]{Vergani2011a}
Vergani, S., Piranomonte, S., Petitjean, P., {et~al.} 2011, Astron. Nachr.,
  332, 292

\bibitem[{Vergani {et~al.}(2017)Vergani, Palmerio, Salvaterra, Japelj,
  Mannucci, Perley, D’Avanzo, Krühler, Puech, Boissier, Campana, Covino,
  Hunt, Petitjean, \& Tagliaferri}]{Vergani2017}
Vergani, S.~D., Palmerio, J., Salvaterra, R., {et~al.} 2017, \aap, 599, A120

\bibitem[{Verhamme {et~al.}(2015)Verhamme, Orlitová, Schaerer, \&
  Hayes}]{Verhamme2015}
Verhamme, A., Orlitová, I., Schaerer, D., \& Hayes, M. 2015, \aap, 578, A7

\bibitem[{Verhamme {et~al.}(2017)Verhamme, Orlitová, Schaerer, Izotov,
  Worseck, Thuan, \& Guseva}]{Verhamme2017}
Verhamme, A., Orlitová, I., Schaerer, D., {et~al.} 2017, \aap, 597, A13

\bibitem[{Verhamme {et~al.}(2008)Verhamme, Schaerer, Atek, \&
  Tapken}]{Verhamme2008}
Verhamme, A., Schaerer, D., Atek, H., \& Tapken, C. 2008, \aap, 491, 89

\bibitem[{Verhamme {et~al.}(2006)Verhamme, Schaerer, \& Maselli}]{Verhamme2006}
Verhamme, A., Schaerer, D., \& Maselli, A. 2006, \aap, 460, 397

\bibitem[{Vernet {et~al.}(2011)Vernet, Dekker, D’Odorico, Kaper, Kjaergaard,
  Hammer, Randich, Zerbi, Groot, Hjorth, Guinouard, Navarro, Adolfse, Albers,
  Amans, Andersen, Andersen, Binetruy, Bristow, Castillo, Chemla, Christensen,
  Conconi, Conzelmann, Dam, De~Caprio, De~Ugarte~Postigo, Delabre,
  Di~Marcantonio, Downing, Elswijk, Finger, Fischer, Flores, François,
  Goldoni, Guglielmi, Haigron, Hanenburg, Hendriks, Horrobin, Horville, Jessen,
  Kerber, Kern, Kiekebusch, Kleszcz, Klougart, Kragt, Larsen, Lizon, Lucuix,
  Mainieri, Manuputy, Martayan, Mason, Mazzoleni, Michaelsen, Modigliani,
  Moehler, Møller, Norup~Sørensen, Nørregaard, Péroux, Patat, Pena, Pragt,
  Reinero, Rigal, Riva, Roelfsema, Royer, Sacco, Santin, Schoenmaker, Spano,
  Sweers, Ter~Horst, Tintori, Tromp, van Dael, van~der Vliet, Venema, Vidali,
  Vinther, Vola, Winters, Wistisen, Wulterkens, \& Zacchei}]{Vernet2011}
Vernet, J., Dekker, H., D’Odorico, S., {et~al.} 2011, \aap, 536, A105

\bibitem[{Vernet {et~al.}(2009)Vernet, Kerber, Mainieri, Rauch, Saitta,
  D'Odorico, Bohlin, Ivanov, Lidman, Mason, Smette, Walsh, Fosbury, Goldoni,
  Groot, Hammer, Kaper, Horrobin, Kjaergaard-Rasmussen, \& Royer}]{Vernet2009}
Vernet, J., Kerber, F., Mainieri, V., {et~al.} 2009, Proc. IAU, 5, 535

\bibitem[{{Vielfaure} {et~al.}(2020){Vielfaure}, {Vergani}, {Japelj}, {Fynbo},
  {Gronke}, {Heintz}, {Malesani}, {Petitjean}, {Tanvir}, {D'Elia}, {Kann},
  {Palmerio}, {Salvaterra}, {Wiersema}, {Arabsalmani}, {Campana}, {Covino}, {De
  Pasquale}, {de Ugarte Postigo}, {Hammer}, {Hartmann}, {Jakobsson},
  {Kouveliotou}, {Laskar}, {Levan}, \& {Rossi}}]{Vielfaure2020}
{Vielfaure}, J.~B., {Vergani}, S.~D., {Japelj}, J., {et~al.} 2020, \aap, 641,
  A30

\bibitem[{Vreeswijk {et~al.}(2004)Vreeswijk, Ellison, Ledoux, Wijers, Fynbo,
  Møller, Henden, Hjorth, Masi, Rol, Jensen, Tanvir, Levan, Castro~Cerón,
  Gorosabel, Castro-Tirado, Fruchter, Kouveliotou, Burud, Rhoads, Masetti,
  Palazzi, Pian, Pedersen, Kaper, Gilmore, Kilmartin, Buckle, Seigar, Hartmann,
  Lindsay, \& van~den Heuvel}]{Vreeswijk2004}
Vreeswijk, P.~M., Ellison, S.~L., Ledoux, C., {et~al.} 2004, \aap, 419, 927

\bibitem[{Vreeswijk {et~al.}(2013)Vreeswijk, Ledoux, Raassen, Smette, De~Cia,
  Woźniak, Fox, Vestrand, \& Jakobsson}]{Vreeswijk2013}
Vreeswijk, P.~M., Ledoux, C., Raassen, A. J.~J., {et~al.} 2013, \aap, 549, A22

\bibitem[{Vreeswijk {et~al.}(2006)Vreeswijk, Smette, Fruchter, Palazzi, Rol,
  Wijers, Kouveliotou, Kaper, Pian, Masetti, Frontera, Hjorth, Gorosabel, Piro,
  Fynbo, Jakobsson, Watson, O'Brien, \& Ledoux}]{Vreeswijk2006}
Vreeswijk, P.~M., Smette, A., Fruchter, A.~S., {et~al.} 2006, \aap, 447, 145

\bibitem[{{White}(2020)}]{White2020}
{White}, N.~E. 2020, arXiv e-prints, arXiv:2003.01592

\bibitem[{Wiseman {et~al.}(2017)Wiseman, Perley, Schady, Prochaska,
  de~Ugarte~Postigo, Krühler, Yates, \& Greiner}]{Wiseman2017b}
Wiseman, P., Perley, D.~A., Schady, P., {et~al.} 2017, \aap, 607, A107

\bibitem[{{Wisotzki} {et~al.}(2016){Wisotzki}, {Bacon}, {Blaizot},
  {Brinchmann}, {Herenz}, {Schaye}, {Bouch{\'e}}, {Cantalupo}, {Contini},
  {Carollo}, {Caruana}, {Courbot}, {Emsellem}, {Kamann}, {Kerutt}, {Leclercq},
  {Lilly}, {Patr{\'\i}cio}, {Sandin}, {Steinmetz}, {Straka}, {Urrutia},
  {Verhamme}, {Weilbacher}, \& {Wendt}}]{Wisotzki2016}
{Wisotzki}, L., {Bacon}, R., {Blaizot}, J., {et~al.} 2016, \aap, 587, A98

\bibitem[{{Wofford} {et~al.}(2013){Wofford}, {Leitherer}, \&
  {Salzer}}]{Wofford2013}
{Wofford}, A., {Leitherer}, C., \& {Salzer}, J. 2013, \apj, 765, 118

\bibitem[{Yang {et~al.}(2016)Yang, Malhotra, Gronke, Rhoads, Dijkstra, Jaskot,
  Zheng, \& Wang}]{Yang2016}
Yang, H., Malhotra, S., Gronke, M., {et~al.} 2016, \apj, 820, 130

\bibitem[{{Yang} {et~al.}(2017){Yang}, {Malhotra}, {Gronke}, {Rhoads},
  {Leitherer}, {Wofford}, {Jiang}, {Dijkstra}, {Tilvi}, \& {Wang}}]{Yang2017}
{Yang}, H., {Malhotra}, S., {Gronke}, M., {et~al.} 2017, \apj, 844, 171

\bibitem[{Zitrin {et~al.}(2015)Zitrin, Labbé, Belli, Bouwens, Ellis,
  Roberts-Borsani, Stark, Oesch, \& Smit}]{Zitrin2015}
Zitrin, A., Labbé, I., Belli, S., {et~al.} 2015, \apj, 810, L12

\end{thebibliography}
%-------------------------------------------------------------------
%-------------------------------------------------------------------
\begin{appendix} %First appendix
\onecolumn

%%%%%%%%%%%%%%%%%%%%%%%%%%%%%%%%%%%%%%%%%%%%%%%%%%%%%%%%%%%%%%%%%%%%%%%%%%%%%%%%%%%%%
%\newpage
\section{Tables of the properties of the objects in the samples} \label{tablerecap}
The following tables present the LGRB host galaxies and afterglows from the TOUGH and XHG samples, the literature, and this work, for which the \Lya emission line has been detected or an upper limit has been estimated.
{\it GRB} and {\it Redshift} are for the name of the LGRB and its redshift. When observations of the host galaxy or the optical afterglow are available, 
we provide the name of the spectrographs used in the columns {\it Hosts} and {\it OA}, respectively. 
"no" indicates that no observation is available. "F1"/"F2" is for VLT/FORS1/2 and "xsh" for VLT/X-shooter spectrograph. 
{\it \Lya Host} and {\it \Lya OA} inform about the detection of the \Lya line in the host or afterglow spectra, respectively. No information is provided when no spectra are available to verify the presence of the line.
{\it F(\Lya)} and L(\Lya) correspond to the \Lya line flux and luminosity (respectively) retrieved from the literature (see references in column {\it Refs}) or derived in this work.
{\it EW(\Lya)} corresponds to the rest-frame \Lya equivalent width. 
{\it \NHI} is for the neutral hydrogen column density determined from the GRB afterglow.
$M_{UV}$ is the UV magnitude of the LGRB host galaxy. 
$M_*$ is its stellar mass.
{\it R} is for the apparent magnitude of the host in the \textit{R} band.
\textit{SFR} is the star-formation rate.
$\log(sSFR/yr^{-1})$ is the specific star-formation rate.
\textit{E(B-V)} is the host extinction.
$12+\log(O/H)$ is the oxygen abundance.
\textit{Z} is the metallicity determined from the absorption lines detected in the GRB afterglow spectrum.
\\
{\bf References:} (1): \citet{MilvangJensen2012}; (2): This work; (3): \citet{Schulze2015}; (4): \citet{Perley2016b}; (5): \citet{Tanvir2019}; (6):\citet{Kruhler2015}; (7): \citet{Palmerio2019}; (8): \citet{Selsing2019}; (9): \citet{Vielfaure2020}; (10): \citet{Jakobsson2003}; (11): \citet{Fynbo2005}; (12): \citet{Vreeswijk2004}; (13): \citet{Jakobsson2004b}; (14): \citet{DAvanzo2010}; (15): \citet{Perley2013}; (16): \citet{DeUgartePostigo2005}; (17): \citet{Kruehler2017}; (18): \citet{Jakobsson2006}; (19): \citet{Bolmer2019}; (20): \citet{Heintz2019a}; (21): \citet{Cucchiara2015}; (22): \citet{Savaglio2006}; (23): \citet{Vreeswijk2006}; (24): \citet{Prochaska2007}; (25): \citet{Thone2011}; (26): \citet{Eliasdottir2009}; (27): \citet{Chen2007}.

\newpage

\begin{landscape}
\begin{table*}[!ht]
\caption{List of the LGRBs in the TOUGH sample with \text{\Lya}-emission detection (in bold) or upper limit. }                 
\large
\centering
 \begin{tabular}{l*{10}{c}r}
 
\hline\hline
 \rule[0.2cm]{0cm}{0.2cm}GRB & Redshift & Host & OA & \Lya Host  & \Lya OA   & EW(\Lya) & F(\Lya)  & L(\Lya) & Refs \\
    &  &  &  &   & \   & [\AA] & [$\rm 10^{-17}\ erg\ s^{-1}\ cm^{-2}$]  & [$\rm 10^{42}\ erg\ s^{-1}$] & \\
\hline

{\bf 050315}	 & 1.9500    & F1       & no        & yes        & ...       & $9.2 \pm 2.8$	& $2.34 \pm 0.68$       & $0.64 \pm 0.19$   & (1) \\
050401	 & 2.8983	 & F1       & F2        & no         & no        & $<16.3$          & $<1.12        $       & $<0.80        $   & (1) \\
050714B	 & 2.4383	 & xsh      & F1        & no         & no        & $<17.08 $        & $<0.74        $       & $<0.36        $   & (2) \\
050730	 & 3.9686	 & F1       & UVES      & no         & no        & ...              & $<0.87        $       & $<1.32        $   & (1) \\
050819	 & 2.5042	 & xsh      & no        & no         & ...       & $<9.30 $         & $<1.74        $       & $<0.91        $   & (2) \\
050820A	 & 2.6147	 & F1       & UVES      & no         & no        & $<8.1$           & $<1.01     	$       & $<0.56        $   & (1) \\
050908	 & 3.3467	 & F1       & F1        & no         & no        & ...              & $<0.64     	$       & $<0.64        $   & (1) \\
050915A	 & 2.5273	 & xsh      & F1        & no         & no        & $<9.45$          & $<0.61        $       & $<0.33        $   & (2) \\
050922C	 & 2.1992	 & F1       & AlFOSC    & no         & no        & ...              & $<1.93     	$       & $<0.70        $   & (1) \\
051001	 & 2.4296	 & xsh      & F1        & no         & no        & $<7.21$          & $<0.84        $       & $<0.41        $   & (2) \\
060115	 & 3.5328	 & F1       & F1        & no         & no        & ...              & $<1.18        $       & $<1.35        $   & (1) \\
060526	 & 3.2213	 & F1       & F1        & no         & no        & $<19.5$          & $<0.79        $       & $<0.73        $   & (1) \\
060604	 & 2.1357	 & F1, xsh  & AlFOSC    & no         & no        & $<12.1$          & $<0.90     	$       & $<0.31        $   & (1) \\
{\bf 060605}	 & 3.7730    & F1       & PMAS      & yes        & no        & $33.7 \pm 10.5$  & $1.70 \pm 0.27$       & $2.28 \pm 0.36$   & (1) \\
060607A	 & 3.0749	 & F1       & UVES      & no         & no        & ...              & $<0.73        $       & $<0.60        $   & (1) \\
{\bf 060707}	 & 3.4240    & F1, xsh  & F1        & yes        & no        & $11.2 \pm 2.3$	& $1.65 \pm 0.31$       & $1.75 \pm 0.33$   & (1) \\
{\bf 060714}	 & 2.7108    & F1       & F1        & no         & yes       & ... ($<26.3$)           & 1.73 ($<8.1$)              & 1.10 ($<0.49$)              & (1), (18) \\
060805A	 & 2.3633	 & xsh      & no        & no         & ...       & $<14.84$         & $<0.77        $       & $<0.35        $   & (2) \\
060814	 & 1.9223  	 & xsh      & F1, F2    & no         & no        & $<7.12$          & $<2.33        $       & $<0.64        $   & (2) \\
{\bf 060908}	 & 1.8836    & F1       & F1, F2    & yes        & no        & $40.4 \pm 6.7$	& $7.78 \pm 0.95$       & $1.94 \pm 0.24$   & (1) \\
061110B	 & 3.4344	 & F1       & F1        & no         & no        & $<10.7$          & $<0.66        $       & $<0.71        $   & (1) \\
070103	 & 2.6208	 & xsh      & F2        & no         & no        & $<19.40$         & $<1.42        $       & $<0.83        $   & (2) \\
{\bf 070110}	 & 2.3523    & F1, xsh  & F1        & yes        & yes       & $31.8 \pm 4.3$   & $4.0 \pm 0.4  $       & $1.73 \pm 0.17$   & (1) \\
070129	 & 2.3384	 & xsh      & F1, F2    & no         & no        & $<13.92$         & $<1.94        $       & $<0.86        $   & (2) \\
070224	 & 1.9922	 & xsh      & no        & no         & ...       & $<75.8$          & $<2.12        $       & $<0.64        $   & (2) \\
070328	 & 2.0627	 & xsh      & no        & no         & ...       & $<9.52$          & $<0.87        $       & $<0.28        $   & (2) \\
070419B	 & 1.9586	 & xsh      & F1        & no         & no        & $<90.06$         & $<2.15        $       & $<0.62        $   & (2) \\
{\bf 070506}	 & 2.3090    & F1       & F1        & yes        & no        & $32.3 \pm 11.8$  & $1.39 \pm 0.35$       & $0.57 \pm 0.14$   & (1) \\
070611	 & 2.0394    & F1       & F2        & no         & no        & ...              & $<0.96        $       & $<0.29        $   & (1) \\
{\bf 070721B}	 & 3.6298    & F1       & F2        & yes        & yes       & $32.5 \pm 8.0$	& $1.12 \pm 0.16$       & $1.37 \pm 0.19$   & (1) \\
070802	 & 2.4541    & F1, xsh  & F2        & no         & no        & $<8.1$           & $<0.89        $       & $<0.43        $   & (1) \\

\hline
\end{tabular}
\tablefoot{For GRB\,060714 the values reported correspond to the \Lya detection from the afterglow spectrum \citep{Jakobsson2006} but the upper limits derived from the host galaxy \citep{MilvangJensen2012} are given between brackets.}
\end{table*}
\end{landscape}

%%%%%%%%%%%%%%%%%%%%%%%%%%%%%%%%%%%%%%%%%%%%%%%%%%%%%%%%%%%%%%%%%%%%%%%%%%%%%%%%%%%%%%%%%%%%%%%%%%%%%
\newpage

\begin{landscape}
\begin{table*}[!ht]
\footnotesize
\caption{Properties of the LGRBs in the TOUGH sample with \text{\Lya}-emission detection (in bold) or upper limit.}                
\centering
 \begin{tabular}{l*{12}{c}r}
 
\hline\hline
 \rule[0.2cm]{0cm}{0.2cm}GRB	 & \NHI	 & $\rm M_{UV}$  & $\rm M_*$  & \textit{R}  & Refs  & SFR  & $\rm \log(sSFR/yr^{-1})$  & E(B-V)  &$\rm 12+\log(O/H)$	 & Z  & Refs \\
 	 & [$\rm \log($\NHI$/cm^{-2}$]	 & [mag] & [$\rm \log(M_*/M_{\odot}$)]  & [mag]  &   & [$\rm M_{\odot}\ yr^{-1}$]  &   &   & 	 &   &  \\
\hline

{\bf 050315}	 & ...	            & $-20.08 \pm 0.14$   & 9.7         & $24.51 \pm 0.15$       & (3),(4)       & ...                     & ...                       & ...                     & ...                       & ...                   & \\
050401	 & $22.60 \pm 0.30$	& $-19.28 \pm 0.31$   & 9.61        & $26.19 \pm 0.31$       & (5),(3)       & ...                     & ...                       & ...                     & ...                       & $-1.05 \pm 0.18$	    & (24) \\
050714B	 & ...              & $-19.61 \pm 0.19$   & ...         & $25.51 \pm 0.20$       & (3)           & $12.9^{+14.0}_{-5.3}$   & ...                       & $0.21^{+0.28}_{-0.21}$  & ...                       & ...                   & (6)  \\
050730	 & $22.10 \pm 0.10$	& $>-17.21$           & $<9.46$     & $>27.2$                & (5),(1)       & ...                     & ...                       & ...                     & ...                       & $-1.96 \pm 0.11$      & (21) \\
050819	 & ...              & $-21.13 \pm 0.09$   & ...         & $23.99 \pm 0.09$       & (3)           & $22^{+426}_{-15}$       & ...                       & $0.34^{+0.84}_{-0.34}$  & ...                       & ...                   & (6)  \\
050820A	 & $21.10 \pm 0.10$	& $-18.93 \pm 0.06$   & 8.95        & $24.8$                 & (5),(1),(17)  & 0.9                     & $-9.00$                   & ...                     & ...                       & $-0.76 \pm 0.13$	    & (6),(21) \\
050908	 & $17.60 \pm 0.10$	& $-18.18 \pm 0.27$   & $<9.91$     & $27.7$                 & (5),(1)       & ...                     & ...                       & ...                     & ...                       & ...                   &  \\
050915A	 & ...              & $-20.47 \pm 0.15$   & 10.66       & $24.70 \pm 0.16$       & (3),(17)      & 1.9                     & $-10.39$                  & $0.84^{+0.61}_{-0.60}$  & ...                       & ...                   & (6)  \\
050922C	 & $21.55 \pm 0.10$	& $>-17.95$           & $<9.01$     & $>26.29$               & (5),(1)       & ...                     & ...                       & ...                     & ...                       & $-1.88\pm 0.14$       & (21) \\
051001	 & ...              & $-20.55 \pm 0.12$   & 10.08       & $24.53 \pm 0.13$       & (3),(4)       & $110^{+124}_{-59}$      & $-8.04^{+0.33}_{-0.33}$   & $0.58^{+0.28}_{-0.28}$  & ...                       & ...                   & (6) \\
060115	 & $21.50 \pm 0.10$	& $-18.61 \pm 0.27$   & 9.43        & $27.1$                 & (5),(1)       & ...                     & ...                       & ...                     & ...                       & $>-1.53$              & (21) \\
060526	 & $20.00 \pm 0.15$	& $>-17.36$           & 9.30        & $>27.0$                & (5),(1)       & ...                     & ...                       & ...                     & ...                       & $-1.09 \pm 0.24$	    & (25) \\
060604	 & ...              & $-19.23 \pm 0.17$   & ...         & $25.62 \pm 0.18$       & (3)           & $7.2^{+9.4}_{-3.6}$     & ...                       & $0.38^{+0.32}_{-0.27}$  & $8.10^{+0.28}_{-0.35}$    & ...                   & (6) \\
{\bf 060605}	 & $18.90 \pm 0.40$	& $-17.94 \pm 0.20$   & $<9.97$     & $>26.5$                & (5),(1)       & ...                     & ...                       & ...                     & ...                       & ...                   & \\
060607A	 & $16.95 \pm 0.03$	& $>-15.52$           & $<9.45$     & $>28.05$               & (5),(3)       & ...                     & ...                       & ...                     & ...                       & ...                   & \\
{\bf 060707}	 & $21.00 \pm 0.20$	& $-20.78 \pm 0.06$   & 9.99        & $25.01 \pm 0.06$       & (5),(3)       & $19.9^{+48.0}_{-14.3}$  & $-8.69^{+0.53}_{-0.56}$   & ...                     & ...                       & $\geq -1.69$                   & (6),(21)  \\
{\bf 060714}	 & $21.80 \pm 0.10$	& $-18.88 \pm 0.28$   & 9.25        & $26.46 \pm 0.28$       & (5),(3)       & ...                     & ...                       & ...                     & ...                       & $\geq -0.97$                   & (21)\\
060805A	 & ...              & $-19.79 \pm 0.13$   & ...         & $25.26 \pm 0.14$       & (3)           & $9.0^{+3.9}_{-2.5}$     & ...                       & $0.00^{+0.16}_{-0.00}$  & ...                       & ...                   & (6)  \\
060814	 & ...              & $-21.46 \pm 0.10$   & 10.00       & $22.96 \pm 0.11$       & (3),(17)      & $47.5^{+72.5}_{-15.6}$  & $-8.32^{+0.40}_{-0.17}$   & $0.17^{+0.39}_{-0.17}$  & $8.46^{+0.10}_{-0.16}$    & ...                   & (7),(6) \\
{\bf 060908}	 & ...	            & $-18.93 \pm 0.17$   & 9.15        & $25.66 \pm 0.18$       & (3),(4)       & ...                     & ...                       & ...                     & ...                       & ...                   & \\
061110B	 & $22.35 \pm 0.10$	& $-19.82 \pm 0.29$   & $<9.47$     & $26.0$                 & (5),(1)       & ...                     & ...                       & ...                     & ...                       & $\geq -1.84$                   & (21) \\
070103	 & ...              & $-21.03 \pm 0.14$   & ...         & $24.21 \pm 0.14$       & (3)           & $43^{+162}_{-17}     $  & ...                       & $0.00^{+0.50}_{-0.00}$  & ...                       & ...                   & (6)  \\
{\bf 070110}	 & $21.70 \pm 0.10$	& $-19.81 \pm 0.11$   & $<9.16$     & $25.19 \pm 0.11$       & (5),(3)       & $8.9^{+10.9}_{-2.8}  $  & $<-7.86$                  & $0.00^{+0.38}_{-0.00}$  & ...                       & $\geq -1.32$                   & (6),(21)  \\
070129	 & ...              & $-20.75 \pm 0.11$   & 10.15       & $24.23 \pm 0.12$       & (3),(4)       & $20^{+28}_{-7}       $  & $-8.85^{+0.38}_{-0.19}$   & $0.17^{+0.35}_{-0.17}$  & ...                       & ...                   & (6)  \\
070224	 & ...              & $-18.71 \pm 0.29$   & ...         & $26.02 \pm 0.31$       & (3)           & $3.2^{+6.5}_{-2.3}   $  & ...                       & ...                     & ...                       & ...                   &  \\
070328	 & ...              & $-20.17 \pm 0.12$   & 9.65        & $24.55 \pm 0.13$       & (3),(17)      & $8.4^{+130.0}_{-4.2} $  & $-8.76^{+1.25}_{-0.33}$   & $0.16^{+0.39}_{-0.16}$  & ...                       & ...                   & (6)  \\
070419B	 & ...              & $-19.44 \pm 0.19$   & 9.84        & $25.20 \pm 0.20$       & (3),(4)       & $21^{+35}_{-11}      $  & $-8.52^{+0.42}_{-0.32}$   & $0.56^{+0.39}_{-0.30}$  & ...                       & ...                   & (6)  \\
{\bf 070506}	 & $22.00 \pm 0.30$	& $-18.80 \pm 0.21$   & ...         & $26.21 \pm 0.22$       & (5),(3)       & ...                     & ...                       & ...                     & ...                       & $\geq -0.65$                   & (21) \\
070611	 & $21.30 \pm 0.20$	& $>-17.52$           & ...         & $>27.27$               & (5),(3)       & ...                     & ...                       & ...                     & ...                       & ...                   &  \\
{\bf 070721B}	 & $21.50 \pm 0.20$	& $-18.39 \pm 0.44$   & $<9.42$     & $27.5$                 & (5),(1)       & ...                     & ...                       & ...                     & ...                       & $\geq -2.14$                   & (21) \\
070802	 & $21.50 \pm 0.20$	& $-19.85 \pm 0.20$   & 9.69        & $25.25 \pm 0.21$       & (5),(3)       & $24^{+11}_{-8}$         & $-8.31^{+0.16}_{-0.17}$   & $0.31^{+0.12}_{-0.12}$  & ...                       & $-0.46 \pm 0.38$  & (6),(26) \\

\hline
\end{tabular}
\end{table*}
\end{landscape}

% %%%%%%%%%%%%%%%%%%%%%%%%%%%%%%%%%%%%%%%%%%%%%%%%%%%%%%%%%%%%%%%%%%%%%%%%%%%%%%%%%%%%%%%%%%%%%%%%%%%%%%%%%%%%%%%%%%%%%%%%%%%%%%%%%%%%%%%%%%%%%%%%%%       
\newpage

\begin{landscape}
\begin{longtable}{l*{9}{c}r}
\caption{List of the LGRBs in the XHG sample with \text{\Lya}-emission detection (in bold) or upper limit determined in this work.} 
\label{TOUGH_LAEs}
\\
\hline\hline
  \rule[0.2cm]{0cm}{0.2cm}GRB & Redshift & Host & OA & \Lya Host  & \Lya OA   & EW(\Lya) & F(\Lya)  & L(\Lya) \\
  &  &  &  &   & \   & [\AA] & [$\rm 10^{-17}\ erg\ s^{-1}\ cm^{-2}$]  & [$\rm 10^{42}\ erg\ s^{-1}$] \\
\hline
\endfirsthead
\caption{continued.}\\
\hline\hline
 \rule[0.2cm]{0cm}{0.2cm}GRB & Redshift & Host & OA & \Lya Host  & \Lya OA   & EW(\Lya) & F(\Lya)  & L(\Lya) \\
    &  &  &  &   & \   & [\AA] & [$\rm 10^{-17}\ erg\ s^{-1}\ cm^{-2}$]  & [$\rm 10^{42}\ erg\ s^{-1}$] \\
\hline
\endhead
\hline
\endfoot

050714B	 & 2.4383	 & xsh, F1  & no        & no           & ...       & $<17.1	      $ & $<0.74	     $      & $<0.36    	 $    \\
050819	 & 2.5042	 & xsh      & no        & no           & ...       & $<9.3	      $ & $<1.74	     $      & $<0.91    	 $    \\
050915A	 & 2.5275	 & xsh      & F1        & no           & no        & $<9.5	      $ & $<0.61	     $      & $<0.33    	 $    \\
051001	 & 2.4295	 & xsh, F1  & no        & no           & ...       & $<7.2	      $ & $<0.84	     $      & $<0.41    	 $    \\
060204B	 & 2.3393	 & xsh      & no        & no           & ...       & $<36.2	      $ & $<8.73	     $      & $<3.88    	 $    \\
060604	 & 2.1355	 & xsh, F1  & AlFOSC    & no           & no        & $<11.7	      $ & $<0.87	     $      & $<0.31    	 $    \\
060707	 & 3.4240    & xsh, F1  & F1        & no           & no        & ...       	  & $<0.65	     $      & $<0.73    	 $    \\
060805A	 & 2.3633	 & xsh      & no        & no           & ...       & $<14.8	      $ & $<0.78	     $      & $<0.35    	 $    \\
060814	 & 1.9223	 & xsh      & F1, F2    & no           & no        & $<7.1	      $ & $<2.33	     $      & $<0.64    	 $    \\
{\bf 060926}	 & 3.2090    & xsh      & F1        & yes          & yes       & $37.0 \pm 7.0$ & $5.30 \pm 0.40 $      & $5.05 \pm 0.38 $    \\
061202	 & 2.2543	 & xsh      & no        & no           & ...       & $<16.7	      $ & $<1.76	     $      & $<0.71    	 $    \\
070103	 & 2.6208	 & xsh, F2  & no        & no           & ...       & $<19.4	      $ & $<1.42	     $      & $<0.83    	 $    \\
{\bf 070110}	 & 2.3523    & xsh      & F2        & yes          & yes       & $33.0 \pm 8.0$ & $2.80 \pm 0.50 $      & $1.26 \pm 0.23 $    \\
070129	 & 2.3384	 & xsh      & F1, F2    & no           & no        & $<13.9	      $ & $<1.94	     $      & $<0.86    	 $    \\
070224	 & 1.9922	 & xsh      & no        & no           & ...       & ...   	      & $<2.12	     $      & $<0.64    	 $    \\
070328	 & 2.0627	 & xsh      & no        & no           & ...       & $<9.5	      $ & $<0.87	     $      & $<0.28    	 $    \\
070419B  & 1.9586	 & xsh      & F1        & no           & no        & ...   	      & $<2.15	     $      & $<0.62    	 $    \\
070521 	 & 2.0865	 & xsh      & no        & no           & ...       & ...    	  & $<1.54	     $      & $<0.52    	 $    \\
070802	 & 2.4538	 & xsh, F1  & F2        & no           & no        & $<9.7	      $ & $<0.71	     $      & $<0.36    	 $    \\
071021	 & 2.4515	 & xsh      & no        & no           & ...       & ...    	  & $<1.41	     $      & $<0.70    	 $    \\
080207	 & 2.0856	 & xsh      & no        & no           & ...       & $<16.1	      $ & $<1.06	     $      & $<0.36    	 $    \\
080602	 & 1.8204	 & xsh      & no        & no           & ...       & $<10.3	      $ & $<3.98	     $      & $<0.96    	 $    \\
080605	 & 1.6410	 & xsh      & F2        & no           & no        & $<18.2	      $ & $<15.29        $      & $<2.86    	 $    \\
080804	 & 2.2059	 & xsh      & UVES      & no           & no        & $<17.5	      $ & $<0.82	     $      & $<0.32    	 $    \\
081210	 & 2.0631	 & xsh      & no        & no           & ...       & $<22.1	      $ & $<1.84	     $      & $<0.60    	 $    \\
081221	 & 2.2590	 & xsh      & no        & no           & ...       & $<27.3	      $ & $<0.56	     $      & $<0.23    	 $    \\
090113	 & 1.7494	 & xsh      & no        & no           & ...       & ...   	      & $<4.13	     $      & $<0.90    	 $    \\
090201	 & 2.1000	 & xsh      & no        & no           & ...       & ...    	      & $<2.78	     $      & $<0.95    	 $    \\
090323	 & 3.5832	 & xsh      & F2        & no           & no        & $<19.0	      $ & $<4.24	     $      & $<5.26    	 $    \\
{\bf 100424A}	 & 2.4656    & xsh      & no        & yes          & ...       & ...            & $3.40 \pm 0.50 $      & $1.72 \pm 0.22 $    \\
110818A  & 3.3609	 & xsh      & xsh       & no           & no        & ...    	      & $<0.47	     $      & $<0.50    	 $    \\
111123A	 & 3.1513	 & xsh      & xsh       & no           & no        & $<6.8	      $ & $<1.15	     $      & $<1.05    	 $    \\
120118B	 & 2.9428	 & xsh      & no        & no           & ...       & $<4.6	      $ & $<2.37	     $      & $<1.84    	 $    \\
120119A	 & 1.7291	 & xsh      & xsh       & no           & no        & ...    	      & $<10.95        $      & $<2.33    	 $    \\
120624B	 & 2.1974	 & xsh      & no        & no           & ...       & ...            & $<0.64	     $      & $<0.25    	 $    \\
120815A	 & 2.3587	 & xsh      & xsh       & no           & no        & $<15.2       $ & $<1.89	     $      & $<0.86    	 $    \\
130131B	 & 2.5393	 & xsh      & no        & no           & ...       & $<6.2        $ & $<1.49	     $      & $<0.81    	 $    \\
\end{longtable}
\end{landscape}

% %%%%%%%%%%%%%%%%%%%%%%%%%%%%%%%%%%%%%%%%%%%%%%%%%%%%%%%%%%%%%%%%%%%%%%%%%%%%%%%%%%%%%%%%%%%%%%%%%%%%%%%%%%%%%%%%%%%%%%%%%%%%%%%%%%%%%%%%%%%%%%%%%%   
\newpage

\begin{footnotesize}              
\begin{landscape}
\begin{longtable}{l*{12}{c}r}
\caption{Properties of the LGRBs in the XHG sample with \text{\Lya}-emission detection (in bold) or upper limit.} 
\\
\hline\hline
 \rule[0.2cm]{0cm}{0.2cm}GRB	 & \NHI	 & $\rm M_{UV}$  & $\rm M_*$  & \textit{R}  & Refs  & SFR  & $\rm \log(sSFR/yr^{-1})$  & E(B-V)  &$\rm 12+\log(O/H)$	 & Z  & Refs \\
 	 & [$\rm \log($\NHI$/cm^{-2}$]	 & [mag] & [$\rm \log(M_*/M_{\odot}$)]  & [mag]  &   & [$\rm M_{\odot}\ yr^{-1}$]  &   &   & 	 &   &  \\
\hline
\endfirsthead
\caption{continued.}\\
\hline\hline
 \rule[0.2cm]{0cm}{0.2cm}GRB	 & NHI	 & $\rm M_{UV}$  & $\rm M_*$  & \textit{R}  & Refs  & SFR  & $\rm \log(sSFR/yr^{-1})$  & E(B-V)  &$\rm 12+\log(O/H)$	 & Z  & Refs \\
 	 & [$\rm \log($\NHI$/cm^{-2}$]	 & [mag] & [$\rm \log(M_*/M_{\odot}$)]  & [mag]  &   & [$\rm M_{\odot}\ yr^{-1}$]  &   &   & 	 &   &  \\
\hline
\endhead
\hline
\endfoot

050714B	 & ...              & $-19.61 \pm 0.19$	  & ...              & $25.51 \pm 0.20$	    & (3)      & $12.9^{+14.0}_{-5.3}  $      & ...                        & $0.21^{+0.28}_{-0.21}$     & ...                      & ...               & (6) \\
050819	 & ...              & $-21.13 \pm 0.09$	  & ...              & $23.99 \pm 0.09$ 	& (3)      & $22^{+426}_{-15}      $      & ...                        & $0.34^{+0.84}_{-0.34}$     & ...                      & ...               & (6) \\
050915A	 & ...              & $-20.47 \pm 0.15$	  & 10.66            & $24.70 \pm 0.16$ 	& (3),(17) & $1.9                  $      &  -10.39                    & $0.84^{+0.61}_{-0.60}$     & ...                      & ...               & (6) \\
051001	 & ...              & $-20.55 \pm 0.12$	  & 10.08            & $24.53 \pm 0.13$ 	& (3),(4)  & $110^{+124}_{-59}     $      & $-8.04^{+0.33}_{-0.33}$    & $0.58^{+0.28}_{-0.28}$     & ...                      & ...               & (6) \\
060204B	 & ...              & ...                 & 9.81             & ...                  & (17)     & $78^{+85}_{-34}       $      & $-7.92^{+0.32}_{-0.25}$    & $0.34^{+0.29}_{-0.23}$     & ...                      & ...               & (6) \\
060604	 & ...              & ...                 & ...              & $25.62 \pm 0.18$ 	& (3)      & $7.2^{+9.4}_{-3.6}    $      & ...                        & $0.38^{+0.32}_{-0.27}$     & $8.10^{+0.28}_{-0.35}$   & ...               & (6) \\
060707	 & $21.00 \pm 0.20$	& $-20.78 \pm 0.06$   & 9.99	         & $25.01 \pm 0.06$ 	& (5),(3)  & $19.9^{+48.0}_{-14.3} $      & $-8.69^{+0.53}_{-0.56}$    & ...                        & ...                      & $\geq -1.69$               & (6),(21) \\
060805A	 & ...              & $-19.79 \pm 0.13$   & ...              & $25.26 \pm 0.14$ 	& (3)      & $9.0^{+3.9}_{-2.5}    $      & ...                        & $0.00^{+0.16}_{-0.00}$     & ...                      & ...               & (6) \\
060814	 & ...              & $-21.46 \pm 0.10$   & $10.03 \pm 0.10$ & $22.96 \pm 0.11$ 	& (3),(4)  & $47.5^{+72.5}_{-15.6} $      & $-8.35^{+0.40}_{-0.17}$    & $0.17^{+0.39}_{-0.17}$     & $8.46^{+0.10}_{-0.16}$   & ...               & (6),(7) \\
{\bf 060926}	 & $22.60 \pm 0.15$	& $-21.59 \pm 0.05$	  & 10.71	         & ...          & (5)      & $26^{+47}_{-17}       $      & $-9.29^{+0.44}_{-0.45}$    & ...                        & ...	                   & $\geq -1.32$               & (6),(21) \\
061202	 & ...              & ...                 & 9.64             & ...                  & (17)     & $43^{+60}_{-22}       $      & $-8.01^{+0.38}_{-0.31}$    & $0.58^{+0.34}_{-0.27}$     & ...                      & ...               & (6) \\
070103	 & ...              & $-21.03 \pm 0.14$	  & ...              & $24.21 \pm 0.14$ 	& (3)      & $43^{+162}_{-17}      $      & ...                        & $0.00^{+0.50}_{-0.00}$     & ...                      & ...               & (6) \\
{\bf 070110}	 & $21.70 \pm 0.10$	& $-19.81 \pm 0.11$	  & $< 9.16$         & $25.19 \pm 0.11$     & (5)      & $8.9^{+10.9}_{-2.8}   $      & $>-8.37$                   & $0.00^{+0.38}_{-0.00}$	    & ...	                   & $\geq-1.32$               & (6),(21) \\
070129	 & ...              & $-20.75 \pm 0.11$   & 10.15            & $24.23 \pm 0.12$ 	& (3),(4)  & $20^{+28}_{-7}        $      & $-8.85^{+0.38}_{-0.19}$    & $0.17^{+0.35}_{-0.17}$     & ...                      & ...               & (6) \\
070224	 & ...              & $-18.71 \pm 0.29$   & ...              & $26.02 \pm 0.31$ 	& (3)      & $3.2^{+6.5}_{-2.3}    $      & ...                        & ...                        & ...                      & ...               & (6) \\
070328	 & ...              & $-20.17 \pm 0.12$   & 9.65             & $24.55 \pm 0.13$ 	& (3),(17) & $8.4^{+130.7}_{-4.2}  $      & $-8.73^{+1.23}_{-0.30}$    & $0.16^{+0.39}_{-0.16}$     & ...                      & ...               & (6) \\
070419B  & ...              & $-19.44 \pm 0.19$	  & 9.84             & $25.20 \pm 0.20$ 	& (3),(4)  & $21^{+35}_{-11}       $      & $-8.52^{+0.42}_{-0.32}$    & $0.56^{+0.39}_{-0.30}$     & ...                      & ...               & (6) \\
070521 	 & ...              & ...                 & 10.5             & ...               	& (17)     & $26^{+34}_{-17}       $      & $-9.09^{+0.37}_{-0.47}$    & $0.71^{+0.05}_{-0.04}$     & ...                      & ...               & (6) \\
070802	 & $21.50 \pm 0.20$	& $-19.85 \pm 0.20$	  & 9.69             & $25.25 \pm 0.21$ 	& (5),(3)  & $24^{+11}_{-8}        $      & $-8.31^{+0.16}_{-0.17}$    & $0.31^{+0.12}_{-0.12}$     & ...                      & $-0.46 \pm 0.63$               & (6),(20) \\
071021	 & ...              & ...                 & 10.10            & ...                  & (17)     & $32^{+20}_{-12}       $      & $-8.59^{+0.21}_{-0.20}$    & $0.19^{+0.16}_{-0.17}$     & ...                      & ...               & (6) \\
080207	 & ...              & ...                 & 11.09            & $25.84 \pm 0.18$     & (17)     & $77^{+86}_{-38}       $      & $-9.20^{+0.32}_{-0.29}$    & $0.66^{+0.28}_{-0.25}$     & $8.74^{+0.15}_{-0.15}$   & ...               & (6) \\
080602	 & ...              & ...                 & $9.40 \pm 0.10$  & ...                  & (7)      & $>48$                        & $> -7.72$                  & $0.58^{+0.29}_{-0.26}$     & $8.69^{+0.12}_{-0.21}$   & ...               & (6),(7) \\
080605	 & ...              & ...                 & $9.60 \pm 0.10$  & ...                  & (7)      & $42.5^{+30.5}_{-18.2} $      & $-7.97^{+0.23}_{-0.24}$    & $0.26^{+0.11}_{-0.10}$     & $8.47^{+0.04}_{-0.04}$   & ...               & (6),(7) \\
080804	 & $21.30 \pm 0.15$	& ...                 & 9.28             & ...                  & (5)      & $15.2^{+41.2}_{-8.7}  $      & $-8.10^{+0.57}_{-0.37}$    & $0.38^{+0.51}_{-0.35}$     & ...                      & $-0.75 \pm 0.16$               & (6),(21) \\
081210	 & ...              & ...                 & 10.07            & ...                  & (17)     & $15.3^{+111.7}_{-7.0} $      & $-8.89^{+0.92}_{-0.27}$    & $0.13^{+0.61}_{-0.13}$     & ...                      & ...               & (6) \\
081221	 & ...              & ...                 & 10.08            & ...                  & (17      & $35^{+106}_{-22}      $      & $-8.54^{+0.61}_{-0.43}$    & $0.31^{+0.55}_{-0.31}$     & ...                      & ...               & (6) \\
090113	 & ...              & ...                 & 9.90             & ...                  & (17)     & $17.9^{+10.1}_{-4.8}  $      & $-8.65^{+0.19}_{-0.14}$    & $0.01^{+0.22}_{-0.01}$     & ...                      & ...               & (6) \\
090201	 & ...              & ...                 & 10.90            & ...                  & (4)      & $48^{+30}_{-14}       $      & $-9.22^{+0.21}_{-0.15}$    & $0.11^{+0.19}_{-0.11}$     & ...                      & ...               & (6) \\
090323	 & $20.75 \pm 0.10$	& $-21.60 \pm 0.18$	  & 10.30            & ...                  & (5),(17) & $24^{+53}_{-17}       $      & $-8.92^{+0.50}_{-0.53}$    & ...                        & ...                      & ...               & (6) \\
{\bf 100424A}	 & ...	            & ...                 & ...              & $> 24$                & (8)      & $21^{+20}_{-8}        $      & ...                        & $0.13^{+0.18}_{-0.13}$	    & $7.93^{+0.25}_{-0.18}$   & ...               & (6) \\
110818A  & $21.90 \pm 0.40$	& $-21.68 \pm 0.05$	  & 10.50            & ...                  & (5),(17) & $44^{+62}_{-26}       $      & $-8.86^{+0.38}_{-0.39}$    & ...                        & $8.25^{+0.17}_{-0.25}$   & ...               & (6) \\
111123A	 & ...              & $-22.05 \pm 0.03$	  & 11.17            & ...                  & (3),(17) & $77^{+163}_{-52}      $      & $-9.28^{+0.49}_{-0.49}$    & ...                        & $8.01^{+0.28}_{-0.28}$   & ...               & (6) \\
120118B	 & ...              & ...                 & ...              & ...                  &          & $28^{+21}_{-11}       $      & ...                        & $0.00^{+0.16}_{-0.00}$     & $7.89^{+0.23}_{-0.17}$   & ...               & (6) \\
120119A	 & $22.60 \pm 0.20$	& ...                 & 9.93             & ...                  & (5),(17) & $43^{+24}_{-14}       $      & $-8.30^{+0.19}_{-0.17}$    & $0.35^{+0.16}_{-0.14}$     & $8.60^{+0.14}_{-0.14}$   & $-0.96 \pm 0.28$    & (6),(20) \\
120624B	 & ...              & ...                 & 10.84            & ...                  & (17)     & $30^{+73}_{-13}       $      & $-9.36^{+0.53}_{-0.24}$    & $0.21^{+0.50}_{-0.21}$     & $8.43^{+0.20}_{-0.27}$   & ...               & (6) \\
120815A	 & $22.05 \pm 0.10$	& ...                 & ...              & ...                  & (5)      & $2.3^{+2.7}_{-1.0}    $      & ...                        & $0.06^{+0.34}_{-0.06}$     & ...                      & $-1.15 \pm 0.12$  & (6),(20) \\
130131B	 & ...              & ...                 & ...              & ...                  &          & $8.0^{+13.4}_{-5.0}   $      & ...                        & ...                        & ...                      & ...               & (6) \\

\end{longtable}
\end{landscape}
\end{footnotesize}

% %%%%%%%%%%%%%%%%%%%%%%%%%%%%%%%%%%%%%%%%%%%%%%%%%%%%%%%%%%%%%%%%%%%%%%%%%%%%%%%%%%%%%%%%%%%%%%%%%%%%%%%%%%%%%%%%%%
\newpage

\begin{landscape}
\begin{table*}[!ht]
\caption{List of the LGRBs from the literature (other than TOUGH and XHG samples) and this work with \text{\Lya}-emission detection. GRBs 070223 and 080810 are not reported in this table despite the fact that they are reported in Appendix \ref{Tab_recap} because no information on \Lya flux is available for these two GRB host galaxies.} 

\centering
 \begin{tabular}{l*{10}{c}r}
 
\hline\hline
 \rule[0.2cm]{0cm}{0.2cm}GRB & Redshift & Host & OA & \Lya Host  & \Lya OA   & EW(\Lya) & F(\Lya)  & L(\Lya) & Refs \\
     &  &  &  &   & \   & [\AA] & [$\rm 10^{-17}\ erg\ s^{-1}\ cm^{-2}$]  & [$\rm 10^{42}\ erg\ s^{-1}$] & \\
\hline
971214	& 3.4200    & Keck      & no        &   yes     & ...       & 13.0 	            & $0.62 \pm 0.07$   & $0.66 \pm 0.07$     & (1)      \\
000926	& 2.0400    & AlFOSC    & NOT       &   yes	    & no        & $71.0 \pm 18.0 $  & $14.90\pm 1.10$   & $4.51 \pm 0.33$     & (1)      \\
011211	& 2.1434    & xsh       & F1        & 	yes 	& no	    & $14.0 \pm 3.0	 $  & $1.60 \pm 0.20$   & $0.58 \pm 0.07$     & (2)      \\
021004	& 2.3298    & xsh       & UVES      &   yes	    & yes 	    & $105.0 \pm 10.0$  & $16.90\pm 0.30$   & $7.45 \pm 0.18$     & (2)      \\
030323	& 3.3720    & no        & F2        & 	...     & yes	    & $108.0 \pm 38.0$	& $1.20 \pm 0.10$   & $1.23 \pm 0.10$     & (1)      \\
030429	& 2.6600    & no        & F1        & 	...     & yes	    & ... 	            & 0.31              & $0.19 \pm 0.00$     & (1)      \\
061222A	& 2.0880    & LRIS	    & no        & 	yes	    & ...       & 31.0 	            & 16.80             & $5.39 \pm 0.00$     & (1)      \\
071031	& 2.6918    & no        & F2        & 	...     & yes	    & ... 	            & $2.36 \pm 0.27$   & $1.41 \pm 0.16$     & (1)      \\
081121	& 2.5134    & xsh       & LDSS3     & 	yes	    & no	    & ... 	            & $4.90 \pm 0.50$   & $2.60 \pm 0.26$     & (2)      \\
081222	& 2.7700    & xsh       & GMOS-S    & 	yes	    & no	    & ... 	            & $1.60 \pm 0.40$   & $1.07 \pm 0.27$     & (2)      \\
090205	& 4.6500    & no        & F1        & 	...	    & yes	    & ... 	            & $2.36 \pm 0.49$   & $5.16 \pm 1.08$     & (1)      \\
121201A	& 3.3830    & no        & xsh       &   ...     & yes	    & ... 	            & $2.14 \pm 0.36$   & $2.31 \pm 0.39$     & (2)      \\
150915A	& 1.9680    & no        & xsh       & 	...	    & yes	    & ... 	            & $5.00 \pm 1.10$   & $1.46 \pm 0.32$     & (2)      \\
151021A	& 2.3300    & no        & xsh       & 	...	    & yes	    & ... 	            & $10.50\pm 0.90$   & $4.63 \pm 0.40$     & (2)      \\
170202A	& 3.6450    & no        & xsh       &   ...     & yes	    & ... 	            & $2.70 \pm 0.60$   & $3.49 \pm 0.78$     & (2)      \\
191004B	& 3.5055    & xsh       & xsh       &   yes     & yes	    & $7.4 \pm 2.6$	    & $1.00 \pm 0.10$   & $1.18 \pm 0.18$     & (9)      \\

\hline

\end{tabular}
\end{table*}
\end{landscape}

% %%%%%%%%%%%%%%%%%%%%%%%%%%%%%%%%%%%%%%%%%%%%%%%%%%%%%%%%%%%%%%%%%%%%%%%%%%%%%%%%%%%%%%%%%%%%%%%%%%%%%%%%%%%%%%%%%%
\newpage

\begin{landscape}
\begin{table*}[!ht]
\footnotesize
\caption{Properties of the LGRBs from the literature (other than TOUGH and XHG samples) and this work with \text{\Lya}-emission detection.}                
\centering
 \begin{tabular}{l*{12}{c}r}
 
\hline\hline
 \rule[0.2cm]{0cm}{0.2cm}GRB	 & \NHI	 & $\rm M_{UV}$  & $\rm M_*$  & \textit{R}  & Refs  & SFR  & $\rm \log(sSFR/yr^{-1})$  & E(B-V)  &$\rm 12+\log(O/H)$	 & Z  & Refs \\
 	 & [$\rm \log($\NHI$/cm^{-2}$]	 & [mag] & [$\rm \log(M_*/M_{\odot}$)]  & [mag]  &   & [$\rm M_{\odot}\ yr^{-1}$]  &   &   & 	 &   &  \\
\hline

971214	& ...	            & ...                       & ...       & ...               &           & ...              & ...           & $0.44^{+0.06}_{-0.03}$  & ...     & ...                & (15)\\
000926	& $21.30 \pm 0.25$	& $-20.40^{+0.07}_{-0.07}$	& 9.64      & ...               & (5)       & ...              & ...           & $0.19^{+0.13}_{-0.09}$  & ...     & $-0.13 \pm 0.21$   & (15),(22)\\
011211	& $20.40 \pm 0.20$	& $-19.97^{+0.10}_{-0.10}$	& 8.0       & $24.95 \pm 0.11$  & (5),(10)  & $<2.0$           & $<-7.70$      & $0.03^{+0.03}_{-0.03}$  & ...     & $-0.90 \pm 0.50$   & (2),(10),(23)\\
021004	& $19.50 \pm 0.50$	& $-20.56^{+0.10}_{-0.10}$	& 9.45      & $24.39 \pm 0.04$  & (5),(11)  & $6.7 \pm 0.8$    & $>-8.62$      & $0.02^{+0.03}_{-0.03}$  & ...     & $-1.30 \pm 0.50$   & (2),(16),(27)\\
030323	& $21.90 \pm 0.07$	& $-18.47^{+0.10}_{-0.10}$  & $<9.85$   & $28.0 \pm 0.3$    & (5),(12)  & ...              & ...           & ...                     & ...     & $-1.26 \pm 0.20$   & (12)\\
030429	& $21.60 \pm 0.20$	& ...                       & ...       & $>26.3$           & (5),(13)  & ...              & ...           & ...                     & ...     & $\geq -1.13$                & (21)   \\
061222A	& ...               & ...                       & 9.55      & ...               & (4)       & ...              & ...           & $0.00^{+0.00}_{-0.00}$  & ...     & ...                & (15)\\
071031	& $22.15 \pm 0.05$	& ...                       & ...       & ...               & (5)       & ...              & ...           & ...                     & ...     & $-1.85 \pm 0.12$   & (21)\\
081121	& ...               & ...                       & 9.24      & ...               & (4)       & $>7.0$           & $>-8.35$      & ...                     & ...     & ...                & (2)\\
081222	& $20.80 \pm 0.20$	& $-22.43^{+0.0}_{+0.0}$    & 9.61      & ...               & (5)       & ...              & ...           & ...                     & ...     & ...                &    \\
090205	& $20.73 \pm 0.05$	& $-21.26^{+0.13}_{+0.13}$	& $<10.7$   & $26.4 \pm 0.3$    & (5),(14)  & ...              & ...           & ...                     & ...     & $\geq -0.57$       &  (21)  \\
121201A	& $22.00 \pm 0.20$	& $-20.84^{+0.21}_{+0.21}$	& ...       & ...               & (5)       & ...              & ...           & ...                     & ...     & ...                &    \\
150915A	& $21.20 \pm 0.30$	& ...                       & ...       & ...               & (5)       & $13.0 \pm 0.6$   & ...           & ...                     & ...     & ...                & (2)\\
151021A	& $22.20 \pm 0.20$	& ...                       & ...       & ...               & (5)       & ...              & ...           & ...                     & ...     & $-0.98 \pm 0.07$   & (19)\\
170202A	& $21.55 \pm 0.10$	& ...                       & ...       & ...               & (5)       & ...              & ...           & ...                     & ...     & $-1.28 \pm 0.09$   & (19)\\
191004B	& $17.20 \pm 0.15$	& $-19.10^{+0.0}_{+0.0}$    & ...       & $>26.6$           & (5),(9)   & $2.85 \pm 1.85$  & ...           & $0.01^{+0.01}_{-0.01}$  & ...     & ...                & (9)\\

\hline
\end{tabular}
\end{table*}
\end{landscape}

\newpage
\section{Table summarizing the LAE-LGRB census.} \label{Tab_recap}

\begin{table*}[ht!]
    \caption{\label{29LAEs} List of the 29 LGRBs with \text{\Lya} emission line detected in the host galaxy or optical afterglow spectra.}
    \label{Tab_recap}    
    \centering
    \resizebox{\textwidth}{!}{%
     \begin{tabular}{l*{9}{c}r}
         \hline\hline
\rule[0.2cm]{0cm}{0.2cm}GRB & Redshift	& Host & OA & $\textrm{Ly}\alpha$ Host & $\textrm{Ly}\alpha$ OA & Flux($\textrm{Ly}\alpha$) & Sample & Refs \\
\hline         
\rule[0.2cm]{0cm}{0.2cm}971214 & 3.4200	&	Keck	& no &	yes	&   -	&	$0.6 \pm 0.1$	            & - & (1), (2), (3), (8)\\
000926 	& 2.0400	&	AlFOSC	    & NOT       &   yes	    & no	&	$14.9 \pm 1.1$                  & - & (4), (8) \\
011211 	& 2.1434	&	xsh	        & F1        & 	yes 	& no	&	$1.6 \pm 0.2$	                & - & (2) \\
021004 	& 2.3298	&	xsh	        & UVES      &   yes	    & yes 	&   $16.9 \pm 0.3$	                & - & (2) \\
030323 	& 3.3720	&	no	        & F2        & 	-	    & yes	&	$1.2 \pm 0.1$	                & - & (2), (6),(8) \\
030429 	& 2.6600	&	no	        & F1        & 	-	    & yes	&	0.3 ($2\sigma$)	                & - & (7) \\
050315 	& 1.9500	&	F1	        & no        & 	yes	    & -	    &	$2.3 \pm 0.7$	                & TOUGH & (8) \\
060605 	& 3.7730	&	F1	        & PMAS      & 	yes	    & no	&	$1.7 \pm 0.3$	                & TOUGH & (8) \\
060707 	& 3.4240	&	F1, xsh	    & F1        & 	yes	    & no	&	$1.7 \pm 0.3$	                & TOUGH, XHG$^a$ & (8), (10) \\
060714 	& 2.7108	&	F1	        & F1        & 	no	    & yes	&	1.7 	                        & TOUGH$^b$ & (8), (10), (11) \\
060908 	& 1.8836	&	F1	        & F1, F2    & 	yes	    & no	&	$7.8 \pm 1.0$	                & TOUGH & (8)\\
060926 	& 3.2090	&	xsh	        & F1        & 	yes	    & yes	&	$5.3 \pm 0.4$	                & XHG & (8), (10) \\
061222A & 2.0880	&	LRIS	    & no        & 	yes	    & -	    &	16.8	                        & - & (8), (12) \\
070110 	& 2.3523	&	F1, xsh	    & F1        & 	yes	    & yes	&	$4.0 \pm 0.4$	                & TOUGH, XHG & (8), (10) \\
070223 	& 1.6295	&	LRIS	    & no        & 	yes	    & -	    &	-	                            & - & (9) \\
070506 	& 2.3090	&	F1	        & F1        & 	yes	    & no	&	$1.4 \pm 0.4$	                & TOUGH & (8), (10) \\
070721B & 3.6298	&	F1	        & F2        & 	yes	    & yes	&	$1.1 \pm 0.2$	                & TOUGH & (8), (10) \\
071031 	& 2.6918	&	no	        & F2        & 	-	    & yes	&	$2.4 \pm 0.3$	                & - & (8), (10) \\
080810 	& 3.3604	&	LRIS        & AlFOSC    & 	yes	    & yes	&	-	                            & - & (10), (23) \\
081121 	& 2.5134	&	xsh	        & LDSS3     & 	yes	    & no	&	$4.9 \pm 0.5$	                & - & (20), (21) \\
081222 	& 2.7700	&	xsh	        & GMOS-S    & 	yes	    & no	&	$1.6 \pm 0.4$	                & - & (17), (20) \\
090205 	& 4.6500	&	no	        & F1        & 	-	    & yes	&	$2.4 \pm 0.5$	                & - & (8), (13) \\
100316A & 3.1600	&	no          & OSIRIS    & 	no	    & yes	&	-	                            & - & (17) \\
100424A	& 2.4656	&	xsh	        & no        & 	yes	    & -	    &	$3.4 \pm 0.5$	                & XHG & (15), (18) \\
121201A & 3.3830	&	no	        & xsh       & -         & yes	&	$2.1 \pm 0.4$	                & XAFT & (16), (19) \\
150915A & 1.9680	&	no	        & xsh       & 	-	    & yes	&	$5.0 \pm 1.1$	                & XAFT & (16) \\
151021A & 2.3300	&	no	        & xsh       & 	-	    & yes	&	$10.5 \pm 0.9$	                & XAFT & (16) \\
170202A & 3.6450    &	no	        & xsh       & -         & yes	&	$2.7 \pm 0.6$	                & XAFT & (16) \\
\rule[-0.2cm]{0cm}{0.2cm}191004B 	& 3.5055    & xsh	& xsh & yes & yes	&	$1.0 \pm 0.1$	        & - & (20), (22) \\         
\hline  
    \end{tabular} }
 \tablefoot{\textit{GRB} and \textit{Redshift} are for the name of the LGRB and its redshift. When observations of the host galaxy or the optical afterglow are available, 
we provide the name of the spectrographs used in the columns \textit{Hosts} and \textit{OA}, respectively. 
"no" indicates that no observation is available. "F1"/"F2" is for VLT/FORS1/2 and "xsh" for VLT/X-shooter spectrograph. 
\textit{\Lya Host} and \textit{\Lya OA} inform about the detection of the \Lya line in the host or afterglow spectra, respectively. No information is provided when no spectra are available to verify the presence of the line.
\textit{Flux(\Lya)} corresponds to the \Lya line flux, in units of $ 10^{-17}$ \ergscmA, retrieved from the literature (see references in column \textit{Refs}) or derived in this work.
\textit{Samples} indicates to which sample the LGRB belongs. \textit{Refs} is for the references where the \Lya detection is reported. \\
$^a$ GRB\,060707 is part of the XHG sample but the \Lya is not detected in the X-shooter spectrum. \\
$^b$ For GRB\,060714 the \Lya line is only detected at 2.5$\sigma$ in the host galaxy observation of \citet{MilvangJensen2012} but is convincingly detected in the afterglow spectrum presented by \citet{Jakobsson2006}, therefore we consider it as a detection and report the flux measured in the afterglow spectrum corrected for Galactic extinction. \\
\textbf{References:} (1): \citet{Kulkarni1998}; (2): \citet{Fynbo2003}; (3): \citet{Ahn2000}; (4): \citet{Fynbo2002}; (5): \citet{Jakobsson2005}; (6): \citet{Vreeswijk2004}; (7): \citet{Jakobsson2004b}; (8): \citet{MilvangJensen2012}; (9): \citet{Perley2016a}; (10): \citet{Fynbo2009}; (11): \citet{Jakobsson2006}; (12): \citet{Perley2009}; (13): \citet{DAvanzo2010}; (14): \citet{Hartoog2015}; (15): \citet{Kruhler2015}; (16): \citet{Selsing2019}; (17): \citet{Tanvir2019}; (18): \citet{Malesani2013}; (19): \citet{Cucchiara2015}; (20): this work, (21): \citet{Berger2008}, (22): \citet{DElia2019}, (23): \citet{Wiseman2017b} }
\end{table*}

\newpage
\section{Results of the shell model with constraints from the GRB host galaxy.} \label{shell_host} 

\begin{table*}[ht!]
    \caption{Shell-model predictions with constraints on: $\rm z^{HG}$, $\rm FWHM_i^{HG}$(\Lya) and $\rm EW_i^{HG}$(\Lya)}
    \centering
    \footnotesize{
     \begin{tabular}{c | c c c c c c c }
         \hline\hline
          \rule[0.2cm]{0cm}{0.2cm}GRB host 	& $\rm \Delta z$ 	& $\rm log(N_{HI}cm^{-2})$        & $\rm V_{exp}$     & log(T/K)             & $\rm \tau_d$ & $\rm FWHM_i$(\Lya) & $\rm EW_i$(\Lya)  \\
         			& 			    &  & [km s$^{-1}$] &   &                & [km s$^{-1}$]               &  [\AA]               \\
         \hline
         \rule[0.2cm]{0cm}{0.2cm}GRB 011211 & $30^{+45}_{-45}$ & $19.9^{+0.2}_{-0.2}$ & $90^{+21}_{-26}$   &  $4.17^{+0.9}_{-0.9}$   & $0.47^{+0.39}_{-0.27}$  & $50^{+10}_{-10}$ & $33^{+20}_{-16}$  \\

        GRB 021004  						& $5^{+45}_{-45}$  & $19.6^{+0.1}_{-0.1}$  & $120^{+5}_{-6}$   &  $4.96^{+0.17}_{-0.14}$   & $0.399^{+0.114}_{-0.084}$  & $223^{+13}_{-12}$ & $165^{+14}_{-13}$  \\

        GRB 060926  						& $7^{+45}_{-45}$     & $19.8^{+0.1}_{-0.1}$        & $158^{+5}_{-5}$     &  $4.97^{+0.17}_{-0.17}$          & $1.27^{+0.28}_{-0.23}$  & $148^{+7}_{-7}$ & $107^{+3}_{-2}$  \\

        \rule[-0.2cm]{0cm}{0.2cm}GRB 070110 & $3^{+45}_{-45}$  & $19.8^{+0.2}_{-0.1}$ & $92^{+13}_{-11}$     &  $3.69^{+0.90}_{-0.55}$   &$0.083^{+0.138}_{-0.062}$  & $20^{+26}_{-12}$ & $20^{+10}_{-6}$  \\
        \hline
    \end{tabular}
    }
    \begin{center}
     \footnotesize{\textbf{Notes.} Same as Table \ref{tableComp_unconst}.}
    \end{center}
\end{table*}

\begin{figure*}[ht!]
    \centering
    \includegraphics[width=0.8\textwidth]{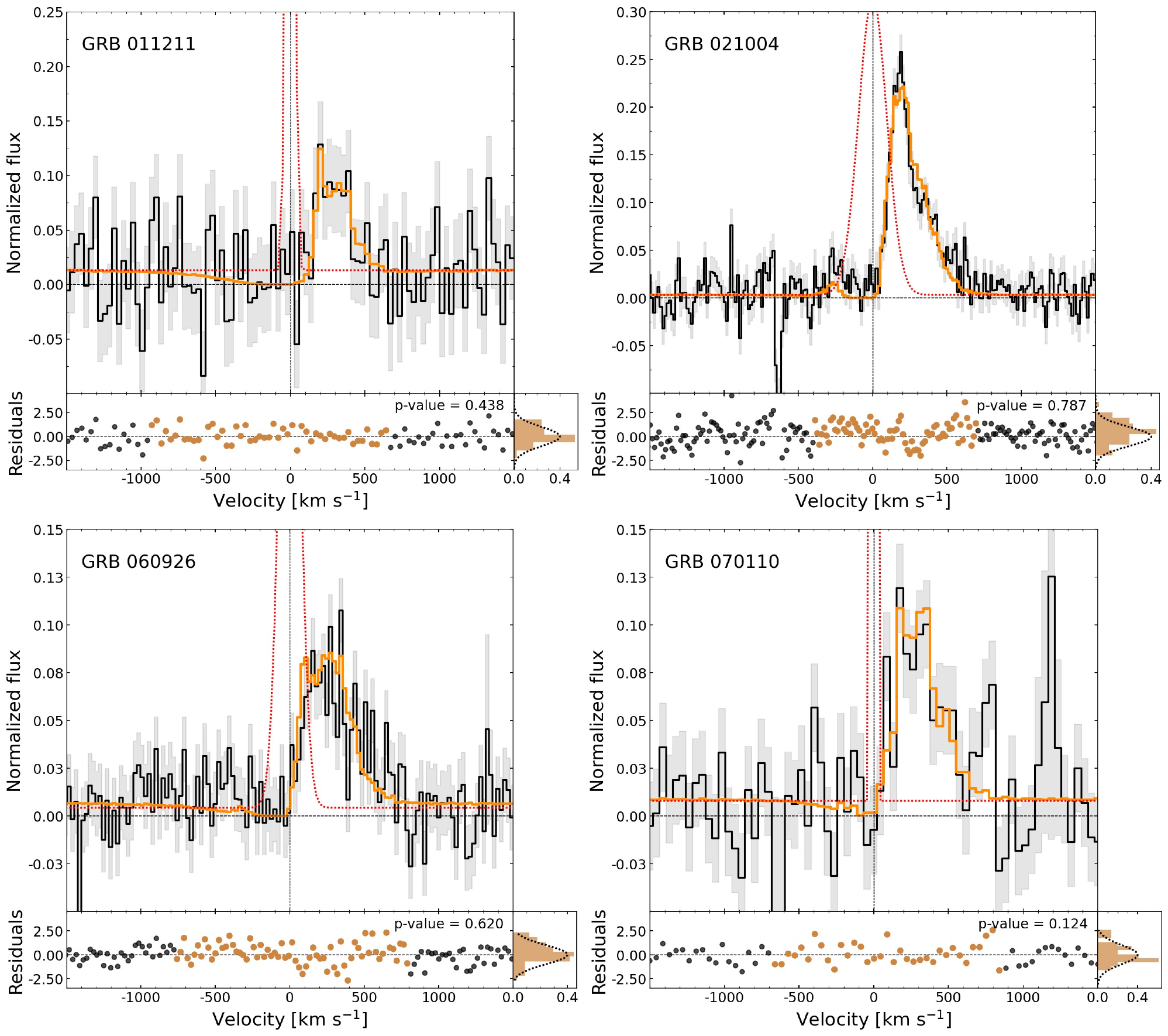}
    \caption{Same as Fig. \ref{unconst} but with the shell model parameters $\rm z^{HG}$, $\rm FWHM_i^{HG}$(\Lya) and $\rm EW_i^{HG}$(\Lya) constrained by the values determined from the observations of the GRB host galaxy.}
\end{figure*}

\newpage
\section{Results of the shell model with constraints from the GRB afterglow.} \label{shell_OA} 

\begin{table*}[ht!]
    \caption{Shell-model predictions with constraints on: $\rm z^{HG}$, \NHIOA, $\rm V_{LIS}^{OA}$ and $\rm \tau_d^{OA}$}
    \centering
    \footnotesize{
     \begin{tabular}{c | c c c c c c c }
         \hline\hline
          \rule[0.2cm]{0cm}{0.2cm}GRB host 	& $\rm \Delta z$ 	& $\rm log(N_{HI}cm^{-2})$        & $\rm V_{exp}$     & log(T/K)             & $\rm \tau_d$ & $\rm FWHM_i$(\Lya) & $\rm EW_i$(\Lya)  \\
         			& 			    &  & [km s$^{-1}$] &   &                & [km s$^{-1}$]               &  [\AA]               \\
         \hline
         \rule[0.2cm]{0cm}{0.2cm}GRB 011211 & $17^{+45}_{-45}$ & $20.2^{+0.1}_{-0.1}$ & $56^{+47}_{-28}$   &  $3.52^{+0.87}_{-0.52}$   & $0.65^{+0.51}_{-0.40}$  & $280^{+140}_{-165}$ & $66^{+60}_{-34}$  \\

        GRB 021004  						& $6^{+45}_{-45}$  & $19.6^{+0.1}_{-0.1}$  & $117^{+5}_{-4}$   &  $4.97^{+0.17}_{-0.15}$   & $0.362^{+0.103}_{-0.075}$  & $269^{+15}_{-20}$ & $162^{+16}_{-15}$  \\

        GRB 060926  						& $60^{+45}_{-45}$     & $21.5^{+0.1}_{-0.1}$        & $3^{+1}_{-1}$     &  $3.04^{+0.12}_{-0.15}$          & $0.003^{+0.004}_{-0.002}$  & $114^{+67}_{-44}$ & $22^{+8}_{-5}$  \\

        \rule[-0.2cm]{0cm}{0.2cm}GRB 070110 & $43^{+45}_{-45}$  & $21.3^{+0.2}_{-0.1}$ & $3^{+1}_{-1}$     &  $3.07^{+0.12}_{-0.17}$   &$0.018^{+0.010}_{-0.007}$  & $10^{+9}_{-4}$ & $62^{+17}_{-17}$  \\
        \hline
    \end{tabular}
    }
    \begin{center}
     \footnotesize{\textbf{Notes.} Same as Table \ref{tableComp_unconst}.}
    \end{center}
\end{table*}

\begin{figure*}[ht!]
    \centering
    \includegraphics[width=0.8\textwidth]{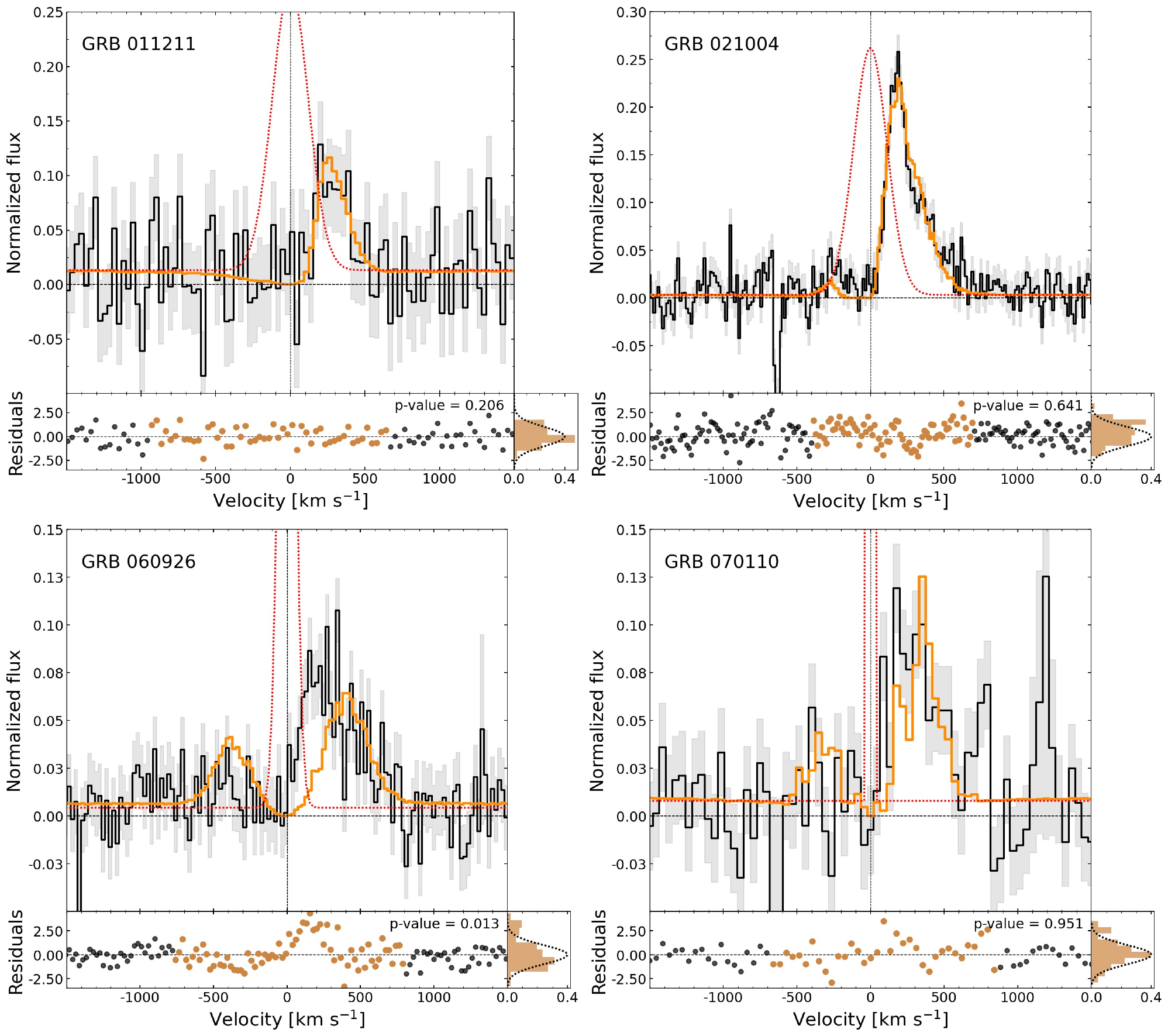}
    \caption{Same as Fig. \ref{unconst} but with the shell model parameters $\rm z^{HG}$, \NHIOA, $\rm V_{LIS}^{OA}$ and $\rm \tau_d^{OA}$ constrained by the values determined from the observations of the GRB afterglow (and systemic redshift from the observations of the host galaxy).}
\end{figure*}

\end{appendix}

\end{document}